\documentclass[12pt,a4paper]{article}

 \renewcommand{\baselinestretch}{1.4}

\usepackage{amsmath}
\numberwithin{equation}{section} 

 \newenvironment{myabstract}[1]{
 \renewcommand{\baselinestretch}{1.4}
 \vspace*{10mm} \large
 \begin{center} {\bf Abstract} \end{center}
 \begin{center} \parbox{14cm}{\small #1}}{\end{center}}

\begin{document}

 \renewcommand{\baselinestretch}{1.7}

 \title{\vspace{-20mm} Tensorial Quantum Gravity and the Cosmological Constant Problem}

 \author{Christophe R\'{e}al \\ 22, rue de Pontoise, 75005 PARIS \vspace{-3mm} \\
 \vspace{4mm}}

\date{November, 2007}

 \maketitle

\vspace{10mm}
 I dedicate this work to Odilia.

 \begin{myabstract}{

In the present article, which is the first part of a work in three
parts, we build an equation of quantum gravity. This equation is
tensorial, is equivalent to general relativity in vacuum, but
differs completely from general relativity inside matter. We can
spot directly in the equation the terms representing the
perturbative quantum corrections to classical gravity and the
nonperturbative quantum corrections. This new equation possesses a
dimensionless gravitational coupling constant, and passes all the
experimental tests that also passes general relativity, because
concerning these tests, the predictions of both theories are
identical. This quantum gravity and general relativity diverge
essentially in the domain of cosmology : we prove that quantum
gravity gives the solution to the whole set of problems left over
by the standard cosmological model based on general relativity.
Essentially we prove that the initial singularity, the big bang,
is smoothed out by quantum gravity, that the flatness problem
finds a precise solution : quantum gravity predicts that $\Omega$
should be just a little more than $1$, which fits perfectly with
the observed value, which is around $1.02$. The cosmological
constant problem also finds its solution since we prove that the
$\Lambda$ term does not come from any dark energy, but comes from
nonperturbative quantum corrections to classical relativity, and
has the exact tiny but strictly positive value needed.
Furthermore, the quantum equation of gravity possesses with no
further efforts features of unification. Indeed, our equation
governs at the same time the large scale of the universe, as did
general relativity, but also the structure of particles.

 }\end{myabstract}

\vspace{10mm}

\part{The quantum equation of gravity}
\section{Introduction}
\subsection{Successes and drawbacks of general relativity}
Would we like to make some comments very much at random on the
successes and drawbacks of general relativity, we first would
emphasize that on one hand this theory passed all experimental
tests, and on the other hand, from a theoretical point of view, we
notice that this theory resists much to being quantized by the
methods in vigor to quantize gauge theories, which describe the
three other interactions. It is especially well known that a
theory, to be renormalizable, must have a dimensionless coupling
constant and that this is not the case of general relativity.
\subsection{General relativity in vacuum}
By having a closer look at general relativity, we notice that the
experimental tests which are known to confirm this theory in fact
only test the equation $R_{ik}=0$, or equivalently only test the
theory in vacuum. The reason for this is that the calculations of
trajectories of massive objects placed in a gravitational field,
in the context of general relativity, use only the equation of
vacuum $R_{ik}=0$, eventually up to the computation of the masses
that generate this gravitational field. When we use only the
equation $R_{ik}=0$, these masses appear as constants of
integration. In other words, the calculation of the deviation of a
beam of light which passes near the sun, but still propagates in
vacuum, the calculation of the advance in the perihelion of the
planet Mercury, which propagates in vacuum too, the calculations
concerning gravitational waves propagating in vacuum and the
approximation to very small fields are all based on the sole
$R_{ik}=0$.
\subsection{General relativity inside matter}
So if the experimental tests designed to test general relativity
only test the vacuum equation $R_{ik}=0$, in which domain of
physics do we effectively use and verify experimentally the
equations of general relativity inside matter? Because general
relativity is a theory based on inertia, and because inertia can
only be defined in vacuum, only general relativity in vacuum is
needed to study physics in a gravitational field, once given this
field. But to compute the masses that generate this gravitational
field, we suppose that the celestial bodies are filled in by a
uniform energy density $\epsilon$, and we use the Schwarzschild
solution to compute these masses : this sole calculation, at the
same time, proves that Newton's theory is retrieved as a limit of
general relativity, and determines the coupling constant of
general relativity relatively to Newton's constant. So, concerning
this problem, if we use only general relativity in vacuum we are
able to do all what we can do with general relativity except that
we cannot compute the masses of the celestial bodies, and that we
cannot retrieve Newton's theory. Or more precisely, we can
retrieve Newton's theory, but we have to put by hand the masses in
our calculations. This problem of computing the masses will find
in our construction a solution directly from the calculation of
the masses of the fundamental particles, as in our second article
$[56]$. For the time being, we can leave this problem and look
towards the other domains of physics where the equations of
general relativity inside matter come into play. As far as black
holes are concerned, what happens inside a black hole is not
easily confronted to experiment. So in fact the only domain where
these equations inside matter are really used in physics and where
a possible confrontation to experiment can be made is cosmology.
So, when we arrive at the conclusion that we should use the
standard cosmological model to probe the equations of general
relativity inside matter, we should not forget either that this
must be done keeping in mind that these equations have never
passed so far any experimental test : we can be assured of general
relativity, but only in vacuum. We believe this is the key
explanation of why gravity has been so difficult to quantize : we
trusted general relativity inside matter because of its successes
in vacuum, but nothing prevents the existence of another theory,
identical to general relativity in vacuum, but differing from it
inside matter, and for example with a dimensionless coupling
constant. We now return to our main idea, that general relativity
inside matter should be experimentally checked by cosmology.
\subsection{The standard cosmological model}
The standard cosmological model possesses great successes and
generate great difficulties too. The computation of the abundances
of the elements, and of the cosmic background radiation, are
successes which seem to indicate that our universe comes from a
hot phase. We find the right abundances provided we suppose a
great deal of non baryonic dark matter in the universe, and
concerning the cosmic background, the behavior of the temperature
$T\sim1/a$ is a key argument to prove the observed perfect black
body behavior of this radiation. But the flatness problem, the
expansion problem, the problem of the initial singularity, the
cosmological constant problem, and the need for so much dark
matter to fit experiment are serious drawbacks of the model. We
emphasize here that the cosmological inflationary models (Guth
1980 $[21]$, Guth 1981 $[22]$, Kazanas 1980 $[30]$, Peebles 1993
$[49]$, Sato 1981 $[59]$, $[60]$), if they solve the flatness
problem, do not solve to our opinion the initial singularity
problem. This because the initial singularity problem is linked to
the beginning of the universe. The fact is that the origin of the
universe, the so called big bang, if it exists, needs a former
cause. And there can be no former cause of the beginning of the
universe. Now, this means exactly that the universe never began,
which is one of the conclusions of the inflationary models. But in
these models, the absence of a beginning is only due to one scalar
particle, and we believe that such a contingent reality cannot be
used to decide of the absence or existence of the beginning of the
universe. Furthermore, this scalar particle is necessarily inside
the universe and its existence is already one of the consequences
of the existence of the universe, so this particle cannot be held
to be the cause of the absence of the beginning of the universe if
it is at the same time a consequence of its existence. Said
differently, from a logical point of view, the absence or presence
of a beginning of the universe is former to its existence, which
is former to the existence of the particle, which for these
reasons cannot be former to the absence of the beginning. In
conclusion, we need to prove the absence of beginning by a
fundamental law of nature, we mean by quantum gravity.
\subsection{Supersymmetry, string theory and dualities}
On the theoretical side, general relativity has become disastrous
when physicists have begun to quantize it. The gravitational
coupling constant is not even dimensionless. Supersymmetry, string
theory and all its avatars born from the dualities, have as a
common origin, the necessity to eliminate all the infinities left
over by the quantization of general relativity. These theories
also have two other common features. First they are very
speculative and it is difficult to check their predictions by
experimental tests, and second they are mathematically involved,
especially because they consist in a try of a mathematical
unification of the interactions. But in this context, unification
has to be made like at random, exactly because experiment cannot
guide us anymore, nor correct our hypotheses. The same happens
with dark matter, some particles are predicted for example by
supersymmetry which are not directly observed, and we have to hope
that these particles will fill the gap between the known baryonic
matter and the calculated total energy density in the universe.
But still, from the hypothesis of the existence of these particles
coming from supersymmetry to the calculation of the density of non
baryonic dark matter, there are large regions of theoretical
reasonings in which experiment cannot enter, as it should, to make
physics go upon a much more secure path.
\subsection{A tensorial equation of quantum gravity}
As we said general relativity has misled us inside matter because
of its successes in vacuum, in such a way that we did not see its
failures. In fact, thinking of the equations of general relativity
inside matter, we immediately notice a specificity of this theory
: as it links the curvature tensor coming from gravity to the
momentum-energy tensor of matter, included the momentum-energy
tensor of electromagnetism, the theory contains necessarily
implicit hypotheses on unification itself. We mean information
about the way the different interactions behave themselves in
respect to each other, and how they operate on fermions. And if
the tensorial equation of gravity goes so far, it probably
contains implicit hypotheses on quantum gravity itself. And this
because the implicit hypotheses made on unification we were just
talking about will have no chance to be right if even quantum
gravity is not taken into account. Now this has the following
consequence : There must exist a tensorial equation describing not
only classical gravity, but also its quantum corrections, so there
exists a tensorial equation of quantum gravity which has
furthermore all probabilities to lead us directly to unification.
\subsection{Quantum gravity built out of experiment}
Contrarily to what happens in string theory, we can build the
equation of quantum gravity and check at each step of the
construction that it is compatible with experiment. First, if the
experimental tests of general relativity only test the equation in
vacuum, we can construct this quantum theory out of general
relativity by changing only the theory inside matter, in such a
way that both theories coincide in the vacuum limit
$\epsilon\rightarrow0$. This way quantum gravity will pass the
same experimental tests that passed general relativity.
Furthermore we know that this quantum gravity should also predict
the existence of a hot universe, and in some way should keep
unchanged the standard calculations for the abundances of the
elements and the cosmic background radiation. This way, all the
successes of the standard cosmological model will be preserved. We
also have the constraints and indications coming from the quantum
regime of the other interactions. So we should construct a quantum
equation of gravity which possesses a dimensionless coupling
constant, and we should be able to read on this tensorial equation
the perturbative and non perturbative quantum corrections to
classical gravity. We also use the fact that cosmology is the
preferred domain of application of tensorial gravity inside
matter, and we now use at our advantage the drawbacks of the
standard cosmological model. For this we adopt a minimal principle
: the quantum equation of gravity should solve at once all the
cosmological problems left over by general relativity with no
additional hypothesis such as dark energy or non baryonic dark
matter. So our simple idea, which turns out to become a method of
investigation, is to construct quantum gravity as a tensorial
equation, by simply correcting general relativity inside matter,
such that all the successes of general relativity are conserved,
and all the problems left over by this theory are automatically
solved, with no additional hypothesis. This way, all experimental
observations made in the domain of cosmology turn out to be
experimental indications on how quantum gravity should be built,
and we can go further upon this secure path which later will lead
us to a unified theory.
\subsection{The specificity of quantum gravity}
Once we make the necessary quantum  corrections to classical
general relativity, we find that as we had predicted, this quantum
gravity will automatically contain unification. Indeed we prove in
the present article that the quantum equations automatically
determine the value of the pressure $p$ which itself controls the
structure of matter, relativistic or non relativistic. For example
the quantum equation predicts for the early universe the value
$p=\epsilon/3$ which has to be put by hand in the standard
cosmological model. So the quantum equation has a first essential
feature of unification : it governs at the same time the large
scale and small scale of the universe.
\subsection{The predictions of quantum gravity}
At the same time, we are going to see in the present article how
the quantum equation automatically gives a solution to the
following cosmological problems. First the flatness problem :
observations tend to give us a value of $\Omega$ around $1$, but
more precisely $\Omega=1.02\pm0.02$ (Bennett and al. 2003 $[2]$).
Quantum gravity predicts a value $\Omega\approx1$ but also
$\Omega>1$. So not only gives quantum gravity the right value of
$\Omega\approx1$, but also it gives account for the fact that
$\Omega$ has been observed to be a little greater than $1$.
Quantum gravity gives a solution to the expansion problem, because
it predicts that $\ddot{a}>0$. Finally quantum gravity gives a
solution to the cosmological constant problem. It gives the origin
of the $\Lambda$ term in the equation : it is the term
corresponding to the non perturbative quantum corrections to
classical gravity, and also it predicts that this term acts as if,
put by hand in the equation as it is in general relativity, it
makes us see the total energy density of matter in the universe,
the apparent energy density $\epsilon_{app}$, multiplied by the
factor $4$, compared to what this energy density $\epsilon$ really
is. We thus have the relation $\epsilon_{app}=4\epsilon$. This
factor $4$ fits strikingly with the observations of Bennett and
al. 2003, $[2]$, which give $\Omega=1.02\pm0.02$ and
$\Omega_{\Lambda}=0.73\pm0.04$. In fact, we have to notice that
these observations are model dependent. Furthermore, in quantum
gravity, the energy density $\epsilon$ and the term $\Omega$ are
not proportional, and we also find :
\begin{equation}\Omega_{\Lambda}=\frac{1}{2}\Omega\end{equation}
which still solves the cosmological constant problem anyway, since
it gives for the value of $\Lambda$ the strictly positive and
incredibly tiny value we were looking for. As a final remark, we
add that quantum gravity, in fact, leads to an entire new class of
cosmological models, quite near each other, and which all display
the same general quantum features. It is then easy to change the
coefficient $1/2$ for another factor, provided we go from a
quantum model to another.
\subsection{Computing the masses}
Our quantum theory in vacuum is equivalent to general relativity
in vacuum. We said in section 1.3 that this situation makes
quantum gravity and general relativity pass the same experimental
tests, except that to retrieve Newton's theory with quantum
gravity, we have to put the masses by hand. In fact, this will
turn at the advantage of quantum gravity, because general
relativity is helpless anyway to compute the masses of the
fundamental particles, whereas quantum gravity will lead to the
web of formulas which were needed to compute the parameters of the
standard model of particle physics. This will be the subject of
our second article, and we just explain here rapidly how these
things are working. To describe a celestial body, there are three
models, the classical, the quantum and the intermediate. The
classical model is adapted to tensorial equations and thus to
general relativity : celestial bodies are made of continuous,
uniform energy densities. The quantum model contradicts by itself
tensorial equations and especially general relativity because
celestial bodies are made of pointlike particles, and the fields
appearing in Einstein's equations can only take two extreme values
: zero quasi everywhere, and infinity at a finite number of
points, where particles are. Clearly we need a third and
intermediate model : a particle is a sphere of very small radius
$r$ and huge energy density $\epsilon$. The intermediate model is
powerful enough to contain both other models. The classical model
is only a approximation of the intermediate model which is far
more precise. This intermediate model contains the quantum model
as a limit, when $r\rightarrow0$, $\epsilon\rightarrow+\infty$,
the mass of the particle being kept fixed. Our conclusion is thus
the following : we can suppose that as far as general relativity
inside matter is concerned, gravity couples to matter only inside
particles. So it appears that in the context of our quantum
equation of gravity, the only masses that we really have to put by
hand are the masses of the particles. We will see in $[56]$ how
they can be computed, but we give here an example of how, from the
point of view of unification, general relativity can be improved
by quantum gravity to make possible the calculation of the
parameters of the standard model.
\subsection{A dimensionless quantum gravitational coupling around unity}
We know that as far as the three other interactions are concerned,
they are governed by gauge theories, and the coupling constants
are dimensionless and around unity, which makes arise the question
to know why gravitation is so tiny. The equations of general
relativity inside matter, we mean their right hand side, possess a
relativistic gravitational coupling constant $\kappa$ which is
linked to Newton's constant $G$ by the formula :
\begin{equation}\kappa=\frac{8\pi G}{c^{4}}\end{equation}
Now we recall that after renormalization, the coupling constants
of the other interactions are not only dimensionless and around
unity, but also depend on energy. So if we suppose for an instant
that the quantum gravitational coupling also depends on energy as
it should, in such a tensorial equation it would depend on the
energy density. Since we look for a dimensionless constant we take
\begin{equation}\kappa=\frac{8\pi
G}{c^{4}}=\frac{\kappa_{0}}{\sqrt{\epsilon}}\end{equation} Then
appear two facts. First $\kappa_{0}$ has become dimensionless, and
second and above all the value of $\kappa_{0}$ should be around
unity. Indeed, we proved in section 1.10 that we can place
ourselves in the case of the intermediate model since this model
contains the others, and furthermore we proved that in this model
gravity couples to matter only inside particles, so the value of
$\epsilon$ in (1.3) should be taken to be the energy density
$\epsilon_{P}$ inside particles. Now what could be the energy
density inside a particle? It should be the greatest energy
density we can think of, which is the energy density corresponding
to the Planck mass in every sphere of radius the Planck length. We
thus find
\begin{equation}\epsilon_{P}=\frac{3M_{P}}{4\pi
L_{P}^{3}}\approx\frac{1}{L_{P}^{4}}=G^{-2}\end{equation} by the
definition of $\epsilon_{P}$, where we left the coefficient
$4\pi/3$ of the volume of a sphere because it is around unity,
where $M_{P}=1/L_{P}$, with $M_{P}$ and $L_{P}$ respectively the
Planck mass and the Planck length. So our conclusion is that with
this value of $\epsilon_{P}$, the quantum coupling constant of
gravity becomes dimensionless, and we have $\kappa_{0}\approx
G\sqrt{\epsilon_{P}}\approx1$, so $\kappa_{0}$ is around unity as
the coupling constants of the other interactions.
\subsection{Tautological signature of unification}
What will appear constantly in $[56]$ and $[57]$, in which we deal
with unification, is that unification has a signature, a mark,
which appears every time it appears, and this mark is tautology.
Here, we just give an example of it. Why is gravitation so tiny?
In fact we should ask why $G$ is so tiny, and from (1.3) with
$\kappa_{0}$ around unity, the question is to know why
$\epsilon_{P}$ is so great. Now considering the masses of the
particles as fixed quantities, $\epsilon_{P}$ is so great because
the radiuses of the particles are so small. Why now the radiuses
of the particles are so small? They appear to us so small because
we are so big compared to them. And we are so big compared to them
simply because we are made of them. Particles are by definition
the smallest pieces of matter, and unification makes appear that
they are effectively very small.
\subsection{Dimensionless cosmology}
If the second part of this work $[56]$ will be devoted to
unification and the calculation of the parameters of the standard
model, in the third part $[57]$ we enter into dimensionless
physics where all units are eliminated. We prove that if the
quantum equation of gravity permits to solve the cosmological
problems left over by general relativity, as it is proved in the
present article, dimensionless cosmology also permits not to loose
the beautiful achievements of the standard cosmological model
which are the calculation of the abundances of the elements and
the explanation of the cosmic background radiation. In fact we can
even improve the standard calculation in two ways : first we can
improve the computed value of the mass fraction $Y\approx0.28$ of
helium and push it into the observed region $0.22\leq Y\leq0.23$
(Pagel and al, 1994 $[47]$). Second, we are able to describe a
method which permits to relax the condition on the energy density
of baryonic dark matter, and this way, we can relax the hypothesis
of the existence of non baryonic dark matter, as well as we can
relax the condition that there are no more than three families of
particles.
\section{The complete deduction of the equation}
\subsection{Introduction} We now apply our ideas to construct
quantum gravity. Applying the minimal principle, we consider all
observations of the cosmos as indications of the tensorial
equation inside matter. Then, we explain how can be constructed
this equation. We will see that considering the experimental data
coming from the observation of our universe, and also the
necessary conditions imposed by the fact that the quantization of
gravity has to obey certain rules that renders it coherent with
the quantization of the other interactions, is far enough to
deduce the rules of tensorial quantum gravity itself.
\subsection{Perturbative corrections}
\subsubsection{Conditions for quantum gravity}
So we first know that quantum gravity should smooth out the
initial singularity. But it has a second and a third conditions to
fulfill, this because as any other interaction, it has to follow
some characteristic features of renormalization. The second
condition is that its coupling constant has to be dimensionless.
Furthermore we know that once renormalized, the coupling constant
of any interaction depends on energy. All the same, the third
condition on the gravitational constant of gravitation is that it
should vary with energy in the quantum regime.
\subsubsection{Dependence of the gravitational constant on energy}
So quantum gravity has to fulfill the former three conditions. In
the early times of the universe, the energy density grew to
$+\infty$. We suspect that a gravitational constant, which is a
sufficiently rapidly decreasing function of energy, and such that
\begin{equation}\lim_{\epsilon\rightarrow+\infty}G(\epsilon)=0\end{equation}
smoothes out the initial singularity. An explanation for this is
the picture of a big crunch : if the universe shrinks to a point
in finite time, we can suspect gravitation to be the cause of this
singularity, because it is an attractive interaction. When the
universe shrinks to a point, $\epsilon\rightarrow+\infty$, and the
former equation (2.1) makes gravitation disappear. We thus expect
the big crunch singularity to be smoothed out. By analogy, we
expect in this case the big bang to be smoothed out too. The
precise proof of this fact will have to wait until we possess the
complete set of equations of quantum gravity. This fact will be
proved rigorously at that time. Now, if we take $\hbar=c=1$, in
the equation
\begin{equation}8\pi
G(\epsilon)=\kappa(\epsilon)=\frac{\kappa_{0}}{\sqrt{\epsilon}}\end{equation}
the new constant $\kappa_{0}$ is dimensionless, and our three
conditions are likely to be satisfied. To keep the most general
equation, we can eventually allow $\kappa_{0}$ to depend on
$\epsilon$ too, to take into account further perturbative
corrections, in this case we keep the notation $\kappa(\epsilon)$.
\subsection{Nonperturbative quantum corrections}
\subsubsection{Conservation of energy}
There is, due to the conservation of energy, a relation of some
kind between perturbative and nonperturbative corrections in
tensorial quantum gravity, as we can see as soon as we write down
the equations of general relativity with a constant of gravitation
depending on $\epsilon$ :
\begin{equation}R_{ik}-\frac{1}{2}Rg_{ik}=\kappa(\epsilon)T_{ik}\end{equation} The
left hand side of this equation, the Einstein's tensor, fulfills
the conservation of energy. Also does the energy-momentum tensor
of matter $T_{ik}$, and we can apply the covariant derivative to
this equation, and contract it with one index $i$, to prove that
$\kappa(\epsilon)$ is independent of $\epsilon$. So, if the
gravitational coupling constant really depends non trivially on
energy, the only possibility is that we missed a term somewhere.
This term corresponds precisely to nonperturbative corrections.
\subsubsection{Nonperturbative corrections}
Looking at what happens in the case of the other interactions, we
see that in these cases there is also nonperturbative terms called
instantons which are obtained by adding a topological term to the
lagrangian. So our entire equation should also contain a
topological term to fulfill all the requirements of an equation
describing the complete quantum regime of gravity. The only
topological term in dimension four is the Gauss-Bonnet term. We
emphasize here that we are not looking for truly gravitational
instantons, but we only retain the idea that we need a topological
term to insert into our equation. We call $\tilde{\Sigma}$ the
Gauss-Bonnet term and define
$\tilde{\Sigma}=4\tilde{\tilde{\Sigma}}$ which will simplify the
calculations : \begin{equation}\tilde{\Sigma} = R^{(4)} -4
R^{(2)}+ R^{2}=4\tilde{\tilde{\Sigma}}\end{equation} where
$R^{(4)}=R^{abcd}R_{abcd}$ and $R^{(2)} =R^{ab}R_{ab}$, $R_{abcd}$
being the Riemann curvature tensor, $R_{ab}$ the Ricci tensor, and
$R$ the scalar curvature. So, putting this term on the left on our
equation we obtain :
\begin{equation}
R_{ik}-\frac{1}{2}Rg_{ik}+\Lambda g_{ik} =\kappa(\epsilon)T_{ik}
\end{equation}
with :
\begin{equation}
\Lambda =
-\theta\tilde{\tilde{\Sigma}}+\Lambda_{0}=-\frac{\theta\tilde{\Sigma}}{4}+\Lambda_{0}
\end{equation}
$\theta$ being constant. The constant $\Lambda_{0}$ is here
inserted for completeness, to write down the most general
equation. When resolving the equation, we will adopt one strategy.
We will impose certain conditions on the solutions and choose
$\Lambda_{0}$ to obtain them. This way $\Lambda_{0}$ will appear
as a kind of constant of integration. This way we will be able to
study and test the general features of our quantum equation. At
the end, we will leave all these models and impose the better
condition $\Lambda_{0}=0$. Now, applying the covariant derivative
on equation (2.5) we see that $\kappa(\epsilon)$ has to depend on
$\epsilon$ because $\tilde{\Sigma}$ is not constant. Exactly, the
conservation of energy gives a precise relation between
perturbative and nonperturbative quantum gravitational
corrections.
\subsubsection{A problem}
We can now write a tensorial equation of quantum gravity inside
matter, but a problem appears. If the gravitational constant
depends on energy, why has this fact never been detected by
experiments in the solar system or even on earth? This new form of
the gravitational constant should change drastically even the
approximate Newton's law. We emphasize that there are numerous
solutions to this problem, and that these solutions all give very
different physics. We have studied them at length in $[58]$, and
we have proved that an answer to this problem necessarily involves
more fundamental principles about unification. We will not need
this complete study in the present work. Here we only consider the
case where the gravitational coupling is spatially constant in the
whole universe, and varies in time proportionally to
$1/\sqrt{\epsilon}$, where $\epsilon$ is the mean energy density
in the universe, as in any other cosmological model with varying
$G$. Here we precise that $\epsilon$ is a global parameter which
is spatially constant, because it is the mean value over large
parts of space of the local parameter $\epsilon$. When, as we do
here, we take for $\epsilon$ the global parameter, the problem of
retrieving Newton's theory and general relativity from our quantum
equation of gravity disappears, our gravitational coupling is
spatially constant, and varies very little with time. So our
quantum equation is very near general relativity, and we do not
need all the technicality of the intermediate model, of the
different pictures for particles, and of the assertions on how far
we can go using only general relativity in vacuum. When $\epsilon$
in the quantum equation is taken to be the local parameter, on the
contrary, all this technicality is necessary, and as we proved in
the introduction, from this point of view, unification looks
completely different. We need then averaging procedures to go from
the equation with local parameters to the equation with global
parameters. We simply remark here that with the equation with
local parameters, the gravitational coupling constant varies
spatially. However, one conclusion made in $[58]$ is that there is
only very little room to allow in our theory spatial variations of
the gravitational coupling constant which could explain the
phenomenon of dark matter in the halos of galaxies by theoretical
means.
\subsection{Generalizing the equation}
\subsubsection{Interpretation of the coefficient in front of the Gauss-Bonnet term}
We called $\theta$ the constant in front of
$\tilde{\tilde{\Sigma}}$, and considerations of dimensions show
that $\theta$ has dimension $[L]^{2}$, where $[L]$ is a length.
This way $\sqrt{\theta}$ has the same dimension as $[L]$. We will
see that after the resolution of (2.5), (2.6), with constant
$\theta$, we find the relation :
\begin{equation}
\theta = \frac{c}{2H}
\end{equation}
which gives, using the present value of $H$, the value :
\begin{equation}
\sqrt{\theta} \cong 1250 Mpc
\end{equation}
So what can be such a huge length in our universe and what is its
physical interpretation? It appears that such a distance can only
be the greatest distance possible, that is, up to a constant of
the order of unity, the radius of the universe itself, but
considered in another model or equation. Another equation, because
in this case, the equation is changed, and the constant term
$\theta$ is replaced by the term $\theta_{0}a^{2}$, where
$\theta_{0}$ is constant and $a$ is the radius of the universe.
\subsubsection{The generalized equation}
To write down the most general equation, we suppose that
$\theta(a)$ is a function of the radius $a$, eventually constant
or proportional to $a^{2}$ itself, but a priori yet undetermined.
The dependence of $\theta$ on $a$ can also be indirect : $\theta$
can depend on $a$ only because it depends on $\epsilon$. Using
dimensions, in this case $\theta$ should be proportional to
$1/\sqrt{\epsilon}$ which just changes a little its behavior,
compared to $\theta\sim a^{2}$. So we begin to notice the
emergence of a entire new class of cosmological models, which are
quite near from each other, with only subtle differences. All
these models will prove themselves to display the same general
features, which could be called quantum features, and the
differences will be numerous enough to make easy the task to find
between all these possibilities, one or several models which fit
with all observed data. Mathematically and for the time being, we
still can write our $\theta$ as a function $\theta(a)$ and we
finally write our most general equation :
\begin{equation}
R_{ik}-\frac{1}{2}Rg_{ik} -\theta(a)
\tilde{\tilde{\Sigma}}g_{ik}=\kappa(\epsilon)T_{ik}
\end{equation}
and :
\begin{equation}
\Lambda=-\theta(a)\tilde{\tilde{\Sigma}}+\Lambda_{0}=-\theta(a)\frac{\tilde{\Sigma}}{4}+\Lambda_{0}
\end{equation}
\subsection{Restoring h and c in the quantum gravitational coupling}
Using (2.9), and since we also know that $T_{0}^{0}=\epsilon$, we
see that :
\begin{equation}\left(R_{0}^{0}-\frac{1}{2}R\right)-\theta\tilde{\tilde{\Sigma}}+\Lambda_{0}=\kappa(\epsilon)\epsilon\end{equation}
We have
$$R_{0}^{0}-\frac{1}{2}R\sim [L]^{-2}$$ where the last symbol means
an equality of dimensions, and where $[L]$ is a length. Of course
we have $[L]\sim c[T]$, where $c$ is the speed of light and $[T]$
is a time. Now $\epsilon [L]^{3}$ is an energy $[E]$, and :
$$\epsilon [L]^{3}\sim [E] \sim \frac{\hbar}{[T]} \sim
\frac{\hbar c}{[L]}$$ so $$\epsilon \sim \frac{\hbar c}{[L]^{4}}$$
At the same time, the left hand side of equation (2.11) has
dimension $[L]^{-2}$, and this is why the whole term on the right
hand side should be proportional to $\sqrt{\epsilon}/\sqrt{\hbar
c}$ in order to give us the dimension :
$$\frac{\sqrt{\epsilon}}{\sqrt{\hbar c}}\sim[L]^{-2}$$ We see that the
coupling $\kappa(\epsilon)$ take the form :
\begin{equation}
\kappa(\epsilon)=\frac{\kappa_{0}}{\sqrt{\hbar c}\sqrt{\epsilon}}
\end{equation}
with $\kappa_{0}$ a dimensionless real number. We will still note
this constant $\kappa_{0}$ when considering $\hbar=c=1$.
\part{The basic set of equations in tensorial quantum gravity}
\section{The quantum equations in cosmology}
\subsection{Introduction}
We now take the quantum equation of gravity in order to apply it
to the Robertson-Walker metric, and find this way the predictions
of quantum gravity concerning cosmology. In this part, we make all
necessary computations to write down the complete set of quantum
equations relating the cosmological parameters, and to prove that
the quantum equation implies by itself the conservation of
entropy. We recall that the quantum equation of gravity takes the
form :
\begin{equation}
R_{ik}-\frac{1}{2}Rg_{ik}+\Lambda g_{ik} =\kappa(\epsilon)T_{ik}
\end{equation}
with :
\begin{equation}
\Lambda = -\theta\tilde{\tilde{\Sigma}}+\Lambda_{0}
\end{equation}
$\theta$ can be constant or can depend of $a$, $\Lambda_{0}$ is
left here for the moment to test solutions, even if we will take
it equal to zero at the end. $\tilde{\tilde{\Sigma}}$ is
proportional to the Gauss-Bonnet term, and equal to :
\begin{equation}
\tilde{\tilde{\Sigma}} = \frac{1}{4}\left(R^{(4)} -4 R^{(2)}+
R^{2}\right)
\end{equation}
where $R^{(4)}=R^{abcd}R_{abcd}$ and $R^{(2)} =R^{ab}R_{ab}$,
$R_{abcd}$ being the Riemann curvature tensor, $R_{ab}$ the Ricci
tensor, and $R$ the scalar curvature.
\subsection{The Robertson-Walker metric, notations}
\subsubsection{The hypothesis of isotropic and homogeneous space}
We first stick to the closed model, we shall see in section 3.5
how can be deduced the equations of the open model from the closed
one. As in any other cosmological model, we make the hypothesis of
isotropic and homogeneous space. We know that this hypothesis
implies that all our variables are averaged ones over large parts
of space. We use coordinates such that free falling matter is at
rest and $t$ is the physical time given by physical free falling
clocks. Our usual unknowns are the radius of the universe $a(t)$,
the pressure $p(t)$ and the energy density $\epsilon(t)$. We let
$\kappa(\epsilon)$ undetermined to test different behaviors of our
solutions. We now write down all results found in Landau, $[33]$,
paragraphs 112 and 113, that will be of use in our present
computation. We use the value of $ds^{2}$ coming from the
Robertson-Walker metric :
\begin{equation}
ds^{2} = c^{2}dt^{2} - a^{2}(t) (d \chi ^{2} + \sin^{2} \chi (d
\theta^{2} + \sin^{2}\theta d\phi^{2}))
\end{equation}
where $r, \theta, \phi$ are the variables of the spherical
coordinates in three dimensions and where $r = a(t) \sin \chi$,
$\chi$ varying from $0$ to $\pi$. Further, we can replace the time
variable $t$, by the dimensionless variable $\eta$, defined by :
\begin{equation}
cdt = a d\eta
\end{equation}
Here we take the convention that an expression as $a'$ means an
$\eta$-derivative and an expression as $\dot{a}$ means a
$t$-derivative. We then obtain :
\begin{equation}
ds^{2} = a^{2}(\eta) (d\eta^{2} - d \chi ^{2} - \sin^{2} \chi (d
\theta^{2} + \sin^{2}\theta d\phi^{2})).
\end{equation}
We write our equations with variables $x^{0}, x^{1}, x^{2}, x^{3}$
being $\eta, \chi, \phi, \theta$. We have from the previous
equation :
\begin{equation}
g_{00} =a^{2};
g_{11}=-a^{2};g_{22}=-a^{2}\sin^{2}\chi;g_{33}=-a^{2}\sin^{2}\chi\sin^{2}\theta
\end{equation}
all non diagonal terms of $g_{ik}$ vanishing.
\subsection{Values of tensors}
We choose to use greek indices to denote space indices varying
from $1$ to $3$, and latin indices to go from $0$ to $3$. We
compute in Part VII the components of the Ricci tensor. These can
also be found in $[33]$  :
\begin{equation}
R_{\alpha 0} = 0
\end{equation}
and :
\begin{equation}
R_{0}^{0} = \frac{3 (a'^{2} - aa")}{a^{4}} = b
\end{equation}
which defines $b$, as well as :
\begin{equation}
R_{\beta \delta} = - \frac{1}{a^{4}} \left(2 a^{2} + a'^{2}
+aa"\right) g_{\beta \delta} = c g_{\beta \delta}
\end{equation}
which defines $c$. Now the scalar curvature :
\begin{equation}
R = b + 3 c = - \frac{6}{a^{3}}(a+a")=d
\end{equation}
which defines $d$. In Parts VI and VII we define the tensor :
\begin{equation}
\Sigma_{ik}={\tilde{\Sigma}}_{ik}-\frac{1}{4}\tilde{\Sigma}g_{ik}
\end{equation}
where :
\begin{equation}
{\tilde{\Sigma}}_{ik}={R_{i}}^{abc}R_{kabc}- 2 R_{iakb}R^{ab} - 2
R_{ia}{R_{k}}^{a} + R_{ik}R
\end{equation}
and where $\tilde{\Sigma}$ is the trace of
${\tilde{\Sigma}}_{ik}$. We have the relations
$$\tilde{\Sigma}=R^{(4)}-4R^{(2)}+R^{2}$$ and
$$\tilde{\tilde{\Sigma}}=\frac{1}{4}\tilde{\Sigma}$$
In this Part VII we compute :
\begin{equation}
{\tilde{\Sigma}}_{\alpha 0} = 0
\end{equation}
we also compute :
\begin{equation}
{\tilde{\Sigma}}_{00}=\frac{b}{3}(3c-b)g_{00}
\end{equation}
and :
\begin{equation}
{\tilde{\Sigma}}_{\alpha \beta}=\frac{b}{3}(3c-b)g_{\alpha \beta}
\end{equation}
Finally, we also prove in Part VII that ${\tilde{\Sigma}}_{ik}$ is
diagonal and that $\Sigma_{ik}$, being at the same time diagonal
and of vanishing trace, verifies $\Sigma_{ik}=0$ in the case of
the homogeneous and isotropic model. This was known from
topological arguments, and the arguments we give in Part VI. This
situation is especially interesting because it acts as a
theoretical check of all the calculations, in Part VII, which lead
to (3.14), (3.15) and (3.16). We straightforwardly deduce :
\begin{equation}\tilde{\Sigma}=4\frac{b}{3}(3c-b)\end{equation}
and
\begin{equation}\tilde{\tilde{\Sigma}}=\frac{b}{3}(3c-b)\end{equation}
\subsection{The equations}
\subsubsection{The equations for the Gauss-Bonnet term}
We start from (3.1) :
$$R_{ik}-\frac{1}{2}g_{ik}+\Lambda g_{ik}=\kappa(\epsilon)
T_{ik}$$ and make operate a covariant derivative, as as in the
conservation of energy :
\begin{equation}
\nabla^{i}\left(R_{ik}-\frac{1}{2}g_{ik}\right)+\partial_{k}\Lambda=\partial_{i}\kappa(\epsilon){T^{i}}_{k}+\kappa(\epsilon)
\nabla^{i}T_{ik}
\end{equation}
and we are left with :
\begin{equation}
\partial_{k}\Lambda = \partial_{i}\kappa(\epsilon){T^{i}}_{k}
\end{equation}
but we have $\partial_{\alpha}\epsilon=0$ so
$\partial_{\alpha}\kappa(\epsilon) = \kappa'(\epsilon)
\partial_{\alpha}\epsilon=0$. As well ${T^{i}}_{k}$ is diagonal,
and $T^{0}_{0}=\epsilon$. So we find :
\begin{equation}
\dot{\Lambda}=\kappa'(\epsilon)\epsilon \dot{\epsilon}
\end{equation}
and :
\begin{equation}
\partial_{\alpha}\Lambda = \partial_{\alpha}\kappa = 0
\end{equation}
Reading the value of ${\tilde{\Sigma}}_{ik}$ in (3.15) and in
(3.16), we can write :
\begin{equation}
{\tilde{\Sigma}}_{k}^{i}=\frac{b}{3}(3c-b)\delta_{k}^{i}
\end{equation}
Now using (3.2) and (3.18) we have :
\begin{equation}
\Lambda=\theta \frac{b}{3}(b-3c)+\Lambda_{0}
\end{equation}
\subsubsection{Computation of b and c} Using the usual $\hbar=c=1$,
from (3.5) it appears that $a'=a\dot{a}$ and also that
$a"=a^{2}\ddot{a}+a{\dot{a}}^{2}$. Using (3.9), it yields :
\begin{equation}
b=\frac{-3\ddot{a}}{a}
\end{equation}
It is also straightforward to compute $c$ : from (3.10), using the
former relations for $\dot{a}$ and $\ddot{a}$, we find :
$$R_{\alpha}^{\alpha}=c=-\frac{1}{a^{4}}\left(2a^{2}+a'^{2}+aa"\right)=-\frac{1}{a^{4}}\left(2a^{2}+2a^{2}{\dot{a}}^{2}+a^{3}\ddot{a}\right)$$
and we have :
\begin{equation}
c=-\left(\frac{2}{a^{2}}+2\left(\frac{\dot{a}}{a}\right)^{2}+\left(\frac{\ddot{a}}{a}\right)\right)
\end{equation}
\subsubsection{The equations of the quantum theory} We also have another
equation coming from the conservation of entropy in the universe,
which is computed in $[33]$, and takes the form :
\begin{equation}
\dot{\epsilon}=-3(\epsilon+p)\frac{\dot{a}}{a}
\end{equation}
As far our equations are concerned, the first equation of movement
is :
\begin{equation}
\left(R_{0}^{0}-\frac{1}{2}R\right)+\Lambda=\kappa(\epsilon)
T_{0}^{0}=\kappa(\epsilon)\epsilon
\end{equation}
with :
$$R_{0}^{0}-\frac{1}{2}R=b-\frac{1}{2}(b+3c)=\frac{b-3c}{2}$$
so we finally obtain :
\begin{equation}
\frac{1}{2}(b-3c)+\Lambda =\kappa(\epsilon)\epsilon
\end{equation}
We have as well a second equation of movement which is :
\begin{equation}
\left(R_{\alpha}^{\alpha}-\frac{1}{2}R\right)+\Lambda
=\kappa(\epsilon)T_{\alpha}^{\alpha}= -\kappa(\epsilon)p
\end{equation}
with :
$$R_{\alpha}^{\alpha}-\frac{1}{2}R=c-\frac{1}{2}(b+3c)=-\frac{b+c}{2}$$
so we finally get :
\begin{equation}
\frac{b+c}{2}-\Lambda =\kappa(\epsilon)p
\end{equation}
\subsection{From the closed model to the open model} To deduce the
equations for the open model from the closed one, we need to apply
the rules given in Landau, $[33]$, paragraph 113. These rules
state that to go from closed to open, we have to replace $\eta,
\chi, a$ by $i\eta, i\chi, ia$, and since we also have
$cdt=ad\eta$, these rules also imply to replace $t$ by $-t$, which
in particular must be done in time derivatives. In other words
each time derivative must be affected by a extra minus sign.
Looking at the values of $b$ and $c$ we have just established, it
is clear that $b$ remains unchanged :
\begin{equation}b=\frac{-3\ddot{a}}{a}\end{equation}
Indeed $\ddot{a}$ is multiplied by $i$ in the open model, because
we have one $i$ for $a$ and two minus signs for each of the two
time derivatives, which finally cancel. $a$ in the denominator is
multiplied by $i$ making $b$ remaining unchanged. In $c$, we see
that a term like $a^{2}$ is multiplied by a minus sign, because
$a$ was multiplied by $i$, also $\dot{a}/a$ is multiplied by a
minus sign, and its square remains unchanged. We note $K=+1$ for
the closed case and $K=-1$ for the open case and we deduce
straightforwardly that in both cases :
\begin{equation}
c=-\left(\frac{2K}{a^{2}}+2\left(\frac{\dot{a}}{a}\right)^{2}+\left(\frac{\ddot{a}}{a}\right)\right)
\end{equation}
\section{Conservation of entropy}
\subsection{Dependence of the three equations}
We know that in the case of constant $\kappa(\epsilon)=8\pi G$,
the equations of movement imply automatically the conservation of
entropy. Now we would like to prove this fact in our present case
with varying $\kappa(\epsilon)$. For this, we have to prove that
the three equations (3.27), (3.29) and (3.31) are dependent.
Adding (3.29) and (3.31) we obtain :
\begin{equation}
b-c=\kappa(\epsilon)(p+\epsilon)
\end{equation}
and we can multiply (3.29) by $2$ and derive :
\begin{equation}
\dot{b}-3\dot{c}+2\dot{\Lambda}=2\kappa'(\epsilon)\dot{\epsilon}\epsilon+2\kappa(\epsilon)\dot{\epsilon}
\end{equation}
and we know that (3.21) is the contraction of the covariant
derivative of the equation of motion, so that it is available as
the law of conservation of energy, besides (3.29) and (3.31).
Thus, we can use it in our present computation and find :
\begin{equation}
\dot{b}-3\dot{c}=2\kappa(\epsilon)\dot{\epsilon}
\end{equation}
Here we see that, in our calculations, we took the derivation of
one equation, which is (3.29). So the system we obtain now is only
equivalent to the first one up to a constant of integration, which
of course involves $\Lambda_{0}$. As a check, we see that
$\Lambda$ does not appear anymore in the last three equations, it
has been eliminated by combination or derivation. We then are left
to prove that (4.1) and (4.3) produce by themselves (3.27). First
sticking to the proof that these three equations are dependent, we
shall prove later that effectively the first two imply the third.
The small difference between these two statements is that when
three equations are dependent, two of them imply the third, but we
do not necessarily know which they are. To prove that the three
equations are dependent, we know that we have these three
equations for three unknown functions $\epsilon, p, a$ and we can
use two of these equations to eliminate $\epsilon$ and $p$. We
thus are left to prove that the third is an identity on $a$. We
have
\begin{equation}
3\kappa(\epsilon)(\epsilon+p)\frac{\dot{a}}{a}=-\kappa(\epsilon)\dot{\epsilon}=3(b-c)\frac{\dot{a}}{a}
\end{equation}
the first equality comes from (3.27) and the second from (4.1),
and we also have :
\begin{equation}
\frac{\dot{b}-3\dot{c}}{2}=\kappa(\epsilon)\dot{\epsilon}=-3(b-c)\frac{\dot{a}}{a}
\end{equation}
using first (4.3) and second (4.4). So we are left with :
\begin{equation}
\dot{b}-3\dot{c}=6(c-b)\frac{\dot{a}}{a}
\end{equation}
Now we prove that (4.6) is trivial and the proof is finished. From
(3.32) and (3.33) we have :
\begin{equation}
b-3c=-\frac{3\ddot{a}}{a}+3\left(\frac{2K}{a^{2}}+2{\left(\frac{\dot{a}}{a}\right)}^{2}+\frac{\ddot{a}}{a}\right)=6\left(\frac{K+{\dot{a}}^{2}}{a^{2}}\right)
\end{equation}
and
\begin{equation}
c-b=-2\left(\frac{K}{a^{2}}+{\left(\frac{\dot{a}}{a}\right)}^{2}-\frac{\ddot{a}}{a}\right)
\end{equation}
If we note
\begin{equation}\beta=\frac{b-3c}{6}=\left(\frac{K+{\dot{a}}^{2}}{a^{2}}\right)\end{equation}
(4.8) and (4.9) show us that (4.6) is equivalent to :
$$\dot{\beta}=(c-b)\frac{\dot{a}}{a}=-2\left(\frac{K+{\dot{a}}^{2}-a\ddot{a}}{a^{2}}\right)\frac{\dot{a}}{a}$$
or to :
$$a\dot{\beta}=-2\left(\beta-\frac{\ddot{a}}{a}\right)\dot{a}$$
so finally equivalent to :
$$a\dot{\beta}+2\beta \dot{a}=2\frac{\dot{a}\ddot{a}}{a}$$
or to :
$$a^{2}\dot{\beta}+2a\dot{a}\beta =2\dot{a}\ddot{a}$$
or to :
$$
\dot{\widehat{(\beta a^{2})}}=2\dot{a}\ddot{a}
$$
Now
$$\beta a^{2}=\left(\frac{K+{\dot{a}}^{2}}{a^{2}}\right)a^{2}=K+{\dot{a}}^{2}$$
and (4.6) is proved to be trivial. \subsection{Formal proof of the
conservation of entropy from the quantum equation} We now give the
formal proof that the two quantum equations effectively lead to
the conservation of entropy. We follow the former proof : When we
look at what has been needed to prove (4.6), we see that only
(3.32) and (3.33) were required, two equations which are pure
identities. So (4.6) is a pure identity following from the
definitions of $b$ and $c$. We suppose the two equations of
movement (3.29) and (3.31), we suppose also the third equation
obtained from the conservation of energy, that is we suppose
(3.21). The first two yield directly (4.1) and with the help of
the third we establish (4.3) as it has been done in the former
section. In (4.5), they are three expressions. (4.3) gives the
equality between the first two, and (4.6) the equality between the
first and the third. So (4.5) is established. In (4.4) they are
three expressions, (4.1) gives the equality between the first and
the third, and (4.5) gives the equality between the second and the
third. So (4.5) is completely established, and the equality of the
first two expressions of this formula is exactly (3.27), that is
to say the conservation of entropy.
\subsection{Conservation of energy of matter in the quantum
context} The condition (3.21) is the equation obtained after
taking the covariant derivative of our quantum equation. It comes
from the quantum equation, the conservation of energy of
Einstein's tensor, and the conservation of energy of the
stress-energy tensor of the matter fields. In general relativity,
with or without the cosmological constant, (3.21) is automatically
verified because in this case we have
\begin{equation}\dot{\Lambda}=\kappa'(\epsilon)=0\end{equation}
So in general relativity, the conservation of energy of the matter
fields is in fact put by hand in the equation by imposing that the
cosmological constant necessitates to be constant. This condition
in the quantum context is replaced by condition (3.21) which was
used to prove the conservation of entropy. In fact, in the
tensorial context, the conservations of entropy and of energy of
the matter fields are two equivalent conditions. This property
could not be stated in the context of general relativity, because
since the conservation of energy was automatic, the conservation
of entropy appeared only automatically verified, but not
necessarily linked with the conservation of energy.
\subsection{Equivalence between the conservations of energy and
entropy} We suppose our two equations of movement, (3.29) and
(3.31), plus the conservation of entropy (3.27). The derivation of
(3.29) implies directly (4.2), which proves that the conservation
of energy (3.21) is now equivalent to (4.3). Still, (4.6) is an
identity, always available. (4.5) possesses three expressions,
(4.6) implies the equality between the first and the third, the
conservation of entropy and (4.1) prove the equality between the
the second and the third, and (4.1) is implied by our two
equations of movement. So these last two plus the conservation of
entropy imply (4.5) completely. The equality of the first two
expressions in (4.5) is exactly (4.3). So we have established the
fact that if we suppose the two equations of movement, the
conservation of entropy and the conservation of energy of the
matter fields are equivalent.
\section{Conclusion : a basic system of equations}
We conclude this calculation by writing down a set of equations
equivalent to the whole set of the quantum equations of gravity.
We know that we first have the cosmological constant term, which
is no more constant,
\begin{equation}\Lambda=\theta\frac{b}{3}(b-3c)+\Lambda_{0}\end{equation}
We put now this value of $\Lambda$ in our equation (3.21):
\begin{equation}\dot{\Lambda}=\dot{\theta} \frac{b}{3}(b-3c)+\theta \frac{\dot{b}}{3}(b-3c)+\theta
\frac{b}{3}\dot{\widehat{(b-3c)}}=\kappa'(\epsilon)\epsilon\dot{\epsilon}
\end{equation}
Equation (5.2) is the law of conservation of energy of the matter
fields, to be imposed on our quantum equation itself. We write now
equation (4.3) :
\begin{equation}
\dot{b}-3\dot{c}=2\kappa(\epsilon)\dot{\epsilon}
\end{equation}
(5.2) and (5.3) give (4.2), which is the derivation of one of the
two equations of movement, namely (3.29). So equations (5.2) and
(5.3) are equivalent to the conservation of energy and to the
equation of movement (3.29) concerning only energy, up to a
constant, since (4.2) is only the time derivative of (3.29). If we
need to determine this constant, which happens to be
$\Lambda_{0}$, we need to apply (3.29) itself, which is :
\begin{equation}\frac{1}{2}(b-3c)+\Lambda=\kappa(\epsilon)\epsilon\end{equation}
Finally, there is another equation of movement, to determine $p$,
which is (3.31) :
\begin{equation}\frac{b+c}{2}-\Lambda=\kappa(\epsilon)p\end{equation}
\part{The first equation of quantum gravity}
\section{Introduction : smoothing out the initial singularity}
We call : first equation quantum gravity, the equation when the
parameter $\theta$ is constant, and in this first part of the
computation, we thus suppose $\theta$ constant. For the time
being, we leave $\Lambda_{0}$ undetermined, as a constant of
integration, and we will see that one value of this constant gives
us, in the case of the early universe, classical inflation,
defined as the exponential growth of $a$, and at the same time,
the standard properties of all cosmological parameters, except $a$
of course. We recall that at the end, in Parts IV and V, $\theta$
will be varying and $\Lambda_{0}$ will be put equal to zero. In
this part, we stick to the case
\begin{equation}\kappa(\epsilon)=\frac{\kappa_{0}}{\sqrt{\epsilon}}\end{equation}
where $\kappa_{0}$ is strictly constant, and where we have
supposed $\hbar=c=1$. More precisely, we are going to prove that
the quantum equation, for $\theta$ constant and strictly positive,
implies that the whole set of cosmological parameters is the same
as in the standard cosmological model applied to the early
universe, except $a$ which has now an exponential growth. This
means that the quantum equation in this case gives simultaneously
the relations $a(t)\sim e^{\chi t}$, $p>0$, $p=\epsilon/3$, and
$\epsilon\sim1/a^{4}$. So we retrieve all features of the standard
early universe except that we have smoothed out the initial
singularity. Another feature of the quantum equation is that it
implies by itself the $p=\epsilon/3$ relation which is not simply
put by hand anymore, but is really a non trivial consequence of
the quantum theory. We emphasize this point because it means the
following conclusion : \textbf{\emph{one sole quantum equation of
gravity governs at the same time the cosmological parameters of
the universe, exactly as did the standard cosmological model, and
the structure of fundamental particles, giving the right relation
between $p$ and $\epsilon$, whereas, furthermore, it smoothes out
the initial singularity with no further hypothesis.}}
\section{Integration of the equations}
We first stick to the computation of $\epsilon$, and leave for the
time being the equation for $p$. We are left with two equations
which are (5.2) and (5.3). We use :
$$\kappa(\epsilon)=\frac{\kappa_{0}}{\sqrt{\epsilon}}$$
with constant $\kappa_{0}$. We find, from (5.2), since $\theta$ is
constant :
\begin{equation}
\theta \frac{\dot{b}}{3}(b-3c)+\theta
\frac{b}{3}\dot{\widehat{(b-3c)}}=-\kappa_{0}\frac{\dot{\epsilon}}{2\sqrt{\epsilon}}
\end{equation}
and from (5.3) :
\begin{equation}
\dot{\widehat{(b-3c)}}=2\kappa_{0}\frac{\dot{\epsilon}}{\sqrt{\epsilon}}
\end{equation}
We combine these equations and find :
\begin{equation}
\theta \frac{\dot{b}}{3}(b-3c)+\theta
\frac{b}{3}\dot{\widehat{(b-3c)}}=-\frac{\dot{\widehat{(b-3c)}}}{4}
\end{equation}
which, for $\theta\neq0$, gives :
$$\theta \frac{\dot{b}}{3}(b-3c)+\theta \left(\frac{b}{3}+\frac{1}{4\theta}\right) \dot{\widehat{(b-3c)}}=0$$
or :
\begin{equation}\theta\dot{\widehat{\left(\frac{b}{3}+\frac{1}{4\theta}\right)}}(b-3c)+\theta
\left(\frac{b}{3}+\frac{1}{4\theta}\right)\dot{\widehat{(b-3c)}}=0\end{equation}
and we finally obtain :
\begin{equation}
\left(\frac{b}{3}+ \frac{1}{4\theta}\right)(b-3c)=C
\end{equation}
where $C$ can be chosen at will since any $C$ gives (7.4). In fact
this is only true because we have let $\Lambda_{0}$ unspecified as
a constant of integration, which has permitted us to use the
derivative (5.2) of one of our equations of motion, instead of the
equation (5.4) itself. We still have obtained a system equivalent
to the quantum equation, up to $\Lambda_{0}$ of course. Now it is
clear that two different choices of $C$ should give us two
different choices of $\Lambda_{0}$. Of course the simplest case is
$C=0$. In the next two sections, we always adopt the choice $C=0$.
\section{The quantum features of the closed early universe, positive case}
\subsection{Introduction}
We call positive case the case in which the Gauss-Bonnet parameter
$\theta$ is positive. We now make this special hypothesis on our
equation, which is that $\theta>0$. This hypothesis rules out the
open model as showed below. This condition $\theta>0$ is necessary
if we want our quantum equation to display the smoothing out of
the initial singularity. As far as the particular case of constant
$\theta$ is concerned, we shall prove in the next section that the
condition $\theta<0$ is compatible with both the closed and open
models, but keeps the initial singularity.
\subsection{Inflation}
From $C=0$ in (7.5), we deduce :
\begin{equation}
\frac{b}{3}=-\frac{1}{4\theta}
\end{equation}
because from (4.7) :
$$(b-3c)=6\left(\frac{K+\dot{a}^{2}}{a^{2}}\right)>0$$
Indeed, the last expression is strictly positive, and thus
nonzero, because we shall prove in the following, using (8.9),
that we are necessarily in the closed model, and thus we have
$K=+1$. Furthermore, since $$b=-\frac{3\ddot{a}}{a}$$ we find :
\begin{equation}
\frac{\ddot{a}}{a}=+\frac{1}{4 \theta}
\end{equation}
With $\theta>0$, we finally obtain the inflation solution :
\begin{equation}
a(t)=a_{0}e^{\chi t}
\end{equation}
\subsection{Value of the Gauss-Bonnet parameter}
We call $\theta$ the Gauss-Bonnet parameter. Now, from the former
solution, we get easily
$$\frac{\ddot{a}}{a}=\chi^{2}$$ and find the value of $\chi$
in our solution :
\begin{equation}
\chi=\frac{1}{2\sqrt{\theta}}
\end{equation}
To reinsert in this equation the constants $\hbar$ and $c$ we use
that $\theta \sim [L]^{2}$ while $\chi \sim [T]^{-1} \sim
c{[L]^{-1}}$, so
\begin{equation}
\chi=\frac{c}{2\sqrt{\theta}}
\end{equation}
\subsection{The values of c, c-b and b-3c} Now we compute all our
functions, to determine, first the behavior of $\epsilon$, second
the value of $p$. We also compute the value of $\theta$ and verify
directly, as a check, that all our quantum equations are satisfied
by our solutions. $\epsilon$ and $p$ are given by (3.27) and (4.1)
:
$$\dot{\epsilon}=-3(\epsilon+p)\frac{\dot{a}}{a}$$
and $$b-c=\kappa(\epsilon) (p+\epsilon)$$ but as we know that
$$\frac{\dot{a}}{a}=\chi$$ we obtain :
\begin{equation}
\dot{\epsilon}=-3\chi (\epsilon +p)
\end{equation}
We also have :
\begin{equation}
b=-3\frac{\ddot{a}}{a}=-3\chi^{2}
\end{equation}
and
\begin{equation}
c=-\left(\frac{2K}{a^{2}}+2{\left(\frac{\dot{a}}{a}\right)}^{2}+\left(\frac{\ddot{a}}{a}\right)\right)=-\left(\frac{2K}{a^{2}}+3\chi^{2}\right)
\end{equation}
So
\begin{equation}
c-b=-\frac{2K}{a^{2}}=-\kappa(\epsilon)(p+\epsilon)
\end{equation}
and
\begin{equation}
b-3c=6\left(\frac{K}{a^{2}}+\chi^{2}\right)
\end{equation}
We recall that for the closed model we have $K=+1$, whereas the
open model, which gives $K=-1$, is ruled out by the relation
(8.9). Indeed $K=-1$ corresponds to the wrong sign in this
equation. We emphasize the fact that what is only ruled out in
this case is the combination : open model plus constant $\theta$.
This matter could as well be interpreted as a clue that $\theta$
is varying. We shall see just below that this hypothesis is
confirmed by other arguments. Nevertheless, we will see again,
studying the abundance of the elements, that the open model
doesn't fit well with the quantum equation. We stick until the end
of this section to the closed model. To put the value of
$\Lambda_{0}$ back in the equation, we need to find this constant
in terms of $\theta$ or $\kappa_{0}$, but not $\chi$, which is a
variable which belongs to the set of solutions.
\subsection{The value of the energy density}
We combine (8.6) and (8.9) to find :
$$\frac{\dot{\epsilon}}{3\chi}=-(\epsilon+p)=-\frac{2}{\kappa(\epsilon)a^{2}}=-\frac{2\sqrt{\epsilon}}{\kappa_{0}a^{2}}$$
so :
$$\frac{\dot{\epsilon}}{2\sqrt{\epsilon}}=\dot{\widehat{(\sqrt{\epsilon})}}=-\frac{3\chi}{\kappa_{0}{a_{0}}^{2}}e^{-2\chi t}$$
we integrate this equation and find :
\begin{equation}
\sqrt{\epsilon}=\frac{3}{2\kappa_{0}{a_{0}}^{2}}e^{-2\chi
t}=\frac{3}{2\kappa_{0}a^{2}}
\end{equation}
where we have chosen the simplest constant of integration without
investigating all solutions. We see then that we obtain the
behavior for $\epsilon$ which has also been found in the standard
cosmological model. Indeed, we have
$$\epsilon a^{4} =\frac{9}{4{\kappa_{0}^{2}}}=Cte$$
Reestablishing the values of $\hbar$ and $c$ :
\begin{equation}
\epsilon a^{4} =\frac{9\bar{h}c}{4{\kappa_{0}^{2}}}=Cte
\end{equation}
\subsection{The value of the pressure from the quantum equation} Now we know
that the behavior of $a$ is $a=a_{0}e^{\chi t}$. This gives, using
(8.12), the behavior of $\epsilon$ :
$\epsilon=\epsilon_{0}e^{-4\chi t}$. We thus have :
\begin{equation}
\dot{\epsilon}=-4\chi \epsilon
\end{equation}
and from (8.6) we see that :
$$\epsilon+p=-\frac{\dot{\epsilon}}{3\chi}=\frac{4}{3}\epsilon$$
We conclude what we wanted :
\begin{equation}
p=\frac{\epsilon}{3}
\end{equation} As claimed, this equation has not been put by hand but is the
a consequence of the quantum equation of gravity itself.
\section{The quantum features of the early
universe in the negative case} We suppose in this section that
$\theta<0$, and $\theta$ still constant. We prove that in the open
case, the quantum equation still leads to the behavior of
$\epsilon$ as in the standard cosmological model, and furthermore
that it still contains the information on the structure of matter.
In other words, we prove in this open case that the quantum
equation leads to the relation $p=\epsilon/3$. We then look for
the expression of $\Lambda_{0}$ that renders the quantum equation
possible in the open model and find that it has to be proportional
to $H$.
\subsection{Behavior of the radius of the unverse}
Again, we start from (7.5) and obtain :
\begin{equation}
\frac{b}{3}=-\frac{1}{4 \theta}
\end{equation}
and, since $$b=-\frac{3\ddot{a}}{a}$$ we find :
\begin{equation}
\frac{\ddot{a}}{a}=+\frac{1}{4 \theta}
\end{equation}
With $\theta<0$, supposing that $a=0$ for $t=0$, we obtain the
solution :
\begin{equation}
a(t)=a_{0}\sin{\chi t}
\end{equation}
\subsection{Value of Gauss-Bonnet parameter} Now from this solution, we get easily
$$\frac{\ddot{a}}{a}=-\chi^{2}$$ and find the value of $\chi$
in our solution :
\begin{equation}
\chi=\frac{1}{2\sqrt{-\theta}}
\end{equation}
To reinsert in this equation the constants $\hbar$ and $c$ we use
that $(-\theta) \sim [L]^{2}$ while $\chi \sim [T]^{-1} \sim
c{[L]^{-1}}$, so
\begin{equation}
\chi=\frac{c}{2\sqrt{-\theta}}
\end{equation}
\subsection{The values of c, c-b and b-3c}
As in the former section, we compute all our functions, to
determine, first the value of $p$, then the behavior of
$\epsilon$. Then we will compute the value of $\theta$ and verify
directly, as a check, that all our quantum equations are satisfied
by our solutions. $\epsilon$ and $p$ are given by (3.27) and (4.1)
:
$$\dot{\epsilon}=-3(\epsilon+p)\frac{\dot{a}}{a}$$
and $$b-c=\kappa(\epsilon) (p+\epsilon)$$ but as we know that
\begin{equation}H=\frac{\dot{a}}{a}=\chi\frac{\cos\chi t}{\sin\chi
t}=\chi\cot\chi t\end{equation} we have :
\begin{equation}
\dot{\epsilon}=-3\chi (\epsilon +p)\cot\chi t
\end{equation}
We also have :
\begin{equation}
b=-3\frac{\ddot{a}}{a}=3\chi^{2}
\end{equation}
and
\begin{equation}
c=-\left(\frac{2K}{a^{2}}+2{\left(\frac{\dot{a}}{a}\right)}^{2}+\left(\frac{\ddot{a}}{a}\right)\right)
=-\left(\frac{2K}{a^{2}}+2\chi^{2}\cot^{2}\chi t-\chi^{2}\right)
\end{equation}
So
$$b-c=\frac{2K}{a^{2}}+2\chi^{2}\cot^{2}\chi
t+2\chi^{2}=\frac{2K}{a^{2}}+\frac{2\chi^{2}}{\sin^{2}\chi t}$$
\begin{equation}=\frac{2\left(K+(a_{0}\chi)^{2}\right)}{a^{2}}=\kappa(\epsilon)(p+\epsilon)\end{equation}
We note $\mu=K+(a_{0}\chi)^{2}$ and find that
\begin{equation}b-c=\frac{2\mu}{a^{2}}=\kappa(\epsilon)(p+\epsilon)\end{equation}
We also have :
\begin{equation}
b-3c=6\left(\frac{K}{a^{2}}+\chi^{2}\cot^{2}\chi t\right)
\end{equation}
\subsection{The value of the energy density}
We had :
$$\dot{\epsilon}=-3\chi(\epsilon+p)\cot\chi
t=\frac{b-c}{\kappa(\epsilon)}(-3\chi)\cot\chi t$$ So we have :
$$\dot{\epsilon}=\frac{2\mu}{\kappa_{0}a^{2}}(-3\chi)\sqrt{\epsilon}\cot\chi t$$
and :
$$\frac{\dot{\epsilon}}{\sqrt{\epsilon}}=2\dot{\widehat{(\sqrt{\epsilon})}}=\frac{3\mu}{\kappa_{0}{a_{0}}^{2}}\frac{-2\chi\cos\chi t}{\sin^{3}\chi t}$$
we integrate this equation and find:
\begin{equation}
\sqrt{\epsilon}=\frac{3\mu}{2\kappa_{0}{a_{0}}^{2}}\frac{1}{\sin^{2}\chi
t}=\frac{3\mu}{2\kappa_{0}{a}^{2}}
\end{equation}
where we have chosen the simplest constant of integration without
investigating all solutions. We see then that we obtain again the
same standard behavior for $\epsilon$ :
$$\epsilon a^{4} =\frac{9\mu^{2}}{4{\kappa_{0}^{2}}}=Cte$$
Reestablishing the values of $\hbar$ and $c$ :
\begin{equation}
\epsilon a^{4} =\frac{9\mu^{2}\bar{h}c}{4{\kappa_{0}^{2}}}=Cte
\end{equation}
\subsection{Consistency with both the open and closed models}
From the former calculation, we see that the only condition that
needs to be realized to make things possible is $\mu>0$. In the
closed model this condition is always verified, whereas in the
open model, since $K=-1$, it is verified provided we have
$a_{0}\chi>1$. We notice that the term $a_{0}\chi$ is the value of
$\dot{a}$ when $t=0$ and $a=0$. This condition has been smoothed
compared to the situation in the standard cosmological model where
$\dot{a}\rightarrow+\infty$ near the initial singularity. This is
because the quantum equation naturally tends to smooth
singularities. In the open model, with $\theta$ being constant and
negative, it cannot do this job too properly, but it still
improves the behavior of the cosmological parameters.
\subsection{The pressure from the quantum equation}
We retrieve the usual behavior of $\epsilon$ :
$\epsilon=\epsilon_{0}a^{-4}$ so we have :
\begin{equation}
\dot{\epsilon}=-4\epsilon\frac{\dot{a}}{a}
\end{equation}
and from (3.27) we see that :
$$\epsilon+p=-\frac{\dot{\epsilon}a}{3\dot{a}}=\frac{4}{3}\epsilon$$
We conclude :
\begin{equation}
p=\frac{\epsilon}{3}
\end{equation} As claimed, this equation has not been put by hand but is the
the result of the quantum equation of gravity itself.
\section{Introduction to the flatness and cosmological constant problems in the context of quantum gravity}
\subsection{Quantum gravity implies that the universe had no beginning}
The equation of quantum gravity, in the case of the early
universe, and in the particular case in which $\theta$ is a
positive constant, leads to an exponential growth for the
cosmological parameter $a$ of the Robertson-Walker metric. This
satisfies the principle which states that this smoothing out of
the initial singularity should be the consequence of a fundamental
law of nature. We recall that we stated this principle because
this smoothing out was equivalent to the fact that the universe
had no beginning. This absence of beginning was needed because
such a beginning would have no former cause.
\subsection{The flatness problem}
We know that the quantum equation displays also characteristic
features of the standard cosmological model, like the behavior of
$p$ and $\epsilon$ :
\begin{equation}p=\frac{\epsilon}{3}\sim\frac{1}{a^{4}}\end{equation}
As a preliminary exercise, we can look at what gives us the
equation of the standard cosmological model when the quantum
behavior of $a$ is put in it. General relativity gives the
equation :
\begin{equation}\frac{K}{a^{2}}+H^{2}=\frac{8\pi
G\epsilon}{3}\end{equation} Concerning the present universe, we
should take $t$ quite large. So we look at the former equation
when $t\rightarrow+\infty$. With the exponential growth of $a$
coming from quantum gravity, we see that, in the last equation :
\begin{equation}\frac{K}{a^{2}}=o(H^{2})\end{equation}
We mean by this notation that two functions $f(t)$ and $g(t)$
verify $f(t)=o(g(t))$, if and only if
$$\lim_{t\rightarrow+\infty}\frac{f(t)}{g(t)}=0$$
We then obtain that the usual ratio $\Omega$ tends to $1$ when
$t\rightarrow+\infty$ :
\begin{equation}\Omega=\frac{8\pi
G}{3H^{2}}\rightarrow1\end{equation} This is an even better
behavior of $\Omega$ than in the classical inflation model, since
this time $\Omega$ tends smoothly to its observed value. We
emphasize that this argument is only a guess since it uses the
equation of general relativity which is not part of the quantum
theory. We shall prove in this part that in fact the case of
constant $\theta$ does solve the flatness problem, but with
different arguments for the cases $\theta>0$ and $\theta<0$.
\subsection{The cosmological constant problem}
After this, we will ask ourselves if quantum gravity can solve the
cosmological constant problem too. We will see that even in the
case of positive constant $\theta$, this problem already finds a
solution. In the present part, in order to understand the general
features of how these various cosmological problems can be solved
by quantum gravity, we stick to the well known case of constant
$\theta$, case in which all values of all cosmological parameters
can be computed exactly. We first compute the value of
$\Lambda_{0}$ in this model, we then study the expansion, flatness
and cosmological constant problems in this same context.
\section{Value of the constant term}
We call $\Lambda_{0}$ the constant term. In this section we prove
that the values of the constant $\Lambda_{0}$ in both cases, when
$\theta$ is a positive constant, and when $\theta$ is a negative
constant, are the same, in the sense that this value is entirely
determined by $\theta$, and that in both cases, the formulas
giving $\Lambda_{0}$ from $\theta$ are identical.
\subsection{The positive case : value of the constant term}
We recall that in this case, we are automatically in the closed
model and that $K=+1$. We compute, using (5.1) and (8.7) :
\begin{equation}
\Lambda=\frac{\theta
b}{3}(b-3c)+\Lambda_{0}=-\theta\frac{\ddot{a}}{a}(b-3c)+\Lambda_{0}=-\theta\chi^{2}(b-3c)+\Lambda_{0}
\end{equation} We know from (8.4) that $\theta\chi^{2}=1/4$ and we obtain, using (8.10) too, for $K=+1$ :
\begin{equation}\Lambda=\frac{(3c-b)}{4}+\Lambda_{0}=-\frac{3}{2}\left(\frac{1}{a^{2}}+\chi^{2}\right)+\Lambda_{0}
\end{equation}
We know from previous calculations, in particular from (3.29),
that :
\begin{equation}
R_{0}^{0}-\frac{1}{2}R=\frac{(b-3c)}{2}=3\left(\frac{1}{a^{2}}+\chi^{2}\right)
\end{equation}
The first of our equations (3.28) was :
\begin{equation}
R_{0}^{0}-\frac{1}{2}R+\Lambda=\kappa(\epsilon)\epsilon
\end{equation}
so we find :
\begin{equation}\frac{3}{2}\left(\frac{1}{a^{2}}+\chi^{2}\right)+\Lambda_{0}=\kappa_{0}\sqrt{\epsilon}\end{equation}
and using the expression of $\epsilon$ found in (8.11), we obtain
:
\begin{equation}
\Lambda_{0}=-\frac{3}{2}\chi^{2}
\end{equation}
We find :
\begin{equation}
\Lambda_{0}=-\frac{3}{2}\chi^{2}=-\frac{3}{8\theta}
\end{equation}
We also can write the value of $\Lambda_{0}$ in terms of Hubble's
constant :
\begin{equation}
\Lambda_{0}=-\frac{3}{2}\chi^{2}=-\frac{3}{2}H^{2}
\end{equation}
which is of the order of $H^{2}$, so tiny enough to be plausible.
Furthermore we see that the quantum equation determines by itself
the value of the cosmological constant term and gives exactly the
value we expect, we mean of the order of $H^{2}$. We shall see in
the next part that indeed, our equation solves the cosmological
problem. So finally, what conclusion can we draw about this
constant $\Lambda_{0}$? In the case of constant $\theta$ and
$\theta>0$, we put in the equation the $\Lambda_{0}$ term, as a
$\Lambda$ term, which is always possible, since $\Lambda_{0}$ is
constant. We can determine the value of $\Lambda_{0}$, using a
principle of pure logic : if $\epsilon$ is the mean energy density
in the universe, the conservation of energy implies that
$\epsilon\rightarrow0$ when $a\rightarrow+\infty$. Comparing this
condition to equation (11.5), it yields :
$$\Lambda_{0}=-\frac{3}{2}\chi^{2}$$ Using the
parameters of the equation itself, $\kappa_{0}$ and $\theta$,
$\Lambda_{0}$ has the value :
$$\Lambda_{0}=-\frac{3}{8\theta}$$ Finally with this value, we
find $$\Lambda_{0}=-\frac{3}{2}H^{2}$$ which has the right sign
and the right order of magnitude to be a solution to the
cosmological constant problem.
\subsection{The negative case: value of the constant term}
We still consider the case of constant $\theta$, but now
$\theta<0$, and we compute, using (5.1), (9.2) and (9.4) :
\begin{equation}
\Lambda=\frac{\theta
b}{3}(b-3c)+\Lambda_{0}=-\theta\frac{\ddot{a}}{a}(b-3c)+\Lambda_{0}=\theta\chi^{2}(b-3c)+\Lambda_{0}
\end{equation} We know that $\theta\chi^{2}=-1/4$ and we obtain, using (9.12) :
\begin{equation}\Lambda=\frac{(3c-b)}{4}+\Lambda_{0}=-\frac{3}{2}\left(\frac{K}{a^{2}}+\chi^{2}\cot^{2}\chi
t\right)+\Lambda_{0}
\end{equation}
We know, from (3.28) and (3.29), that :
\begin{equation}
R_{0}^{0}-\frac{1}{2}R=\frac{(b-3c)}{2}=3\left(\frac{K}{a^{2}}+\chi^{2}\cot^{2}\chi
t\right)
\end{equation}
Using (3.28) :
\begin{equation}
R_{0}^{0}-\frac{1}{2}R+\Lambda=\kappa(\epsilon)\epsilon
\end{equation}
we find :
\begin{equation}
\frac{3}{2}\left(\frac{K}{a^{2}}+\chi^{2}\cot^{2}\chi
t\right)+\Lambda_{0}=\kappa_{0}\sqrt{\epsilon}\end{equation} and
using the expression of $\epsilon$ found in (9.13) :
$$\kappa_{0}\sqrt{\epsilon}=\frac{3\mu}{2a^{2}}$$
we obtain :
\begin{equation}
\frac{K}{a^{2}}+\chi^{2}\frac{\cos^{2}\chi t}{\sin^{2}\chi
t}+\frac{2}{3}\Lambda_{0}=\frac{\mu}{a^{2}}=\frac{K}{a^{2}}+\frac{(a_{0}\chi)^{2}}{a^{2}}=\frac{K}{a^{2}}+\frac{\chi^{2}}{\sin^{2}\chi
t}\end{equation} We conclude :
\begin{equation}
\Lambda_{0}=\frac{3}{2}\chi^{2}
\end{equation}
for both models, closed and open. We find :
\begin{equation}
\Lambda_{0}=\frac{3}{2}\chi^{2}=-\frac{3}{8\theta}
\end{equation}
which is the same formula than in the $\theta>0$ model. We cannot
write the value of $\Lambda_{0}$ in terms of Hubble's constant
anymore, because now this one varies with time whereas
$\Lambda_{0}$ is a true constant, obtained directly from the
parameters of the equation.
\section{The expansion problem}
\subsection{The problem}
 The expansion problem is explained in detail for
example in Peacock 1999, 2005 $[48]$, Chapter 11, section 11.1 :
"... Nevertheless it is the only level of explanation that
classical cosmology offers : the universe expands now because it
did so in the past. Although it is not usually included one might
thus with justice add an "expansion problem" as perhaps the most
fundamental in the catalogue of classical cosmological problems.
Certainly, early generations of cosmologists were convinced that
some specific mechanism was required in order to explain how the
universe was set in motion."
\subsection{The quantum behavior of the radius of the universe}
In classical cosmology, the parameter $a$ has such a behavior that
$\ddot{a}<0$. In these conditions, the deceleration parameter
\begin{equation}q=-\frac{\ddot{a}a}{\dot{a}^{2}}\end{equation}
is positive. In the seventies, one could read (Weinberg $[67]$)
that the observed value of $q$ is around $1/2$, most certainly
positive. This positive value of $q$, equivalent to a negative
$\ddot{a}$, is at the origin of the expansion problem. An
expanding universe means $\dot{a}>0$, and the solution to the
expansion problem is equivalent to the understanding of the
condition $\dot{a}>0$, with no further hypothesis on the initial
conditions concerning $\dot{a}$. The condition $\ddot{a}<0$
implies initial conditions where $\dot{a}$ was even greater. In
the quantum regime, where $q=-1$, the expansion problem is solved
at once : with $\theta$ constant and positive we have
$\ddot{a}>0$, the positivity of $\dot{a}$ is explained with no
further hypothesis. Indeed, in the context of the exponential
growth of $a$, the value of $\dot{a}$ tends to zero when
$t\rightarrow-\infty$, and this condition itself is linked to the
absence of beginning we were looking for. So the problem of
quantum gravity is not the expansion problem, which it solves
easily, but the value of $q$. For this reason, we tried to improve
this value by making $\theta$ vary. What we found is that we can
make $q$ tend to zero when $t\rightarrow+\infty$, but still with
$q$ negative. There is no choice of varying $\theta$ that can make
$\ddot{a}$ negative. In other words, the only possibility for
$\ddot{a}<0$ is $\theta$ constant and negative. Finally, except
for this particular case, the solution of the expansion problem is
always provided by the quantum equation. On the other side, the
measured value of $q$ has been revised with time. It appears today
that negative values of $q$ are even more probable, and $q=-1$ is
not ruled out anymore. Different experiments have made appear the
fact that $\ddot{a}$ could be positive. All these experiments of
course play in favor of the quantum equation.
\section{The cosmological constant problem}
\subsection{Comparison of theories : inertia}
In order to study the cosmological constant problem, we compute
the total value of the cosmological term $\Lambda$ in the case of
constant $\theta$. This term comes from the topological
Gauss-Bonnet term and from $\Lambda_{0}$. We see that in our
equation, this term does not come from any effective dark energy,
but from the quantum corrections to classical gravity. The only
unexplained term is $\Lambda_{0}$ but it does not matter since in
the case of varying $\theta$, we shall take $\Lambda_{0}=0$. In
general relativity, to compute the total value of $\Omega$, we see
that the cosmological constant term is negative on the left hand
side of the equation. We pass it on the other side, where it adds
as a positive term to usual matter. This operation done, we have
on the right hand side the total energy density of matter, counted
with the supposed dark energy, and the equation makes it equal of
course to the left hand side. The left hand side in this context
is :
\begin{equation}G^{0}_{0}=R^{0}_{0}-\frac{1}{2}R\end{equation}
We divide this expression by $3H^{2}$ to find $\Omega_{TOT}$,
which is $\Omega$ calculated with all kinds of matter together,
even with dark energy, itself generated by $\Lambda$. We thus find
the following formula for $\Omega_{TOT}$ :
\begin{equation}\Omega_{TOT}=\frac{1}{3H^{2}}\left(R^{0}_{0}-\frac{1}{2}R\right)\end{equation}
We emphasize that the last formula is the one which has to be used
in order to compute the value of $\Omega_{TOT}$ in the context of
the quantum equation of gravity. Indeed, the difference of the
behaviors of general relativity and quantum gravity relatively to
the parameter $\epsilon$, makes difficult to compare the right
hand sides of these two equations. This has a natural explanation,
that we give just right now. First we cannot compute the value
$\kappa_{0}$ so easily. We shall prove in $[58]$ that this is a
characteristic of all tensorial equations to possess a paradox in
respect to their right hand side. Because they are in fact
effective equations, that should be derived from unification, they
still can be interpreted from different points of view, which
changes at least the value we should take for $\kappa_{0}$. These
tensorial equations in fact all belong to theories which treat
gravitation as an inertial interaction. Since Newtonian gravity is
a limit case of general relativity, and since in general
relativity, gravity is a pure effect of inertia, we deduce that in
some sense Newtonian gravity is also purely inertial. Thus, in our
observations of the sky, our experiments, when compared to general
relativity, can only give us information about the properties of
inertia of the celestial bodies, that is to say about the
geodesics of space-time. All this information being locked in
$G_{0}^{0}$. Indeed, the equation of general relativity containing
$G_{0}^{0}$, the sole equation which is used to compute the
masses, to compare general relativity to Newton's theory, and
compute the relativistic gravitational coupling constant from
Newton's $G$. The equation containing $G_{\alpha}^{\alpha}$ is
used only to compute the value of $p$. These arguments give the
principle that the dimensionless term $\Omega_{TOT}$ should be
defined in both general relativity and quantum gravity by the term
controlling the equations of the geodesics in vacuum, so in both
cases by $G_{0}^{0}$, and we find
\begin{equation}\Omega_{TOT}=\frac{1}{3H^{2}}G_{0}^{0}\end{equation}
Using the equation of quantum gravity, we obtain :
\begin{equation}3H^{2}\Omega_{TOT}=G_{0}^{0}=2\kappa_{0}\sqrt{\epsilon}\end{equation}
In the standard equation of general relativity, the cosmological
constant is affected by a minus sign. Our equation, on the left
hand side, contains the $\Lambda$ term with a positive sign, so
our quantum $\Omega_{\Lambda}$, which is the dimensionless energy
density, reads :
\begin{equation}\Omega_{\Lambda}=-\frac{\Lambda}{3H^{2}}\end{equation}
We recall that within the realm of the standard cosmological
model, Bennett an al. 2003, $[2]$, experiments on the cosmic
microwave background radiation imply the fairly precise relation :
\begin{equation}\Omega_{\Lambda}=\frac{3}{4}\Omega_{TOT}\end{equation}
The quantum equation of gravity displays a natural energy density
coming from $\Lambda$ itself, composed of two elements, the
Gauss-Bonnet term and $\Lambda_{0}$. These two terms have
different origins : the topological Gauss-Bonnet term is
interpreted as a non perturbative quantum correction to classical
gravity, $\Lambda_{0}$ is more had hoc, with no interpretation
until now, but we will be able to get rid of this term in the
varying $\theta$ context. For the time being, we just evaluate
these terms to see if they can give account for the supposed dark
energy of the standard cosmological model, and if they resolve the
cosmological constant problem, that this if they yield a strictly
positive value of $-\Lambda$ which is tiny enough to be of the
order of the former value preconized by Bennett and al. We first
analyze the case of positive constant $\theta$.
\subsection{Positive constant Gauss-Bonnet term}
From (3.29) and (13.3) we find :
\begin{equation}3H^{2}\Omega_{TOT}=R^{0}_{0}-\frac{1}{2}R=\frac{b-3c}{2}\end{equation}
Equation (8.10) gives the value :
\begin{equation}\frac{b-3c}{2}=3\left(\frac{1}{a^{2}}+\chi^{2}\right)=3\left(\frac{1}{a^{2}}+H^{2}\right)\end{equation}
So we find :
\begin{equation}\Omega_{TOT}=\frac{1}{\dot{a}^{2}}+1\end{equation}
which is an especially interesting solution to the flatness
problem. We find that when $t\rightarrow+\infty$, we have
$\Omega_{TOT}\rightarrow1$, which is what is observed, plus the
fact that $\Omega_{TOT}>1$, which gives account for the fact that
the observed value is around $1.02$ rather than around $1$. The
term coming from the Gauss-Bonnet expression is, from (5.1) :
\begin{equation}3H^{2}\Omega_{GB}=-\theta\frac{b}{3}(b-3c)\end{equation}
but from (8.1) we obtain
\begin{equation}\theta\frac{b}{3}=-\frac{1}{4}\end{equation}
Finally we find :
\begin{equation}\Omega_{GB}=\frac{1}{2}\Omega_{TOT}\end{equation}
So, the value we find for the cosmological constant term is just
of the right order of magnitude, is explained by the sole quantum
corrections to classical general relativity, and is even very near
the observed value, taking into account that this value is model
dependent. With varying $\theta$ and other behaviors of
$\kappa(\epsilon)$ than the simple
$\kappa(\epsilon)\sim1/\sqrt{\epsilon}$, we possess a entire new
class of cosmological models displaying about the same quantum
features. Clearly the model of constant $\theta$, $\theta>0$,
gives already the whole set of features needed for our quantum
gravity. It gives a solution to the flatness problem, to the
cosmological constant problem, it smoothes out the initial
singularity, and contains in one sole quantum equation of gravity
the information of the largest and smallest scales of the
universe, that is to say the behavior of the radius $a$ of the
universe and the state of matter, relativistic or non
relativistic, via the pressure $p$.
\subsection{Additional remarks}
We analyze now the $\Lambda_{0}$ term : We know, from (11.6), that
\begin{equation}\Lambda_{0}=-\frac{3}{2}H^{2}\end{equation}
So we find :
\begin{equation}\Omega_{\Lambda_{0}}=\frac{1}{2}\end{equation}
This term has again the right sign and is very near the observed
value, in the context of the standard cosmological model. If we
now look at the values of the three different parts of $\Omega$,
we can define $\Omega_{\Lambda}=\Omega_{GB}+\Omega_{\Lambda_{0}}$
and compute :
\begin{equation}\Omega_{\Lambda}=\frac{1/2+\dot{a}^{2}}{1+\dot{a}^{2}}\Omega_{TOT}\end{equation}
We know that in this model, the value of $\dot{a}^{2}$ tends to
infinity when $t\rightarrow+\infty$. This model predicts a quite
interesting situation. In the early universe,
$\dot{a}^{2}\approx0$ and
$$\Omega_{\Lambda}=\frac{1}{2}\Omega_{TOT}$$ Then $\dot{a}^{2}$
regularly increases, and the ratio of $\Omega_{\Lambda}$ to
$\Omega_{TOT}$ increases also with $\dot{a}^{2}$. We could imagine
an actual value of $\dot{a}^{2}$ to be around unity :
$\dot{a}^{2}\approx1$ such that
$$\Omega_{\Lambda}=\frac{3}{4}\Omega_{TOT}$$
This does not fit completely because for $\dot{a}^{2}\approx1$,
$\Omega_{TOT}$ becomes twice too big. This error comes probably
from the value of $\Lambda_{0}$ that should disappear in the case
of varying $\theta$. Later, the value of $\dot{a}^{2}$ should
continue to increase and the former ratio will tend to $1$ when
$t\rightarrow+\infty$ :
$$\Omega_{\Lambda}=\Omega_{TOT}$$
Thus, the universe is evolving to a situation where an apparent
dark energy is growing until being almost all of matter. We find a
situation analogous to an empty closed and expanding universe.
\subsection{The case of constant and negative Gauss-Bonnet term}
We first notice that in the proof of the relation
\begin{equation}\Omega_{GB}=\frac{1}{2}\Omega_{TOT}\end{equation}
we did not use the specific equations concerning the different
signs of $\theta$. Thus this relation is still valid in the
negative case. For the term $\Lambda_{0}$, it is positive from
(11.15) or (11.16), and for this reason has the wrong sign. It
readily appears that the negative case, a priori, does not fit as
easily as the positive case to cosmological observations.
Nevertheless we have :
\begin{equation}\Omega_{TOT}=\frac{1}{3H^{2}}G_{0}^{0}=\left(\frac{K}{a^{2}}+H^{2}\right)=\frac{K}{\dot{a}^{2}}+1\end{equation}
We find again in this case that $\Omega_{TOT}$ should take values
near unity, a little greater than $1$ in the closed model and a
little less than $1$ in the open model.
\part{The generalized quantum equation of gravity}
\section{Equations in the general case}
\subsection{Introduction : the problem of the constant term}
As mentioned earlier, we do not have any interpretation for
$\Lambda_{0}$. Until now, this constant has been very useful
because the case of constant $\theta$ has shown the main features
of the quantum equation of gravity, with only very simple
calculations. Now, in order to get rid of the problem of the
interpretation of $\Lambda_{0}$, we put it equal to zero. For the
computations below, we first leave $\theta(a)$ undetermined, in
order to study the different solutions of our equation.
\subsection{Computation of the equations}
We know that to determine $\epsilon$, there are two equations,
equivalent to our quantum problem, up to the constant of
integration $\Lambda_{0}$, which is now zero. These two equations
are (5.2) and (5.3) : Equation (5.3) is an equation in which
$\Lambda$ has been eliminated, and is still valid with varying
$\theta$, because $\theta$ appears only in $\Lambda$ :
\begin{equation}
\dot{b}-3\dot{c}=2\kappa(\epsilon)\dot{\epsilon}=2\kappa_{0}\frac{\dot{\epsilon}}{\sqrt{\epsilon}}
\end{equation}
(5.2) yields :
\begin{equation}
\dot{\theta}\frac{b}{3}(b-3c)+\theta
\frac{\dot{b}}{3}(b-3c)+\theta
\frac{b}{3}\dot{\widehat{(b-3c)}}=\kappa'(\epsilon)\epsilon\dot{\epsilon}=-\kappa_{0}\frac{\dot{\epsilon}}{2\sqrt{\epsilon}}
\end{equation}
In order to impose $\Lambda_{0}=0$, we have to use (5.4), as
explained in section 5 :
\begin{equation}
\frac{1}{2}(b-3c)+\Lambda=\kappa(\epsilon)\epsilon=\kappa_{0}\sqrt{\epsilon}
\end{equation}
We can now impose $\Lambda_{0}=0$ and the definition of $\Lambda$
becomes, from (5.1) :
\begin{equation}
\Lambda=\theta\frac{b}{3}(b-3c)
\end{equation}
Using (4.7), (3.32) and (3.33) are equivalent to :
\begin{equation}b=-\frac{3\ddot{a}}{a}\end{equation} and
\begin{equation}b-3c=6\left(\frac{K}{a^{2}}+H^{2}\right)\end{equation} We know that
\begin{equation}H=\frac{\dot{a}}{a}\end{equation} is no more constant. From (14.3), we can deduce :
\begin{equation}\frac{1}{2}(b-3c)-\frac{\ddot{a}}{a}\theta(b-3c)=\kappa(\epsilon)\epsilon=\kappa_{0}\sqrt{\epsilon}\end{equation}
Replacing now the value of $(b-3c)$, coming from (14.6), in the
last relation, we obtain :
\begin{equation}
6\left(\frac{K}{a^{2}}+H^{2}\right)\left(\frac{1}{2}-\theta
\frac{\ddot{a}}{a}\right)=\kappa(\epsilon)
\epsilon=\kappa_{0}\sqrt{\epsilon}
\end{equation}
We can also integrate equation (14.1) to find, making use of the
value of $b-3c$ again :
\begin{equation}
6\left(\frac{K}{a^{2}}+H^{2}\right)=2\int\kappa(\epsilon)d\epsilon=4\kappa_{0}\sqrt{\epsilon}+\lambda_{0}
\end{equation}
where $\lambda_{0}$ is a constant of integration. We emphasize
that in the last two equations, the first equalities are the exact
relations for any $\kappa(\epsilon)$, the second being of course
the case $$\kappa(\epsilon)=\frac{\kappa_{0}}{\sqrt{\epsilon}}$$
\subsection{Checking the dependence of the three equations}
We verify, in the case of the last value of $\kappa(\epsilon)$,
and as a check, that the three equations (14.2), (14.9) and
(14.10) are in fact dependent. We combine the first and third
equations (14.2) and (14.10), taking into account that (14.10) and
(14.1) are equivalent, and we find :
\begin{equation}
\dot{\theta}\frac{b}{3}(b-3c)+\theta\frac{\dot{b}}{3}(b-3c)+\theta
\frac{b}{3}\dot{\widehat{(b-3c)}}=-\kappa_{0}\frac{\dot{\epsilon}}{2\sqrt{\epsilon}}=-\frac{1}{4}\times2\kappa_{0}\frac{\dot{\epsilon}}{\sqrt{\epsilon}}
=-\frac{\dot{\widehat{(b-3c)}}}{4}
\end{equation}
which, for $\theta\neq 0$, gives :
$$\dot{\theta}\frac{b}{3}(b-3c)+\theta \frac{\dot{b}}{3}(b-3c)+\theta \left(\frac{b}{3}+\frac{1}{4\theta}\right) \dot{\widehat{(b-3c)}}=0$$
or :
$$\dot{\widehat{\left[\theta\left(\frac{b}{3}+\frac{1}{4\theta}\right)\right]}}(b-3c)+\theta\left(\frac{b}{3}+\frac{1}{4\theta}\right)\dot{\widehat{(b-3c)}}=0$$
and we find :
\begin{equation}
\left(\frac{1}{4}+\theta\frac{b}{3}\right)(b-3c)=K_{0}
\end{equation}
where $K_{0}$ is another constant of integration. We finally find
the equation :
\begin{equation}
6\left(\frac{K}{a^{2}}+H^{2}\right)\left(\frac{1}{4}-\theta\frac{\ddot{a}}{a}\right)=K_{0}
\end{equation}
We are left with three dependent equations (14.9), (14.10) and
(14.13), if and only if the following condition on the constants
of integration is satisfied :
\begin{equation}K_{0}=-\frac{1}{4}\lambda_{0}\end{equation}
To see this, we take (14.9) and add to it equation (14.10)
multiplied by $-1/4$. This result corresponds to what we wanted to
prove : the three equations are dependent, but the constants of
integration can no longer be taken at will.
\section{The three equations of quantum cosmology}
\subsection{The three equations}
We now summarize the former calculations to write down the system
of the three equations of movement, which are equivalent to the
whole set of equations describing quantum gravity. In the former
equation, we take the constants $K_{0}$ and $\lambda_{0}$ equal to
zero to obtain only the simplest solution:
\begin{equation}
\theta(a)\frac{\ddot{a}}{a}=\frac{1}{4}
\end{equation}
and also :
\begin{equation}
\frac{K}{a^{2}}+H^{2}=\frac{2}{3}\kappa_{0}\sqrt{\epsilon}
\end{equation}
We do not forget equation (5.5) which gives us the pressure $p$ :
\begin{equation} \frac{b+c}{2}-\Lambda=\kappa(\epsilon)p\end{equation}
\subsection{The equation for the pressure} We start from the former
relation, taking into account that
$$\kappa(\epsilon)=\frac{\kappa_{0}}{\sqrt{\epsilon}}$$ and also from (14.1), (14.5) and (15.1) :
$$\Lambda=\theta\frac{b}{3}(b-3c)=-\frac{\theta\ddot{a}}{a}(b-3c)=-\frac{1}{4}(b-3c)$$
Using equation (15.3) for $p$, we obtain :
\begin{equation}\frac{b+c}{2}-\Lambda=\frac{b+c}{2}+\frac{b-3c}{4}=\frac{3b-c}{4}=\frac{\kappa_{0}}{\sqrt{\epsilon}}p\end{equation}
We now use the identity :
$$\frac{3b-c}{4}=\frac{1}{12}(b-3c)+\frac{2}{3}b$$
Using equations (14.5) and (14.6) for the values of $b$ and $b-3c$
:
$$b=\frac{-3\ddot{a}}{a}$$ and
$$b-3c=6\left(\frac{K}{a^{2}}+H^{2}\right)$$
we find :
\begin{equation}\frac{1}{2}\left(\frac{K}{a^{2}}+H^{2}\right)-\frac{2\ddot{a}}{a}=\frac{\kappa_{0}}{\sqrt{\epsilon}}p\end{equation}
We now use the other two equations. From (15.2) we have :
\begin{equation}\frac{1}{2}\left(\frac{K}{a^{2}}+H^{2}\right)=\frac{1}{3}\kappa_{0}\sqrt{\epsilon}\end{equation}
From (15.1) we have :
\begin{equation}-\frac{2\ddot{a}}{a}=-\frac{1}{2\theta}\end{equation}
We combine all these relations to obtain :
\begin{equation}\frac{1}{3}\kappa_{0}\sqrt{\epsilon}-\frac{1}{2\theta}=\frac{\kappa_{0}}{\sqrt{\epsilon}}p\end{equation}
and finally :
\begin{equation}p=\frac{\epsilon}{3}-\frac{{\kappa_{0}}^{-1}\sqrt{\epsilon}}{2\theta}\end{equation}
We deduce from this equation that :
\begin{equation}p=\frac{\epsilon}{3}\left(1-\frac{1}{\frac{2}{3}\kappa_{0}\sqrt{\epsilon}\theta(a)}\right)\end{equation}
and also, using (15.2) :
\begin{equation}p=\frac{\epsilon}{3}\left(1-\frac{1}{\left(\frac{K}{a^{2}}+H^{2}\right)\theta(a)}\right)\end{equation}
We prove in the next section that there exists a function
$\theta(a)$ which gives the relation $p=\epsilon/3$ in the early
universe and gives the other relation $p=0$ for the present
universe, as it should.
\subsection{Behavior of the pressure}
We notice here that different choices of $\theta$ in the quantum
equation give different values for the pressure $p$, and
especially we can obtain any behavior of the kind
\begin{equation}p=\lambda\epsilon\end{equation} for any value of
$\lambda$, such that $0\leq\lambda\leq\frac{1}{3}$, provided we
make the right choice of $\theta$. Indeed, from (15.9) and
(15.12), we find :
$$\left(\frac{1}{3}-\lambda\right)\epsilon=\left(\frac{1-3\lambda}{3}\right)\epsilon=\frac{{\kappa_{0}}^{-1}\sqrt{\epsilon}}{2\theta}$$
or :
\begin{equation}\frac{1}{\theta}=(1-3\lambda)\left(\frac{2}{3}\kappa_{0}\sqrt{\epsilon}\right)=(1-3\lambda)\left(\frac{K}{a^{2}}+H^{2}\right)\end{equation}
For example the case $p=0$ is obtained for the choice :
\begin{equation}\theta=\frac{a^{2}}{K+\dot{a}^{2}}\end{equation}
It appears that there are three natural choices for the behavior
of $\theta$: proportional to $a^{2}$, inversely proportional to
$H^{2}$ or inversely proportional to $\kappa_{0}\sqrt{\epsilon}$.
These three choices display different behavior of the pressure
$p$. We have just seen that the last choice displays a behavior of
the kind $$p=\lambda\epsilon$$ with constant $\lambda$. The choice
\begin{equation}\frac{\theta_{1}}{H^{2}}\end{equation}
leads to
\begin{equation}p=\frac{\epsilon}{3}\left(1-\frac{\dot{a}^{2}}{\theta_{1}(1+\dot{a}^{2})}\right)\end{equation}
If we take the value $\theta_{1}=1$, the equation
\begin{equation}\frac{\ddot{a}}{a}=\frac{1}{4\theta(a)}=\frac{\dot{a}^{2}}{4a^{2}}\end{equation}
has a solution $a(t)=kt^{\alpha}$, because putting this value of
$a(t)$ in the equation, we obtain identically :
$\alpha(\alpha-1)=\alpha^{2}/4$ and $\alpha=4/3$. We thus find :
$\dot{a}\rightarrow0$ for the early universe and
$\dot{a}\rightarrow+\infty$ for the late universe. In these
conditions, formula (15.16) gives $p=\epsilon/3$ for the early
universe and $p=0$ for the late universe. In fact, the behavior of
$p$ in this case is not completely satisfying, because looking
more precisely at the formula for $p$, we see that even if $p$
tends to zero when $t\rightarrow+\infty$, it is not decreasing to
zero fast enough. For example, the value $\dot{a}=1$ gives only
$p=\epsilon/6$. Still, we notice here that concerning these
problems, we can improve the predictions of our model by changing
the behavior of two functions : $\theta(a)$ and
$\kappa(\epsilon)$.
\part{Generalized equation : the quantum solution to the cosmological problems}
In this part, we recall that we still take a varying $\theta$, and
at the same time, we impose the condition $\Lambda_{0}=0$. So the
$\Lambda$ term has now a completely determined origin : it
represents the exact Gauss-Bonnet term. In this context, we show
how the quantum equation of gravity can solve a number of
cosmological problems, by a judicious choice of $\theta$. We first
prove that the sign of $\theta$ should be positive, that the
expansion problem has always a solution, for any choice of
$\theta$. We then prove that the flatness and cosmological
constant problems find also their solutions from the quantum
equation. We finally recall that our equation of quantum gravity
leads to three equations for cosmology, (15.1), (15.2) and (15.9).
(15.1) gives the behavior of the parameter $a$ :
\begin{equation}\theta(a)\frac{\ddot{a}}{a}=\frac{1}{4}\end{equation}
The second equation (15.2) gives the relation between Hubble's
constant, the radius $a$ of the universe, and the energy density
$\epsilon$ :
\begin{equation}\frac{K}{a^{2}}+H^{2}=\frac{2}{3}\kappa_{0}\sqrt{\epsilon}\end{equation}
The last equation (15.9) gives the pressure $p$ :
\begin{equation}p=\frac{\epsilon}{3}-\frac{\sqrt{\epsilon}}{2\kappa_{0}\theta}\end{equation}
\section{Constraints on the Gauss-Bonnet parameter and the expansion problem}
\subsection{Returning to the formula for the pressure}
From the former relation
$$p=\frac{\epsilon}{3}-\frac{\sqrt{\epsilon}}{2\kappa_{0}\theta}$$
giving the value of the pressure $p$, we can give an important
constraint on $\theta(a)$. As already noticed, the former equation
for $p$ proves clearly that the value of $p$ depends essentially
on the value of $\theta(a)$, so an appropriate choice of $\theta$
can give us the value we need or want for $p$. Furthermore we see
that the value
\begin{equation}\frac{1}{\theta}=\frac{2}{3}\sqrt{\epsilon}=\frac{K}{a^{2}}+H^{2}\end{equation}
gives $p=0$ exactly. We also see that the quantum theory possesses
a very interesting limit, which is $\theta\rightarrow+\infty$. In
this case the particles are made of perfect relativistic stuff,
since then we have the exact relation :
\begin{equation}p=\frac{\epsilon}{3}\end{equation}
\subsection{The constraint on the Gauss-Bonnet parameter}
We reanalyzed the formula for $p$ because we also have on $p$ the
constraint :
$$0\leq p\leq\frac{\epsilon}{3}$$ The second inequality, compared to
the value of $p$, leads to
\begin{equation}\theta(a)>0\end{equation}
This sign of $\theta$ is a general condition which should always
be valid in the context of the quantum equation of gravity. We
notice that if in the context of constant $\theta$, we have been
able to study the case $\theta<0$, this was only because we
supposed a non vanishing $\Lambda_{0}$. We are now in the case
$\Lambda_{0}=0$, and for this reason a negative sign for $\theta$
is no longer possible.
\subsection{The expansion problem}
We recall that the expansion problem shall be solved once
explained why our universe is in expansion. In other words,
solving the problem is explaining, independently of any choice of
initial conditions for $\dot{a}$ in the early days of the
universe, why we have presently the condition $\dot{a}>0$. We now
turn to the relation :
$$\theta\frac{\ddot{a}}{a}=\frac{1}{4}$$
and see that, since $\theta>0$, and of course, since $a>0$, we
have $\ddot{a}>0$, and $\dot{a}$ is an increasing function of
time. This explains its positivity with no reference to initial
conditions, as proved below.
\subsection{Asymptotic behavior of the time derivative of the radius}
An increasing function like $\dot{a}(t)$ is not, mathematically,
necessarily positive. But, if we suppose that the universe is
necessarily old, the values of $t$ in our equations must be large.
We can suppose, mathematically, for this reason, that the present
regime corresponds to $t\rightarrow+\infty$. We just concluded
that $\theta(a)>0$ and $\ddot{a}(t)>0$. Thus, $\dot{a}(t)$ is
strictly increasing. An increasing function will necessarily have
a limit $\lambda$, when $t\rightarrow+\infty$. If
$\lambda\neq\pm\infty$, we know that we have, when
$t\rightarrow+\infty$ :
\begin{equation}a(t)\sim\lambda t\end{equation}
where the symbol $f(t)\sim g(t)$ is used, here, to signify that
the ratio of these two functions tends to $1$ when
$t\rightarrow+\infty$. When $\lambda<0$, then, in the
$t\rightarrow+\infty$ regime, $a(t)<0$ which is physically
impossible. When $\lambda=-\infty$ the situation is even worse. So
we conclude that $\lambda\geq0$. If we discard the case
$\lambda=0$, we are left with $\lambda>0$, and to reach this
strictly positive limit, the function $\dot{a}(t)$ has to be
strictly positive in the $t\rightarrow+\infty$ regime. We could
even discard the case $\lambda=0$ by mathematical arguments.
Indeed, we prove just below that in fact the value of $\lambda$,
more than a kind of initial condition for $t\rightarrow+\infty$,
is essentially determined by $\theta$ itself. This is another
feature of the quantum equation that initial conditions are
completely determined by the equation itself, and this is another
sign of its quantum nature. Indeed, we know that the Heisenberg's
incertitude relations are based on the principle that to go from
the classical theory to the quantum theory, the classical concept
of initial conditions has to be abandoned, as analyzed Landau.
\subsection{The role of the Gauss-Bonnet parameter}
Let us suppose here, to fix ideas, that $\lambda>0$, or
$\lambda=+\infty$. We have
$$\lim_{t\rightarrow+\infty}a(t)=\lambda$$ and we pose
$$\mu=\frac{\lambda^{2}}{\lambda^{2}+K}$$
Recalling that $K=+1$ and $K=-1$ correspond respectively to the
closed and open models, we have $0<\mu\leq1$ in the closed model
and $\mu\geq1$, or $\mu<0$, in the open model. We can adopt the
convention that $\mu=1$ in the case $\lambda=+\infty$, and this in
both models. We use again the equation :
$$\theta\frac{\ddot{a}}{a}=\frac{1}{4}$$
or equivalently :
$$\ddot{a}=\frac{a}{4\theta}$$ Multiplying this relation by
$\dot{a}$ we find :
\begin{equation}\ddot{a}\dot{a}=\frac{a\dot{a}}{4\theta}\end{equation}
and after integration, with $\dot{a_{0}}$ taken as a constant of
integration :
\begin{equation}\dot{a}^{2}={\dot{a_{0}}}^{2}+\int_{a_{0}}^{a}\frac{ada}{2\theta(a)}\end{equation}
where the condition for $\lambda$ finite is :
\begin{equation}\int_{a_{0}}^{+\infty}\frac{ada}{2\theta(a)}<+\infty\end{equation}
reminding ourselves that $\theta(a)>0$. We now make the hypothesis
that, when $t\rightarrow-\infty$, the universe is inflationary, so
at this value of $t$, it can be written that
$a_{0}=\dot{a}_{0}=0$, and we obtain :
\begin{equation}\lambda=\int_{0}^{+\infty}\frac{ada}{2\theta(a)}\end{equation}
This proves that with the additional assumption of inflation in
the early universe, which was our first postulate, $\lambda$ is
only a characteristic of $\theta(a)$, and is completely determined
by our equation. For the time being, it can be simply noticed that
postulating inflation is also a kind of initial condition, and
that $\lambda$ has in fact a double nature. A part of it, too,
having something to do with a constant of integration. We also
precise that when
$$\int_{a_{0}}^{+\infty}\frac{ada}{2\theta(a)}=+\infty$$ this time
$\lambda=+\infty$ and this relation is independent of any initial
condition : in this case $\lambda$ is entirely determined by
$\theta(a)$. In this case, the condition $\lambda=0$ has been
completely discarded mathematically.
\section{The age of the universe}
\subsection{Introduction} We want to study the condition $tH\sim1$
as $t\rightarrow+\infty$. We know that the relation $t\approx 1/H$
is observed, in the context of the standard model of cosmology
(Bennett and al., 2003, $[2]$). The situation for our quantum
equation is quite different. In the standard model, there are only
a few fixed parameters, and observations of the comic background
radiation have strong implications on the other predictions of the
model. In the quantum regime, we have an entire function
$\theta(a)$ to be determined, which leaves much more
possibilities. This is because the difficulty has been displaced.
With the choice of a entire function, it is easy to explain a lot
of phenomena, but the difficulty, in the quantum regime, is that
the function $\theta(a)$ also should explain how unification comes
into play in the picture. How this is done, is the subject of
$[57]$. In any case, it is interesting to see, at least as an
exercise, if there are choices of $\theta$ that yield the relation
$tH\sim1$ when $t\rightarrow+\infty$.
\subsection{The finite case} We suppose that
$0<\lambda<+\infty$ and we take $t\rightarrow+\infty$, so we know
that $\dot{a}\sim\lambda$ and that
$$H=\frac{\dot{a}}{a}\sim \lambda a^{-1}$$ Now integrating
$\dot{a}\sim\lambda$ in respect to $t$, we find $a(t)\sim\lambda
t$ so we find  $tH\sim1$ as wanted. Thus, any finite value of
$\lambda$ yields directly the observed relation.
\subsection{The infinite case} We have the relation :
\begin{equation}\dot{a}^{2}={\dot{a_{0}}}^{2}+\int_{a_{0}}^{a}\frac{xdx}{2\theta(x)}\end{equation}
We study the case : $\theta(a)\sim K_{n,p} a^{n}(\ln a)^{p}$ when
$a\rightarrow+\infty$.
\subsubsection{The case : n strictly greater than 2}
In order to have an infinite $\lambda$, the integral must be
divergent when $a=+\infty$, so finite $\lambda$ implies the
relation $n\leq2$, and $p\leq1$ if $n=2$. We recall that in any
other case, that is to say for $n>2$ and for $n=2$ with $p>1$,
$\lambda$ is finite and the condition $tH\sim1$ is verified from
the section 17.2.
\subsection{The case : n strictly less than 2}
To simplify the calculation, we stick to the case $p=0$. So it
remains the relation $\theta(x)\sim K_{n}x^{n}$ when
$x\rightarrow+\infty$, and
$\lim_{r\rightarrow+\infty}\dot{a}(t)=+\infty$, because $\lambda$
is infinite. It yields
$$\dot{a}^{2}=\dot{a_{0}}^{2}+\int_{a_{0}}^{a}\frac{xdx}{2\theta(x)}\sim\int_{a_{0}}^{a}\frac{xdx}{2K_{n}x^{n}}$$
Calculating the integral, we find :
$$\dot{a}^{2}\sim\frac{a^{2-n}}{2(2-n)K_{n}}$$ We thus have :
\begin{equation}a^{\frac{n-2}{2}}\dot{a}\sim\frac{1}{\sqrt{2(2-n)K_{n}}}\end{equation}
and :
\begin{equation}a^{n/2}H\sim\frac{1}{\sqrt{2(2-n)K_{n}}}\end{equation}
In the case $n<0$, we obtain, as $a\rightarrow+\infty$, the
condition $H\rightarrow+\infty$, ruled out by the small observed
value of $H$. The case $n=0$ is more interesting, it is analogous
to constant positive $\theta$, which we have already studied.
Sticking to the case $n>0$, and integrating our equation we find :
$$\frac{2}{n}a^{\frac{n}{2}}\sim\frac{t}{\sqrt{2(2-n)K_{n}}}$$
Finally
\begin{equation}tH\sim\frac{2}{n}\sqrt{2(2-n)K_{n}}a^{n/2}H\sim\frac{2}{n}\end{equation}
which rules out combinations $n<2;p=0$, for the condition
$tH\sim1$ is to be verified. Of course, for values of $n$ just a
little less than $2$, we have an approximate relation, since in
these cases $2/n\approx1$. We notice that the former calculations
seem to designate a special value of $n$ of particular interest :
to obtain $tH\sim1$, the former calculation under the hypothesis
$n<2$ led us back to the value $n=2$, which for this reason
appears as a kind of central candidate. This value $n=2$ is also
the value we can guess from dimensional arguments, recalling
$\theta$ has the dimension of a squared length.
\subsection{The case n=2} We now suppose that $\theta(x)\sim
Kx^{2}$ when $x\rightarrow+\infty$, and find that :
$$\dot{a}^{2}=\dot{a_{0}}^{2}+\int_{a_{0}}^{a}\frac{xdx}{2\theta(x)}\sim\int_{a_{0}}^{a}\frac{dx}{2Kx}$$
calculating the integral we find :
$$\dot{a}^{2}\sim\frac{1}{2K}\ln a$$ This yields
\begin{equation}\dot{a}\sim\frac{1}{\sqrt{2K}}\sqrt{\ln
a}\end{equation} or :
$$\frac{\dot{a}}{\sqrt{\ln a}}\sim\frac{1}{\sqrt{2K}}$$ and integrating
this equation :
$$\int_{a_{0}}^{a}\frac{dx}{\sqrt{\ln x}}\sim\frac{t}{\sqrt{2K}}$$
Calculating this integral by the change of variables :
$y=\sqrt{\ln x}$, we find :
$$\int_{a_{0}}^{a}\frac{dx}{\sqrt{\ln x}}=2\int_{\sqrt{\ln a_{0}}}^{\sqrt{\ln a}}e^{y^{2}}dy$$
and we use the relation :
$$\int_{\sqrt{\ln a_{0}}}^{X}e^{y^{2}}dy\sim\int_{1}^{X}e^{y^{2}}dy\sim\frac{e^{X^{2}}}{2X}$$
To prove the last relation, we notice that it can be written
$$\int_{1}^{X}e^{y^{2}}dy=\int_{1}^{X}\frac{2ye^{y^{2}}}{2y}dy$$
and integrating by parts, integrating $2ye^{y^{2}}$ and deriving
$(2y)^{-1}$, we find :
$$\int_{1}^{X}e^{y^{2}}dy=\frac{e^{X^{2}}}{2X}-\frac{e}{2}+\int_{1}^{X}\frac{e^{y^{2}}}{2y^{2}}dy$$
the second integral on the right being negligible compared to the
first on the left. The term $e/2$ is negligible too, because it is
finite compared to integrals which tend to infinity. So we obtain
the relation :
\begin{equation}\frac{t}{\sqrt{2K}}\sim\int_{a_{0}}^{a}\frac{dx}{\sqrt{\ln
x}}=2\int_{\sqrt{\ln a_{0}}}^{\sqrt{\ln
a}}e^{y^{2}}dy\sim\frac{a}{\sqrt{\ln a}}\end{equation} We finally
obtain, using (17.5) and (17.6) :
\begin{equation}tH=t\frac{\dot{a}}{a}\sim\sqrt{2K}\frac{a}{\sqrt{\ln
a}}\times\frac{1}{a}\times\frac{1}{\sqrt{2K}}\sqrt{\ln
a}=1\end{equation} which proves that in the case $n=2;p=0$, the
observed relation between $H$ and $t$ is satisfied. The former
calculation is another hint that the case $n=2$ should be
preferred. However, this central case is the sole case which
exhibits the condition $tH\sim1$, and $\lambda=+\infty$ together.
At the same time, it does not imply that the present value of
$\dot{a}(t)$ is much greater that a number of the order of unity.
Despite the fact that $\dot{a}\rightarrow+\infty$ when
$t\rightarrow+\infty$, the relation
$$\dot{a}\sim\frac{1}{\sqrt{2K}}\sqrt{\ln a}$$ shows that $\dot{a}$ tends
to its limit slowly enough to be even today very far from having
taken great values. This case also gives us hints on how could
work the equation in the very early universe, a fact that is
explained just below.
\section{The big bang and before}
\subsection{The direct calculation}
Formerly, we integrated the relation
$$\theta\frac{\ddot{a}}{a}=\frac{1}{4}$$ to obtain
\begin{equation}\dot{a}^{2}(t)=\dot{a_{pr}}^{2}+\int_{a_{pr}}^{a(t)}\frac{xdx}{2\theta(x)}=\dot{a_{0}}^{2}-\int_{a(t)}^{a_{pr}}\frac{xdx}{2\theta(x)}\end{equation}
Here, we can take for $a_{pr}$ the present value of the radius of
the universe, and $t_{pr}$ is the present value of the
cosmological time. Looking back in time, when the value of $a(t)$
was much smaller, we see that the former relation imposes at any
time :
\begin{equation}\int_{a(t)}^{a_{pr}}\frac{xdx}{2\theta(x)}\leq\dot{a_{pr}}^{2}\end{equation}
because the square $\dot{a}^{2}(t)\geq0$. Now, for values of
$\theta(x)$ such that, for fixed $\alpha$, the integral
\begin{equation}\int_{0}^{\alpha}\frac{xdx}{2\theta(x)}\end{equation}
diverges in the vicinity of $x=0$, the integral in (18.3) tends to
infinity, and (18.2) can no longer be verified. So, for theses
choices of $\theta(x)$, the initial singularity is smoothed out
even more drastically than by an exponential growth of $a(t)$. The
universe seems to have started with a strictly positive radius,
that we note $a_{0}$ from now on. In fact a problem immediately
arises. Going back in time, has this smallest value of the radius
of the universe been reached in finite or infinite time? If it has
been reached in finite time, the value $a_{0}$ is only a minimum
of the function $a(t)$, and we have to suppose that the universe
started with an infinite value of $a(t)$ for
$t\rightarrow-\infty$, then reduced to the minimum value $a_{0}$,
and then grew again to give the universe we know. All these
conclusions are based on the relation $\ddot{a}>0$. If the
smallest value $a_{0}$ is reached in infinite time, we have no
beginning for the universe either, but now with the picture of a
since ever growing universe, from a radius $a_{0}$ at the time
$t\rightarrow-\infty$. We take the value
$\theta(x)=\theta_{0}x^{2}$, where $\theta_{0}$ is a constant. We
see that $$\int_{0}^{\alpha}\frac{xdx}{2\theta(x)}$$ diverges in
the vicinity of $x=0$, so we find that there is a smallest
possible radius $a_{0}$. In both cases, if this value is only a
limit when $t\rightarrow-\infty$, or if this value is only a
minimum of $a(t)$ for a finite value $t_{0}$, we find that
$\dot{a}_{0}=0$. So the relation for $a(t)$ reads :
\begin{equation}\dot{a}^{2}(t)=\int_{a_{0}}^{a(t)}\frac{xdx}{2\theta(x)}\end{equation}
It is then a simple exercise to find that the value $a_{0}$ have
been reached in finite time. In fact there is no choice of
$\theta(x)$ that can change this general fact. Indeed, we know
that
$$\ddot{a}=\frac{a}{4\theta(a)}$$ When approaching the value
$a_{0}$, we can suppose that $\theta(x)$ is an increasing function
of $x$, even in the general case where it is only supposed that
the integral (18.3) diverges. Indeed, for the integral to diverge
in the vicinity of $x=0$, the ratio $\theta(x)/x$ has to tend to
zero, so it is natural to suppose that $\theta(x)$ is an
increasing function of $x$, at least for small values of $x$, and
$x=a_{0}$ is supposed to be a small value of $x$. When we go to
the value $a_{0}$, the function $\theta$ goes to its minimum value
$\tilde{\theta_{0}}$. For example
$\tilde{\theta_{0}}=\theta_{0}a_{0}^{2}$ in the particular
$\theta(x)=\theta_{0}x^{2}$ case. We thus see that $\ddot{a}$
tends to the finite strictly positive value
$a_{0}/4\tilde{\theta_{0}}$. If the value $a_{0}$ were reached for
$t\rightarrow-\infty$, we would have seen $\ddot{a}$ tend to zero,
which is impossible. So for any function $\theta(x)$, $a_{0}$ is
only a minimum reached in finite time. The conclusion is that our
universe had a shrinking phase from $a=+\infty$ when $t=-\infty$
to our big bang, $a=a_{0}$ and $t=0$, and is in a expanding phase
since then and for ever. We will study more precisely what could
have happened before the big bang in $[57]$, but we should be
aware that no definitive conclusion can be made about this period.
The general principle of logic is that it is never possible to
conclude in a region of knowledge where we cannot be contradicted
by experiment. As an exercise, we imagine in the next section a
physical principle, which cannot either be contradicted by
experiment, and which yields opposite conclusions about the big
bang.
\subsection{An indirect calculation}
We suppose now that the function $\theta(x)$ possesses discrete
values. In other words, we make an additional hypothesis on
$\theta$, analogous to the statement that permits to go from
classical values of the energy to quantum values : in classical
physics, the energy takes continuous positive values, whereas in
quantum physics, there is a mass gap, the first values of the
energy being quantized. So we suppose that $\theta(a)$ is defined
by a kind of approximate formula which still is :
$\theta(a)=\theta_{0}a^{2}$, analogous to the classical continuous
values of the energy, but that this formula has to be furthermore
corrected, by quantizing the values of $\theta$,
$\tilde{\theta_{0}}$ being its smallest strictly positive value,
and we note $\tilde{\theta_{1}}$ the smallest value of $\theta(x)$
strictly greater than $\tilde{\theta_{0}}$. What does now happen
for the parameter $a$ in the very early universe? The approximate
value of $\theta$ shows as we said that going back in time the
radius is shrinking to the value $a_{0}$, which is a minimum. Now
we take into account the true discrete values of $\theta$, which,
as we can see, makes $\theta$ become a step function. We see that
as the radius is shrinking, the discrete values of $\theta$ are
going smaller. When $a(t)$, which tends to $a_{0}$, becomes
strictly less than $a_{1}$ corresponding to $\tilde{\theta_{1}}$,
which means $\tilde{\theta_{1}}=\theta_{0}a_{1}^{2}$, the value of
$\theta$ becomes definitively equal to $\tilde{\theta_{0}}$, and
our universe becomes definitively exponentially growing, with a
constant value of Hubble's constant. This results from our
calculations of the constant and strictly positive $\theta$ case
of Part III. So we find another principle, which changes enough
the behavior of $\theta$, to make the big bang look completely
different, in such a manner that no sure conclusion can be made on
this matter which stays unreachable by experiments.
\section{The cosmological constant problem}
\subsection{The problem in the classical context}
As far as the standard model of cosmology is concerned, the
cosmological parameters of the model are measured with a very good
approximation (Bennett and al. 2003, $[2]$). In particular there
are, in this model, two important parameters, the total energy
density $\Omega\approx1$ and the energy density of dark matter
$\Omega_{\Lambda}$, the observed relation being :
\begin{equation}\Omega_{\Lambda}=\frac{3}{4}\Omega\end{equation}
There is a lot of dark energy density, which remains unexplained.
Furthermore, the model uses the equation of general relativity,
with a cosmological constant $\Lambda$ :
\begin{equation}R_{ik}-\frac{1}{2}Rg_{ik}-\Lambda g_{ik}=8\pi GT_{ik}\end{equation}
Now the term $\Omega_{\Lambda}$ is defined by the formula :
\begin{equation}\Omega_{\Lambda}=\frac{\Lambda}{3H^{2}}\approx\frac{3}{4}\end{equation}
The cosmological constant problem is to understand why a constant
like $\Lambda$ should be nonzero, and furthermore should possess
such a tiny strictly positive value :
\begin{equation}\Lambda\approx\frac{9H^{2}}{4}\end{equation}
Finally, we can state the problem in the following way : the
equations of general relativity are in the number of two, one
which gives $\epsilon$, the other gives $p$. We have, furthermore,
the equation of conservation of entropy, so three equations plus
the fact that they are dependent. So we choose two equations, say
the conservation of entropy and :
\begin{equation}R_{0}^{0}-\frac{1}{2}R-\Lambda =8\pi G\epsilon\end{equation}
If we pass the constant $\Lambda$ to the right hand side of the
equation, and insert it in the term in $\epsilon$, we find a new
$\epsilon$, which we could call $\epsilon_{app}$, because it is an
apparent energy density. The value of $\Lambda$ is such that :
\begin{equation}\epsilon_{app}=4\epsilon\end{equation}
As a remark, $\epsilon_{app}$ is the value of the observed energy
density, when the equation without the cosmological term is used,
that is to say we have
$$R_{0}^{0}-\frac{1}{2}R=8\pi G\epsilon_{app}$$
The other equation just gives the relation between $p$ and
$\epsilon$. If we want this conservation of entropy still to be
valid for apparent quantities, we have to pose $$p_{app}=4p$$ but
since in the case of the standard model we have $p=0$, this does
not change anything for the value of the pressure.
\subsection{The quantum equation}
In the context of the quantum equation, we know the origin of the
$\Lambda$ term. We know form (5.1) that :
\begin{equation}\Lambda=\theta(a)\frac{b}{3}(b-3c)+\Lambda_{0}\end{equation}
where we have supposed $\Lambda_{0}=0$, so the whole $\Lambda$
term has an identified origin : it corresponds to topological
corrections to classical gravity. We recall that we had :
$$b=\frac{-3\ddot{a}}{a}$$ and
$$\frac{\ddot{a}}{a}=\frac{1}{4\theta}$$ So we find that
\begin{equation}\Lambda=-\frac{b-3c}{4}\end{equation} whereas the
value of $G_{0}^{0}=R_{0}^{0}-\frac{1}{2}R$ can be read in (3.29)
and (14.6) :
\begin{equation}G_{0}^{0}=\frac{b-3c}{2}=3\left(\frac{K}{a^{2}}+H^{2}\right)\end{equation}
In the quantum context, we also have two equations, plus the
conservation of entropy, and they also are dependent. We can
choose equation (19.9) and the conservation of entropy. Then, we
have to analyze (19.9), and how $\epsilon$ is affected by
forgetting the $\Lambda$ Gauss-Bonnet term. We compute, using
(19.8), (19.9) and (15.2) :
$$G_{0}^{0}+\Lambda=\frac{b-3c)}{4}=\frac{3}{2}\left(\frac{K}{a^{2}}+H^{2}\right)=\kappa_{0}\sqrt{\epsilon}$$
If we forget the $\Lambda$ term in this equation, we have to
replace
\begin{equation}G_{0}^{0}+\Lambda=\kappa_{0}\sqrt{\epsilon}\end{equation}
by :
\begin{equation}G_{0}^{0}=\kappa_{0}\sqrt{\epsilon_{app}}\end{equation}
where $\epsilon_{app}$ is the apparent matter density, exactly as
we did in our analysis of the case of general relativity. The
difference is that now $\epsilon_{app}$ can be calculated from the
quantum equations and compared to the original $\epsilon$.
Equations (19.8) and (19.9) give directly :
\begin{equation}\Lambda=-\frac{1}{2}G_{0}^{0}\end{equation}
Forgetting $\Lambda$ in our equation would have the net effect of
changing
$$G_{0}^{0}+\Lambda=\frac{1}{2}G_{0}^{0}$$ for $G_{0}^{0}$. So we see
that the net effect of forgetting the $\Lambda$-term on the left
hand side of the equation is to multiply the right hand side by
$2$, which has the effect of doubling $\kappa_{0}$, if we
interpret this change in terms of a change of the gravitational
constant. However, if we prefer interpret the change in the
equation as a change in $\epsilon$, we get the right relation :
\begin{equation}\epsilon_{app}=4\epsilon\end{equation} and as
already noticed, $\epsilon_{app}$ is the new apparent matter
density. So the lack of the $\Lambda$-term in our equation makes
us see a density four times bigger than it should. This factor $4$
corresponds to a prediction of the quantum equation, and is equal
to the factor $4$ coming from the observations of the cosmos, in
the context of general relativity. This is striking enough to make
us think that we are on the right track with our equation of
quantum gravity.
\subsection{Complete calculation of the cosmological constant}
If we compute $\Omega_{\Lambda}$ by the method of section 13.1,
which uses the fact that our observations of the values of the
masses in the cosmos are only based on the principle of inertia,
we obtain the relation (13.3) :
\begin{equation}\Omega_{TOT}=\frac{G_{0}^{0}}{3H^{3}}\end{equation}
We recall that we had (19.12) :
\begin{equation}\Lambda=-\frac{1}{2}G_{0}^{0}\end{equation}
and from (19.9) and (15.2) :
\begin{equation}G_{0}^{0}=3\left(\frac{K}{a^{2}}+H^{2}\right)=2\kappa_{0}\sqrt{\epsilon}\end{equation}
So we obtain :
\begin{equation}\Omega_{TOT}=\frac{2\kappa_{0}\sqrt{\epsilon}}{3H^{2}}\end{equation}
We know that $\Lambda$ is negative because it possesses an extra
minus sign compared to the usual $\Lambda$ of general relativity.
Putting all these relations together, we find that our equation
predicts for the usual $\Lambda$ a positive value, verifying :
\begin{equation}\Omega_{\Lambda}=\frac{\Lambda}{3H^{2}}=\frac{1}{2}\Omega_{TOT}\end{equation}
which is clearly in the domain of uncertainties of the
observations, since this domain is determined by the relations
$$-1<\frac{\Lambda}{3H^{2}}<2$$
With $\Omega=1.02$, our $\Lambda$ is just at the center of the
former interval.
\subsection{A remark on the coefficients 2 and 4 of the former sections}
We now observe that the coefficient $4$ between the true physical
and apparent energy densities is only $4$ because it is viewed
from the place of $\epsilon$, under the square root. Of course
this coefficient becomes $2$, viewed from the place of
$\kappa_{0}$, or even from the place of $\Lambda$, that is to say
outside the square root. The interpretation of this factor $4$
depends on how the quantum equation is established in the context
of unification, and depends on the origin of the dependence of the
gravitational coupling $G$ on $\epsilon$. Here we just rapidly
explain how things could go, in a complete unified theory. When we
double $\kappa_{0}$, reestablishing $\hbar$ and $c$, we double in
fact $\kappa_{0}/\sqrt{\hbar c}$. Now, suppose that in a unified
theory, multiplying the gravitational constant by some factor has
the effect of multiplying also $\hbar$ by the same factor. In the
former section, we saw that the effect of $\Lambda$ was to
multiply, not $\kappa_{0}$, but $\kappa_{0}/\sqrt{\hbar c}$, by
$2$. Given our hypotheses, to multiply this term by $2$, we have
to multiply $\kappa_{0}$ by $4$, such that, $\hbar$ being
multiplied by $4$, the complete ratio $\kappa_{0}/\sqrt{\hbar c}$
is only multiplied by $2$. So it can be seen that in the
coefficient $4$ multiplying the energy density, and coming from
the Gauss-Bonnet term, there is most probably a factor $2$ which
is a classical correction to $\epsilon$, and another factor $2$
coming from further, more fundamental, quantum corrections to
$2\epsilon$. Or in other words, there is a factor $2$ coming from
the quantum corrections due to $\Lambda$, and another factor $2$
coming from corrections belonging to unification.
\section{The flatness problem}
\subsection{Value of the time derivative of the radius of the universe}
We have seen that, because $\dot{a}$ is an increasing function of
time, it has to possess a limit $\lambda$ when
$t\rightarrow+\infty$, where $\lambda>0$ is finite or infinite.
Furthermore, we have shown that in the case
$\theta(a)=\theta_{0}a^{2}$, the relation is (17.5) :
$$\dot{a}\sim\frac{1}{\sqrt{2\theta_{0}}}\sqrt{\ln(a/a_{0})}$$ such
that the present value of $\dot{a}$ is still finite, because even
if $\lambda=+\infty$, $\dot{a}$ goes so slowly to infinity, that
it should, in the present universe, take its value around unity.
In any case, we wrote
\begin{equation}\mu=\frac{\lambda^{2}}{\lambda^{2}+K}\end{equation}
with $0\leq\mu\leq1$, for the closed model, which is characterized
by the relation $K=+1$, and $\mu\geq1$ in the open model
corresponding to $K=-1$. We adopt the convention that $\mu=1$ in
the case $\lambda=+\infty$ for both models. However, from now on,
we note $\lambda$ the present value of $\dot{a}$, and as we said
$\lambda$ should be around unity.
\subsection{The classical and quantum flatness problems}
We want to prove that the total energy density of matter,
$\Omega_{TOT}$, that is to say the energy density when dark energy
is taken into account, at least has a present value near unity. In
classical gravity, the value of $\Omega_{TOT}$ takes the form
\begin{equation}\Omega_{TOT}=\frac{8\pi G\epsilon_{app}}{3H^{2}}\end{equation}
In the quantum regime, a reasonable relation between $G$ and
$\kappa_{0}$ is found by comparing general relativity and quantum
gravity, where we now have :
\begin{equation}\frac{K}{a^{2}}+H^{2}=\frac{2}{3}\kappa_{0}\sqrt{\epsilon}\end{equation}
In general relativity the relation was :
\begin{equation}\frac{K}{a^{2}}+H^{2}=\frac{8\pi
G}{3}\epsilon+\frac{\Lambda}{3}\end{equation} as can be seen for
example in Peebles, 1993, $[49]$, equation (5.18). As we know that
the contribution of $\Lambda$ is about three quarters of the total
energy density, sticking on the true $\epsilon$, we find :
\begin{equation}\frac{K}{a^{2}}+H^{2}=\frac{32\pi G}{3}\epsilon\end{equation}
Of course, this relation also results from the relation
$\epsilon_{app}=4\epsilon$, which is a consequence of the quantum
equation of gravity. We thus should have :
\begin{equation}\kappa_{0}=16\pi G\sqrt{\epsilon}\end{equation}
This relation can also be proved by using the principle of
equivalence between gravitation and inertia. Using this principle
we led us to (19.16), the quantum expression for $\Omega_{TOT}$
was found in (19.19). We thus obtain :
\begin{equation}\Omega=\frac{8\pi G\epsilon_{app}}{3H^{2}}=\frac{32\pi G\epsilon}{3H^{2}}=\frac{2\kappa_{0}\sqrt{\epsilon}}{3H^{2}}\end{equation}
and we find (20.6) again. We have to prove that this expression of
$\Omega_{TOT}$ tends to a finite value when $t\rightarrow+\infty$.
We know that the present value of $\dot{a}$ is $\dot{a}=\lambda$.
We thus find
$$\frac{1}{a^{2}}=\frac{H^{2}}{\lambda^{2}}$$ Finally, we obtain :
$$\frac{1}{\mu}H^{2}=\left(\frac{K}{\lambda^{2}}+1\right)H^{2}=\left(\frac{K}{\dot{a}^{2}}+1\right)H^{2}=\frac{K}{a^{2}}+H^{2}=\frac{2}{3}\kappa_{0}\sqrt{\epsilon}$$
such that :
\begin{equation}\frac{1}{\mu}H^{2}=\frac{2}{3}\kappa_{0}\sqrt{\epsilon}\end{equation}
We now use the value of $\Omega_{TOT}$ to find :
\begin{equation}\Omega=\Omega_{TOT}=\frac{2\kappa_{0}\sqrt{\epsilon}}{3H^{2}}=\frac{1}{\mu}\geq1\end{equation}
Here we suppose that we are in the closed model. Indeed, we have
\begin{equation}\frac{1}{\mu}=\frac{K}{\lambda^{2}}+1\end{equation}
in such a way that $1/\mu\approx1$ and $1/\mu\geq1$ if and only if
we are in the closed model. Since the observed value of
$\Omega_{TOT}$ seems to be just a little greater than $1$, we can
conclude we are in the closed model. When $\lambda$ is not used
anymore to note the present value of $\dot{a}$, but rather its
limit when $t\rightarrow+\infty$, our result is not the present
value of $\Omega$ but its limit value. The present value have been
observed to be, in the context of the standard cosmological model
(Bennett and al., 2003) :
\begin{equation}\Omega=1.02\pm0.02\end{equation}
To find $\Omega=1.02$ in the quantum context, we need the present
value of $\dot{a}$ to be
$$\dot{a}_{0}=\lambda=7.07$$ and to find the greatest possibility
$\Omega=1.04$ we need
$$\dot{a}_{0}=\lambda=5$$
A remark can be made : if it can be observed, in our universe,
distances of the order of $200Mpc$, and if the $cH^{-1}$ distance
is about $4000Mpc$, we then are sure that $\dot{a}\geq1/20=0.05$.
That the universe could be one hundred times bigger than this
minimum value does not seem a priori to be ruled out by any
experiment, and only very small values of $\dot{a}$ are ruled out.
\subsection{Value of one coefficient}
We see that our equations have the remarkable property to explain,
first, why the value of $\Omega$ is so near unity, but also that
it is strictly greater than $1$. The observed value of $1.02$ fits
perfectly with our equations, and proves furthermore that we are
in the closed model. In the case where
$\theta(a)=\theta_{0}a^{2}$, we found the relation (17.5) for
$\dot{a}$ :
\begin{equation}\dot{a}=\frac{1}{\sqrt{2\theta_{0}}}\sqrt{\ln(a/a_{0})}\end{equation}
To give an approximate value of $\theta_{0}$, we can make the
hypothesis that the approximate value of $a_{0}$ in the quantum
theory is the value of the radius of the universe in the standard
early phase of the universe. We find a relation of the kind
\begin{equation}\frac{a}{a_{0}}\approx10^{10}\end{equation}
Replacing this value in (20.11) we find $\dot{a}=7.07$ for the
value $\theta_{0}\approx5.3$, so the solution of the flatness and
cosmological constant problems did not make appear a new
$\theta_{0}$ problem, since this time $\theta_{0}$ is around
unity.
\section{Towards unification : the structure of matter and the ratio baryon to photon number}
Analyzing the formula for $p$ in the case
$\theta(a)=\theta_{0}a^{2}$, we find, form (15.11) :
\begin{equation}p=\frac{\epsilon}{3}\left(1-\frac{1}{\left(\frac{K}{a^{2}}+H^{2}\right)\theta(a)}\right)=
\frac{\epsilon}{3}\left(1-\frac{1}{K+\dot{a}^{2}\theta_{0}}\right)\end{equation}
This is clear that this relation has not the right behavior, in
order to predict $p\rightarrow0$ when $t\rightarrow+\infty$, but
it has the right form to predict once again a value of
$\theta_{0}$ near unity. Indeed, the condition $0\leq
p\leq\epsilon/3$ is equivalent to
$$\theta_{0}\geq\frac{1}{K+\dot{a}^{2}}$$
The question here is to know what happens if the $p=0$ condition
cannot be verified anymore for the present universe. This delicate
problem is studied in $[57]$ and $[58]$. We just sketch here the
ideas that will be developed there. There are two ways to get rid
of the relation $p=0$. The first is to consider a particle made of
a small sphere of radius $r$, and huge energy density
$\epsilon_{P}$. Outside particles, in vacuum, there is no
pressure, $p=0$, but there is no energy density either :
$\epsilon=0$. Let us suppose now that the stuff making the
particles is relativistic, in such a way that inside the particle
we have the relation $p=\epsilon_{P}/3$. The values of $p$ and
$\epsilon$ which should be taken in the cosmological equation are
the mean values of $p$ and $\epsilon$, these mean values being
calculated over all parts of space, inside and outside particles.
If, inside particles, we have $p=\epsilon_{P}/3$, outside
particles this relation is still valid because there we have
$p=\epsilon/3=0$. So the mean values of the pressure and the
energy density still will verify the relation $p=\epsilon/3$ in
the cosmological equation. As we can see, the value of $p$ in the
cosmological equation probes the structure of particles in the
intermediate model. In the pointlike particles model, the relation
between $p$ and $\epsilon$ is calculated considering the mean
relative velocities between particles. Now, if one electron for
example, is made of more fundamental particles, which have great
relative velocities between each other, but still are confined
inside the electron, in the same manner that confined quarks live
inside a proton, we find as well in the pointlike model a relation
of the type $p=\epsilon/3$. Such hypotheses, which ultimately
belong to the domain of application of unification, can be proved
exact or at least very probable, because they fit with fundamental
experimental observations as well as with the most theoretical
pictures. From the theoretical point of view, if we imagine these
more fundamental particles, which possess a relativistic speed,
and a nonzero radius, and move, confined, inside the electron, we
notice that these small spheres should be Lorentz contracted along
the direction of their movement, in such a way that they should
loose one dimension. Indeed the radius of the sphere along this
direction should be $r=\sqrt{1-v^{2}/c^{2}}\approx0$, because we
have $v\approx c$. So these more fundamental particles are
membranes as in M-Theory. Such an hypothesis, like $p=\epsilon/3$,
fits with observations, because the value of $p$ does not only
probes the structure of matter. It also gives information on the
the total number $N_{B}$ of baryons in the universe. A relation of
the kind $p=\epsilon/3$ implies, since in the present universe the
energy density of photons is negligible, $p_{B}=\epsilon_{B}/3$,
where $p_{B}$ and $\epsilon_{B}$ are the baryonic pressure and
energy density. Such a relation for baryons, associated with the
conservation of entropy, and beyond their relativistic structure,
gives a behavior of the baryonic energy density of the kind
$\epsilon_{B}\sim1/a^{4}$. It is well known that when the total
number of baryons is constant, the number density is proportional
to $1/a^{3}$. Here, since $\epsilon_{B}\sim1/a^{4}$, we deduce
that the total number of baryons is proportional to $1/a$. If the
present value of the ratio $\eta$ of the total number of baryons
to the total number of photons has the observed value, in the
context of general relativity, of $\eta\approx6.1\times10^{10}$
(Bennett and al. 2003), and if the ratio $a/a_{0}$ of the present
radius of the universe to the radius of the early universe has the
value, computed with the standard cosmological model :
$a/a_{0}\approx10^{10}$, we must conclude that the hypothesis
$N_{B}\sim1/a$ is just fine to obtain that in the early universe,
there were about the same number of photons and baryons. Once
again, such conclusions belong to the domain of unification, and
will be treated at length in $[56]$ and $[57]$.
\section{A limit case of the theory}
There is a limit case of our quantum equation of gravity, which is
the case $\theta\rightarrow+\infty$. Looking at the original
equation of quantum gravity, we see that the term $\theta$
multiplies only the Gauss-Bonnet term. So if we multiply the two
sides of the equation by $1/\theta$ and if we take the limit
$\theta\rightarrow+\infty$, the theory we obtain in this limit
implies the vanishing of the Gauss-Bonnet topological term.
Furthermore, the equation for $a$, which took the form :
$$\frac{\ddot{a}}{a}=\frac{1}{4\theta(a)}$$ now becomes $\ddot{a}=0$
or $\dot{a}=Cte$. We thus find a value of the deceleration
parameter $q=0$. The value of $p$ is interesting since this time,
from (15.9), we find the exact relativistic relation :
$$p=\epsilon/3$$
Also very interesting, in this limit, is the relation for
$\epsilon$ :
\begin{equation}\frac{K}{a^{2}}+H^{2}=\frac{2}{3}\kappa_{0}\sqrt{\epsilon}\end{equation}
If we note $\dot{a}(t)=\lambda$, always possible since
$\dot{a}(t)$ is constant, the relation (22.1) becomes :
\begin{equation}\frac{\lambda'}{a^{2}}=\frac{2}{3}\kappa_{0}\sqrt{\epsilon}\end{equation}
where we have $\lambda'=K+\lambda^{2}$. We find that in this case,
the closed, open and flat model are strictly equivalent, since
they only differ by the value of $\lambda'$, which also depends on
the initial condition $\lambda$, which is the value of $\dot{a}$
at $t=0$. So the flatness problem finds here a complete solution.
The value of $\Omega_{TOT}$ is now constant and equal to :
\begin{equation}\Omega_{TOT}=\left(1+\frac{K}{\lambda^{2}}\right)\end{equation}
which is not necessarily equal to $1$, but still, the values
strictly greater than $1$, strictly less than $1$, or equal to $1$
correspond respectively to the closed, open and flat model. The
formula :
\begin{equation}\Omega_{\Lambda}=\frac{1}{2}\Omega\end{equation}
which corresponds to the behavior
$$\kappa(\epsilon)\sim\epsilon^{-1/2}$$
remains the same for all values of $\theta$. Thus, this relation
is unchanged in the limit $\theta\rightarrow+\infty$. The origin
of this $\Lambda$ is what could be called a Gauss-Bonnet ghost
term. It is a relic of the Gauss-Bonnet term, which does not
vanish in this specific formula when we study the limit theory
$\theta\rightarrow+\infty$, whereas the Gauss-Bonnet term itself
vanishes in the equation of quantum gravity in this limit. Since
in this case, $\Omega_{TOT}$ is constant, the term
$\Omega_{\Lambda}$ is constant too, and the $\Lambda$ term in the
equation is strictly proportional to $H^{2}$. We see that this
limit case gives a perfect solution for the flatness problem as we
said, but also a solution to the cosmological constant problem,
via a $\Lambda$ term having a ghost topological origin. This limit
also gives a perfect relativistic structure for the fundamental
particles, which corresponds to a solution to the problem
corresponding to why the present value of the ratio $\eta$ is so
small. If we want to change the value $\Omega_{\Lambda}$ in (22.4)
for the relation
\begin{equation}\Omega_{\Lambda}=\frac{3}{4}\Omega\end{equation}
with the same ghost topological origin for $\Lambda$, it suffices
to consider the case in which
$$\kappa(\epsilon)\sim\epsilon^{-3/4}$$ Indeed, relations (14.9) and (14.10) give
\begin{equation}6\left(\frac{K}{a^{2}}+H^{2}\right)\left(\frac{1}{2}-\theta\frac{\ddot{a}}{a}\right)=\kappa(\epsilon)\epsilon\end{equation}
and
\begin{equation}6\left(\frac{K}{a^{2}}+H^{2}\right)=2\int\kappa(\epsilon)d\epsilon\end{equation}
Taking now $\kappa(\epsilon)=k\epsilon^{-\alpha}$, we find :
\begin{equation}6\left(\frac{K}{a^{2}}+H^{2}\right)=\frac{2k\epsilon^{1-\alpha}}{1-\alpha}\end{equation}
and we also find, using (22.6) :
\begin{equation}\frac{2k}{1-\alpha}\left(\frac{1}{2}-\theta\frac{\ddot{a}}{a}\right)=k\end{equation}
We obtain :
\begin{equation}\frac{\ddot{a}}{a}=\frac{\alpha}{2\theta}\end{equation}
We then use (14.4) and (14.5) :
\begin{equation}-\Lambda=\theta\frac{\ddot{a}}{a}(b-3c)\end{equation}
and (19.9) :
\begin{equation}G_{0}^{0}=\frac{(b-3c)}{2}\end{equation} so :
\begin{equation}-\Lambda=\alpha G_{0}^{0}\end{equation}
The conclusion is then :
\begin{equation}\Omega_{\Lambda}=\alpha\Omega_{TOT}\end{equation}
So the value $\alpha=3/4$ gives (22.5), but implies a
gravitational coupling which has the dimension of an energy.
\part{Conservation of energy and topology}
\section{Introduction}
\subsection{Some well known facts}
\paragraph{Conservation of energy}
If we look at the Einstein equations $R_{ik}-\frac{1}{2}Rg_{ik}=
\kappa T_{ik}$ of general relativity, we see that the
gravitational part is composed of a tensor $G_{ik}$, which
verifies minimal conditions for the equation possible. The first
condition, which enabled Einstein to find out his tensor, is that
it should be constructed out of second derivatives of the
fundamental variables of the theory, which are the $g_{ik}$.
Mathematically, this means that $G_{ik}$ must be constructed from
the curvature tensor. Looking at the other side of the equation,
we immediately see another necessary condition on $G_{ik}$,
imposed by the law of conservation of energy $\nabla^{i}T_{ik}=0$
on the matter tensor. So the equation is possible if and only if
$\nabla^{i}G_{ik}=0$. In fact, the tensor calculus provides us
with this equation by a formal computation.
\paragraph{Dimension and topology}
However, the tensor $G_{ik}=R_{ik}-\frac{1}{2}Rg_{ik}$ has
dramatically different properties, depending on the dimension of
space-time, and particularly its properties are different in the
case $D=2$ and when $D\geq3$. In dimension $D=2$, the
Hilbert-Einstein action $\int\sqrt{-g}R$ is topological,  and the
Einstein tensor $R_{ik}-\frac{1}{2}Rg_{ik}$ possesses the
condition of conformal invariance, we mean that its trace
vanishes.
\subsection{Constructing other tensors for gravity}
\paragraph{A dimensionless coupling constant}
Using these first remarks, we consider the mathematical problem to
construct all possible tensors $\Sigma_{ik}$, made of the
curvature tensor, and verifying the necessary law of conservation
of energy : $\nabla^{i}\Sigma_{ik}=0$. We will see that, if
$G_{ik}$ is the only tensor made of $R_{ijkl}$, of degree one in
$R_{ijkl}$, and verifying the law of conservation of energy, there
also is a unique tensor made of $R_{ijkl}$, of degree two in
$R_{ijkl}$, and verifying the same law. The essential feature of
this tensor is that possesses, in dimension $D=4$, the properties
of the Einstein tensor in $D=2$. It is conformal invariant, we
mean that its trace vanishes, and it has a dimensionless
gravitational coupling constant. It is clear that it can
conjectured that there exists, for each integer $n$, a unique
tensor made of $R_{ijkl}$, of degree n in $R_{ijkl}$, which is
conformal invariant and which possesses a dimensionless coupling
constant in dimension $D=2n$
\paragraph{When topology appears}
In fact, these tensors of degree $n$ in $R_{ijkl}$ have, in their
respective dimension $D=2n$, another property of Einstein's tensor
in $D=2$ : they are trivial, because they are topological. Thus,
in dimension $D=4$, starting from a tensor of degree $2$ in
$R_{ijkl}$, we can see in the calculation the following striking
property : the sole condition of conservation of energy, makes
appear in our tensor the exact coefficients of the topological
Gauss-Bonnet term. In dimension $D=2n$, the mathematical
conjecture is that the sole equation of conservation of energy
makes appear in the tensor of degree $n$ in $R_{ijkl}$ the
coefficients of the Euler form. Since Donaldson invariants and
then Seiberg-Witten invariants, we know a lot about the relations
between physics and topology. Here, we use such a simple and
direct relation between these two fields to construct another kind
of quantum equation of gravity.
\paragraph{Complex gravity and the other quantum interactions}
In the quantum context, the wavy nature of matter is reflected by
the fact, that in some way, complex field variables come into
play. Looking carefully at a list in which all energy-momentum
tensors, ready to quantization, are written down and put together,
(Grib, Mamayev, Mostepanenko 1992, $[18]$ Part I, Chapter 1), and
by simple inspection, we observe general quantum features : all
these tensors are of degree two in complex field variables and the
doubling is made via complex conjugates. Applying the same rules,
by analogy, to gravity, we arrive at a natural conclusion :
gravity should be complex, we mean $g_{ik}$ should be complex, the
tensor for gravity should be of degree $2$, it thus should be the
complex analog of the vanishing topological real tensor of degree
$2$, which makes appear $D=4$ as a preferred dimension of
space-time. We will not investigate more this complex tensor here,
but if, in the complex case, this tensor is effectively non
vanishing, we believe these links between reality and complexity,
conservation of energy and topology, could be the key to
understand why our world possesses four dimensions.
\section{The tensor of degree two}
\subsection{Einstein's tensor of degree one}
We just remember how we prove the existence and the uniqueness of
$G_{ik}$ of degree one in $R_{ijkl}$. As $G_{ik}$ is of degree one
in $R_{ijkl}$, only can it contain $R_{ik}$ and $R$. So we have
$G_{ik}=R_{ik}+\alpha R g_{ik}$ where $\alpha$ is a constant to be
determined. Using the tensor calculus which gives formally
$\nabla^{i}R_{ik}=\frac{1}{2}\partial_{k}R$, we see that
$\nabla^{i}G_{ik}=(\alpha+\frac{1}{2})\partial_{k}R=0$ if and only
if $\alpha=-\frac{1}{2}$. This gives the existence and the
uniqueness of the tensor, as well as its exact expression. We now
study the case of the degree two.
\subsection{Ingredients for the tensor of degree two}
\paragraph{The method} Many more terms will contain the
tensor of degree two, because in this case, there are the
possibilities of using the four indexed $R_{ijkl}$, with indices
contracted, as in $R_{iabc}{R_{k}}^{abc}$ or as in
$R_{iakb}R^{ab}$. So we first have to determine all possible
terms, and then calculate all the constants appearing in the
linear combination forming our tensor. We recall that we note
$\Sigma_{ik}$ for this tensor. Next, using the law of conservation
of energy for $\Sigma_{ik}$, we show that there is a unique
solution to this set of constants.
\paragraph{The general form of the tensor} To find the components
of $\Sigma_{ik}$ of degree two, we have simply to multiply two
tensors of the form $R_{abcd}$, $R_{ab}$, or $R$ and use as well
$g_{ik}$, where the indices $a,b,c,d$ are chosen to be $i$ or $k$,
or are otherwise contracted. \paragraph{Products containing the
scalar curvature $R$} For a product $R^{2}$ the only possible term
is $R^{2}g_{ik}$, for a product of $R_{ab}$ with $R$, again one
possibility, which is $RR_{ik}$.
\paragraph{Products Ricci-Ricci} For two products of $R_{ab}$, the
indices $i$ and $k$ must belong to different $R_{ab}$, to avoid
the appearance of the contraction $R$, a case already studied, and
using that $R_{ab}$ is symmetric, we get the only
$R_{ia}{R_{k}}^{a}$.
\paragraph{Products Ricci-Riemann} For products of $R_{ab}$ and
$R_{abcd}$, the term $R_{ik}$ cannot appear, otherwise the
contraction of $R_{abcd}$ is $R$. As well, if $R_{ia}$ appears,
using the symmetries in the indices of $R_{abcd}$, we can suppose
that $k$ is the first index. We have then an expression of the
form $R_{ia}R_{kpqr}$, where, among the indices, two are in the up
position, one in the down position, $a$ appears once, in the up
position, to be contracted with the index $a$ of $R_{ia}$, and say
$b$ appears twice among $p$, $q$ and $r$, and is contracted. Then,
in this Riemann tensor $a$ cannot be the second index, otherwise
the contraction over $b$ is zero, so we can suppose $a$ is the
third index, and the contraction over $b$ gives us another Ricci.
So nor $i$ neither $k$ can appear in the Ricci, and we have then
an $R^{ab}$ where $a,b$ are to be contracted with indices of a
Riemann tensor. As $R^{ab}$ is symmetric in $a,b$, it cannot be
contracted with indices $a,b$ placed in an antisymmetric position
in $R_{pqcd}$, and as $R_{pqcd}$ is antisymmetric in the first two
indices and also in the last two, there is one $a$ in the first
two and one $b$ in the last two. Using again the symmetries of the
indices in the Riemann tensor, we can chose $i$ in first place and
$k$ in the third, and we are left with the only possibility
$R^{ab}R_{iakb}$.
\paragraph{Products Riemann-Riemann} For the product of two Riemann
tensors, it is quite direct to see that the only possibility is
${R_{i}}^{abc}R_{kabc}$. First, as before, we can suppose that $i$
is the first index of the first Riemann. Now if $i$ and $k$ appear
only in the first Riemann, $c$ for example appears twice in the
second, giving us zero or Ricci. So $k$ is the first index of the
second Riemann. Now we can chose the first as ${R_{i}}^{abc}$ and
using the antisymmetry of the second tensor in the last two
indices, we can suppose that in it, the last two indices are in
alphabetical order. We are left with $R_{kabc}$, $R_{kbac}$ and
$R_{kcab}$. Using now that in the first Riemann $b$ and $c$ appear
in antisymmetric positions, we have
${R_{i}}^{abc}R_{kbac}=-{R_{i}}^{abc}R_{kcab}$. Using finally the
identity $R_{kabc}-R_{kbac}+R_{kcab}=0$, we see that all possible
tensors can be written only in terms of ${R_{i}}^{abc}R_{kabc}$.
\paragraph{Terms involving the metric tensor} In all this, we have
discarded the possibility of the appearance of $g_{ik}$, but the
same arguments permit to conclude that the only possible terms are
$R^{(4)}g_{ik}$ where $R^{(4)}=R^{abcd}R_{abcd}$, $R^{(2)}g_{ik}$
where $R^{(2)}=R^{ab}R_{ab}$ and of course the $R^{2}g_{ik}$ first
considered. \paragraph{Synthesis} To conclude we get then the most
general tensor $\Sigma_{ik}$ of degree two :
\begin{equation}
\Sigma_{ik}={R_{i}}^{abc}R_{kabc}+\alpha R_{iakb}R^{ab}+\beta
R_{ia}{R_{k}}^{a}+\gamma R_{ik}R+\left(\delta R^{(4)}+\epsilon
R^{(2)}+\eta R^{2}\right)g_{ik}
\end{equation}
\subsection{Three formulas}
In order to calculate the coefficients appearing in
$\nabla^{i}\Sigma_{ik}$, we need a first formula :
\begin{equation}
\nabla^{i}R_{iabc}=\nabla_{b}R_{ac}-\nabla_{c}R_{ab}
\end{equation}
Starting with the Bianchi identity :
\begin{equation}
\nabla_{m}R_{nabc}+\nabla_{c}R_{namb}+\nabla_{b}R_{nacm}=0
\end{equation}
and contracting over m and n, we obtain (24.2) directly :
$$\nabla^{n}R_{nabc}+\nabla_{c}R_{ab}-\nabla_{b}R_{ac}=0 $$ Here, we
adopt the convention that the contraction of the first and the
third indices in the Riemann tensor gives the Ricci tensor, and
then the contraction of the first and fourth indices in the
Riemann tensor gives minus the Ricci tensor, because of the
antisymmetry of third and fourth indices in the Riemann tensor.
This gives the formula (24.2). Now, we calculate the coefficients
of $\nabla^{i}\Sigma_{ik}$ one by one. First we need a second
formula :
\begin{equation}
\nabla^{i}({R_{i}}^{abc}R_{kabc})=(\nabla_{b}R_{ac}){R_{k}}^{abc}-(\nabla_{c}R_{ab}){R_{k}}^{abc}+2R^{iabc}(\nabla_{c}R_{kabi})
\end{equation}
Using the properties of the connection $\nabla$, we find :
\begin{equation}
\nabla^{i}({R_{i}}^{abc}R_{kabc})=(\nabla^{i}{R_{i}}^{abc})R_{kabc}+{R_{i}}^{abc}(\nabla^{i}R_{kabc})
\end{equation}
and using
\begin{equation}(\nabla^{i}{R_{i}}^{abc})R_{kabc}=(\nabla^{i}R_{iabc}){R_{k}}^{abc}\end{equation}
as well as equation (24.2), we obtain immediately that the first
term of the right hand side of (24.5) equals the first two terms
of formula (24.2). So, we only need to prove that the second term
on the right hand side of (24.5) equals the third of formula
(24.4). Now, using the Bianchi identity (24.3), we have :
\begin{equation}{R_{i}}^{abc}(\nabla^{i}R_{kabc})=-{R_{i}}^{abc}\nabla_{b}{R_{kac}}^{i}-R^{iabc}\nabla_{c}R_{kaib}\end{equation}
Using the antisymmetry of the indices $b$ and $c$ in the first
Riemann tensor of the term
$-{R_{i}}^{abc}\nabla_{b}{R_{kac}}^{i}$, we obtain that this term
equals ${R_{i}}^{abc}\nabla_{c}{R_{kab}}^{i}$, which also equals
the term $-R^{iabc}\nabla_{c}R_{kaib}$, using the antisymmetry of
$i$ and $b$ in the second Riemann tensor. Altogether, we see that
formula (24.4) has been proved. Now, we study the term arising in
$\nabla^{i}\Sigma_{ik}$ from the second term of $\Sigma_{ik}$,
asserting the following formula :
\begin{equation}
\nabla^{i}(\alpha R_{iakb}R^{ab})=\alpha
R^{ab}(\nabla_{k}R_{ab})-\alpha R^{ab}(\nabla_{b}R_{ak})-\alpha
(\nabla_{c}R_{ab}){R_{k}}^{abc}
\end{equation} To prove it, we use :
$$\nabla^{i}(R_{iakb}R^{ab})=(\nabla^{i}R_{iakb})R^{ab}+R_{iakb}\nabla^{i}R^{ab}$$
\begin{equation}
=(\nabla^{i}R_{iakb})R^{ab}+{{R^{ia}}_{k}}^{b}\nabla_{i}R_{ab}
\end{equation}
Now, applying to the first term on the right hand side of (24.9)
the identity (24.2), we readily obtain the first two terms on the
right hand side of formula (24.8). But the second term on the
right hand side of (24.9) can be written
$(\nabla_{c}R_{ab}){{R^{ca}}_{k}}^{b}$ and since exchanging the
two pairs of indices in the Riemann tensor does not change its
value, we obtain $(\nabla_{c}R_{ab}){R_{k}}^{bca}$. Now, using the
well known identity
\begin{equation}
R_{iklm}+R_{imkl}+R_{ilmk}=0
\end{equation}
this term becomes
$$-(\nabla_{c}R_{ab}){R_{k}}^{abc}-(\nabla_{c}R_{ab}){R_{k}}^{cab}$$
In this last equation, the second term gives zero because $a,b$
are contracted and appear in symmetric positions in the Ricci
tensor and in antisymmetric positions in the Riemann tensor. This
finishes the proof of formula (24.8).
\subsection{Computing the coefficients of the tensor}
Taking the $\beta$ and $\gamma$ terms of $\Sigma_{ik}$, and
remembering that $\nabla^{i}R_{ik}=\frac{1}{2}\partial_{k}R$, we
find at once :
\begin{equation}
\nabla^{i}(\beta R_{ia}{R_{k}}^{a})=\frac{1}{2} \beta
(\partial^{a}R)R_{ak}+ \beta R^{ia} \nabla_{i} R_{ka}
\end{equation}
and
\begin{equation}
\nabla^{i}(\gamma R_{ik}R)=\frac{1}{2} \gamma (\partial_{k}R)R+
\gamma R_{ik} (\partial^{i}R)
\end{equation}
\paragraph{Computing $\alpha$, $\beta$, $\gamma$}
Looking closely at our equations, we see that the second terms of
(24.8) and (24.11) can be eliminated by the choice $\alpha=\beta$,
and that the first term of (24.11) gives zero, when combined with
the second term of (24.12), provided that we choose the relation
$\beta=-2\gamma$. So, by simple inspection of our equations, we
possess an easy way to calculate our coefficients.
\paragraph{A relation which simplifies the whole calculation}
Turning now to the $\delta$-term, we have :
$$\nabla^{i}\left(R^{(4)}g_{ik}\right)= \nabla^{i}[R_{abcd} R^{abcd}g_{ik}] = 2 (\nabla_{k}R_{abcd}) R^{abcd}$$
which equals :
$$-2R^{abcd}\nabla_{d}R_{abkc}-2R^{abcd}\nabla_{c}R_{abdk}$$
because of (24.3). Both of these terms equal
$2R^{abcd}\nabla_{c}R_{abkd}$, the second because in the second
Riemann tensor, $d$ and $k$ are in antisymmetric positions, and
the first because in the first Riemann tensor, $c$ and $d$ are in
antisymmetric positions. We thus find :
$$\delta\nabla^{i}\left(R^{(4)}g_{ik}\right)= \delta \nabla^{i}[R_{abcd} R^{abcd}g_{ik}]= 4 \delta R^{abcd}\nabla_{c}R_{abkd}$$
$$ =4 \delta \nabla_{c}(R^{abcd}R_{abkd}) -4 \delta R_{abkd}\nabla_{c}R^{abcd}$$
Now happens a considerable simplification, because the first term
of the former equation can be written $4 \delta
\nabla^{c}(R_{abcd}{{R^{ab}}_{k}}^{d})$. Thus, $c$, which is
contracted, can be called $i$, and we can exchange the two pairs
of indices in both Riemann tensors, obtaining : $4 \delta
\nabla^{i}(R_{idab}{R_{k}}^{dab})$. This term is exactly the term
of $\nabla^{i}\Sigma_{ik}$ which corresponds to the first term in
the sum giving $\Sigma_{ik}$. So we find that choosing $\delta = -
\frac{1}{4}$, we eliminate all the terms of (24.4). We are now
left with a very few terms, the first and the third terms on the
right hand side of (24.8), the first term on the right hand side
of (24.12), the last $-4 \delta R_{abkd}\nabla_{c}R^{abcd}$ and
finally the $\epsilon$ and $\eta$ terms of (24.1).
\paragraph{Computation of the $\delta$-term} This term can be
written
$$-4 \delta (\nabla^{c}R_{abcd}){{R^{ab}}_{k}}^{d}= (\nabla_{a}R_{db}-\nabla_{b}R_{ad}){{R^{ab}}_{k}}^{d}$$
using the value of $\delta$ and also formula (24.2). The second
term on the right hand side of this formula is equal to the first,
because in the Riemann tensor, $a$ and $b$ appear in antisymmetric
positions, and we are left with :
$$2(\nabla_{a}R_{bd}){{R^{ab}}_{k}}^{d}=2(\nabla_{a}R_{bd}){R_{k}}^{dab}=2(\nabla_{c}R_{ab}){R_{k}}^{bca}$$
Indeed, we obtain the first equality by exchanging the pairs of
indices in the Riemann tensor, and the second by renaming
contracted indices. We use formula (24.10) to write
${R_{k}}^{bca}=-{R_{k}}^{abc}-{R_{k}}^{cab}$, and we observe that
the second term has $a$ and $b$ in antisymmetric positions, which
gives zero because these indices are contracted with
$\nabla_{c}R_{ab}$. So the calculation of the $\delta$-term of
$\nabla^{i}\Sigma_{ik}$ is finished, and gives us only
$-2(\nabla_{c}R_{ab}){R_{k}}^{abc}$, this term vanishing with the
third term of (24.8) if and only if $\alpha = -2$. Comparing this
result with the other relations obtained for $\beta$ and $\gamma$,
we now find $\beta=-2$, and $\gamma=+1$.
\paragraph{The $\epsilon$-term} We are now ready to study the
$\epsilon$-term. We know that we still have to cancel the first
term of (24.8) and the first term of (24.12).
$$\epsilon \nabla^{i}[R_{ab} R^{ab}g_{ik}] = 2 \epsilon
(\nabla_{k}R_{ab}) R^{ab}$$ cancels directly the first term of
(24.8) provided $2\epsilon=-\alpha$, so $\epsilon=+1$.
\paragraph{The $\eta$-term} The $\eta$-term gives $\eta
\nabla^{i}[R^{2}g_{ik}]=2\eta R(\partial_{k}R)$, cancelling
 the first term of (24.12) provided $2\eta=-\frac{1}{2}\gamma$, which
leads to $\eta = - \frac{1}{4}$, providing us finally a set of
constants for which $\nabla^{i}\Sigma_{ik}=0$. Finally we proved
the statement of existence in the following theorem :
\paragraph{Theorem :}There exists a unique tensor $\Sigma_{ik}$,
constructed from all possible products of degree two of the
Riemann tensor, its contractions, and the metric tensor, which
verifies the law of conservation of energy :
$\nabla^{i}\Sigma_{ik}=0$. This tensor contains effectively all
possible products and has the form :
\begin{equation}
\Sigma_{ik}={R_{i}}^{abc}R_{kabc}- 2 R_{iakb}R^{ab} - 2
R_{ia}{R_{k}}^{a} + R_{ik}R - \frac{1}{4}\left( R^{(4)} -4
R^{(2)}+ R^{2}\right)g_{ik}
\end{equation}
where $R^{(4)}=R^{abcd}R_{abcd}$ and $R^{(2)}=R^{ab}R_{ab}$
\paragraph{Existence} As we said, the existence in the theorem has
been proved before, we just notice that we used for this proof all
identities we know concerning the Riemann tensor and its
contractions.
\paragraph{Uniqueness}Of course, we have also proved that there was
no more possible products which could be ingredients of the tensor
$\Sigma_{ik}$, and that our coefficients formed the complete set
of degrees of freedom of our mathematical problem. Finding these
coefficients has been possible because we could cancel all terms
in $\nabla^{i}\Sigma_{ik}=0$, using the well known relations on
the Riemann tensor. It appears that, as there does not exist any
such other relation on this tensor, available in the generic
situation, this was the unique manner of cancelling these terms,
and that the coefficients of $\Sigma_{ik}$ are unique. Here, we
give a method to obtain an explicit proof of the uniqueness of
$\Sigma_{ik}$ : starting with the value of $\Sigma_{ik}$ with all
its coefficients, at first undetermined, we compute
$\nabla^{i}\Sigma_{ik}$ in different explicit choices of the
metric $g_{ik}$, and each example gives us a linear combination of
our coefficients, that we put equal to zero. So we find a linear
system for these coefficients and with enough choices of different
$g_{ik}$, we obtain enough equations, to prove finally that only
the coefficients of the theorem give zero in the generic
situation.
\section{Higher dimensions, topology and complex gravity}
\subsection{The conjecture in higher dimensions}
From what has been done in the case of degree two, it is easily
guessed what can be done as well in the case of degree $n$. We can
consider a tensor $\Sigma_{ik}$, of degree $n$ in the Riemann
tensor and its contractions, and first find all possible products
of degree $n$ that could appear in $\Sigma_{ik}$. Then, we find
all coefficients by imposing that in $\nabla^{i}\Sigma_{ik}$, all
terms vanish. Looking at the case $n=1$ and $n=2$, it should be
clear that it is a way of proving the following conjecture :
\paragraph{Conjecture :}There is a unique tensor $\Sigma_{ik}$
constructed from all possible products of degree $n$ in the
Riemann tensor and its contractions, constructed with the metric
tensor too, and which verifies the law of conservation of energy :
$\nabla^{i}\Sigma_{ik}=0$. This tensor has the form :$$\Sigma_{ik}
= {\tilde{\Sigma}}_{ik}-\frac{1}{2n}{\tilde{\Sigma}}g_{ik}$$ where
$\tilde{\Sigma}$ is the Euler form in dimension $2n$, as well as
the trace of ${\tilde{\Sigma}}_{ik}$. Furthermore, $\Sigma_{ik}$
vanishes, becomes it comes, using the calculus of variation, from
the topological Euler lagrangian.
\subsection{Topology} The appearance in the tensor of degree $2$ of the
Gauss-Bonnet term
$$\tilde{\Sigma} = R^{(4)} -4 R^{(2)}+ R^{2}$$
authorizes us to conjecture that our tensor completely vanishes in
dimensions four, because it comes from the topological
Gauss-Bonnet action :
$$\int \sqrt{-g}(R^{(4)} -4 R^{(2)}+R^{2})$$
In dimensions different from four, the same tensor, of degree two,
comes from the would-be-a-Gauss-Bonnet action :
$$\int \sqrt{-g}(R^{(4)} -4 R^{(2)}+ R^{2})$$
We thus have found an interesting method to write, from an a
priori trivial topological action, a non trivial equation : start
from the topological action in dimension $n$, go to another
dimension where the same action is not trivial anymore, and use
the calculus of variation to extract the tensorial equation. Then,
take the tensor, and go back to the critical dimension. The
question is : does the tensor obtained in this way should be
discarded as being trivial or is it relevant to describe some kind
of physics? This has been the first route which we used to find
our equation of quantum gravity. Because the gravitational tensor
of degree $2$ first displays a dimensionless coupling constant,
and second fits so well with the energy-momentum tensors of the
other interactions, even if it is identically zero, we though
there should be some kind of physics behind. We finally retained
the idea of keeping only its trace in the equation, which gave the
$\Lambda$ term, the law of conservation of energy being in the
equation of quantum gravity being provided by the variations of
$\kappa(\epsilon)$.
\paragraph{A dimensionless coupling constant}
Forgetting that our tensor vanishes for one moment, we consider
the equation that such a tensor would give :
$$\Sigma_{ik} = \kappa T_{ik}$$ To determine the dimension of the
coupling constant, we look at :
$$\Sigma_{0}^{0} = \kappa T_{0}^{0}$$
Here the Riemann and Ricci tensors, when containing the same
number of up and down indices, as well as the scalar curvature,
have dimension $[L]^{-2}$, where $[L]$ is a length. So,
$\Sigma_{0}^{0}$ has dimension $[L]^{-4}$. Now, $T_{0}^{0}$ equals
$\epsilon$, the energy density of matter, and has dimension, in
dimension $D=4$, $[E][L]^{-3}\sim\hbar[T]^{-1}[L]^{-3}\sim\hbar
c[L]^{-4}$, since energy $[E]$ has dimension $\hbar[T]^{-1}$ and
where of course $[T]$ is a time. Comparing these two results, we
see that
$$ \kappa = \frac{\kappa_{0}}{\bar{h}c}$$
where $\kappa_{0}$ has no dimension at all.
 \subsection{Complex gravity} We know that our tensor $\Sigma_{ik}$
probably vanishes because it is the energy-momentum tensor coming
from a topological lagrangian by the calculus of variations, but
there is another form of this tensor, which at least at first
sight, is not necessarily trivial, and which could prove itself
very interesting. Because it is of second order in the curvature
tensor, $\Sigma_{ik}$ possesses a natural extension to complex
gravity. As in the quantum tensors describing particles of
different spin, we can write down a tensor of degree two by
doubling the curvature terms by complex conjugates. By inspection
of these known quantum tensors, we guess easily the procedure to
follow. Indeed, we pose as new fundamental variables, the complex
metric $g_{ik}$ verifying the condition :
\begin{equation}
 g^{*}_{ki}=g_{ik}
\end{equation}
where $z^{*}$ corresponds to the complex conjugate of $z$, and we
pose the complex tensor :
$$\Sigma_{ik}=\frac{1}{2}{{R^{*}}_{i}}^{abc}R_{kabc}+\frac{1}{2}{{R}_{i}}^{abc}R^{*}_{kabc}-{R^{*}}_{iakb}R^{ab}-
R_{iakb}{R^{*}}^{ab}-{R^{*}}_{ia}{R_{k}}^{a}-{R}_{ia}{R^{*}_{k}}^{a}$$
\begin{equation}
+\frac{1}{2}{R^{*}}_{ik}R +\frac{1}{2}R_{ik}{R^{*}}
 - \frac{1}{4}\left(R^{(4)} -4 R^{(2)}+ R{R^{*}}\right)g_{ik}
\end{equation}
where $R^{(4)}={R^{*}}^{abcd}R_{abcd}$ and $R^{(2)} =
{R^{*}}^{ab}R_{ab}$. Equations (25.1) and (25.2) should normally
imply that $\Sigma_{ik}$ is a real symmetric tensor which verifies
the condition of conservation of energy.
\part{The Gauss-Bonnet term}
\section{Introduction}
In Part VI, we have seen how the tensor $\Sigma_{ik}$, in the case
of real gravity, which, independently of the fact that it
vanishes, can be formally deduced from the conditions that it is
of degree two in the Riemann tensor, and that it satisfies the law
of conservation of energy $\nabla^{i}\Sigma_{ik}=0$. We then saw
that this tensor contains automatically in its trace the
Gauss-Bonnet factor
\begin{equation}
\tilde{\Sigma} = R^{(4)} -4 R^{(2)}+ R^{2}
\end{equation}
where $R^{(4)}=R^{abcd}R_{abcd}$ and $R^{(2)} =R^{ab}R_{ab}$. The
precise form of this tensor is :
\begin{equation}
\Sigma_{ik}={\widetilde{\Sigma}}_{ik}-\frac{1}{4}\widetilde{\Sigma}g_{ik}
\end{equation}
where :
\begin{equation}
{\tilde{\Sigma}}_{ik}={R_{i}}^{abc}R_{kabc}- 2 R_{iakb}R^{ab} - 2
R_{ia}{R_{k}}^{a} + R_{ik}R
\end{equation}
and where we note $\tilde{\Sigma}$ for the trace of
${\tilde{\Sigma}}_{ik}$. In this Part, we effectuate the complete
computation of $\tilde{\Sigma}_{ik}$, in the case of the
Robertson-Walker metric, in order to find $\tilde{\Sigma}$, an
expression that we need because it appears in the quantum equation
of gravity. In doing this, we will as a check verify the identity
:
\begin{equation}
\Sigma_{ik}=0
\end{equation}
\section{The metric}
\subsection{Introduction}
So we compute $\tilde{\Sigma}_{ik}$ in the case of the homogeneous
and isotropic case, and for this we still take a metric of
signature $(+1,-1,-1,-1)$ and indices going from $0$ to $3$. We
note with greek indices $\alpha$, $\beta, \gamma, \delta...$ space
indices going from $1$ to $3$. So we have
\begin{equation}
\eta _{00} = +1 ;  \eta _{0 \alpha} = 0 ;  \eta _{\alpha \beta} =
- \delta _{\alpha \beta}
\end{equation}
where $\eta_{ik}$ is Minkowski metric in four dimensions. We then
derive the Robertson-Walker metric, following Landau, $[33]$,
paragraph 111, the computations of this derivation being necessary
to compute all Riemann components, as it is necessary in order to
obtain $\tilde{\Sigma}_{ik}$.
\subsection{The spatial part of the calculation : the closed model}
We first stick to the closed model, because there are simple
relations to deduce the formulas of the open model from this
particular case. Homogeneity and isotropy of space imply that the
scalar curvature in three dimensions is constant, in the
three-space variables. In fact, the Riemann tensor in three
dimensions has enough symmetries to be computed :
\begin{equation}
P_{\alpha \beta \gamma \delta} = \lambda \left(g_{\alpha \gamma}
g_{\beta \delta} - g_{\alpha \delta} g_{\beta \gamma}\right)
\end{equation}
and in the closed model, we choose this constant $\lambda$ to be
positive. To be very precise the metric tensor appearing in the
last equation should be the euclidian metric tensor of the space
of dimension three, which is the opposite of $g_{ik}$, because
here the restriction of the signature of the space-time of
dimension four is $-1, -1,-1$. Thus, the restriction of the metric
of dimension four on the space of dimension three is the opposite
of the euclidian metric in dimension three. However, as the
components of $g_{\alpha \beta}$ appear multiplied in pairs in
(27.2), this equation is still correct. We need to explain this in
detail, because if this subtlety does not matter for the usual
calculation of the Ricci tensor, it matters here a lot, because it
can make appear extra signs in $\tilde{\Sigma}_{ik}$, in case the
calculation would not be done with care. Exactly, the three
dimensional euclidian metric is
\begin{equation}\gamma_{\alpha\beta}=-g_{\alpha\beta}\end{equation}
and the Riemann tensor in three dimensions is :
\begin{equation}P_{\alpha\beta\gamma\delta}=\lambda\left(\gamma_{\alpha\gamma}\gamma_{\beta\delta}-\gamma_{\alpha\delta}\gamma_{\beta\gamma}\right)\end{equation}
If we now take the Ricci tensor in three space by contracting this
Riemann tensor, we obtain :
\begin{equation}P_{\alpha\beta}=2\lambda\gamma_{\alpha\beta}\end{equation}
This equation, as we said, makes appear an extra sign, when
$g_{\alpha\beta}$ is used instead of $\gamma_{\alpha\beta}$,
because then :
\begin{equation}P_{\alpha\beta}=-2\lambda g_{\alpha\beta}\end{equation}
Finally, the scalar curvature is obtained and its value is :
\begin{equation}P=6\lambda\end{equation}
Now isotropy implies that $g_{0 \alpha} = 0$ otherwise the vector
field $g_{0 \alpha} \neq 0$ would introduce by itself a space
anisotropy. Imposing $g_{00} = 1$ means that we choose the time $
t = \frac{x^{0}}{c}$ in our equations to be the physical time,
that is to say the time showed by physical free falling clocks. We
note $a(t)$ the inverse of the square root of the scalar curvature
in three dimensions, which can be interpreted as the radius of the
universe, we do not note $R$ for this radius to avoid the
confusion with the scalar curvature in four dimensions. Anyway we
then have the relation :
\begin{equation}
\lambda = \frac{1}{a^{2}}
\end{equation}
as can be seen for example in $[33]$. Now, using spherical
coordinates in four dimensions, and choosing a frame which moves,
at every point of space-time, with the physical free falling
matter we find the value of $ds^{2}$ :
\begin{equation}
ds^{2} = c^{2}dt^{2} - a^{2}(t) (d \chi ^{2} + \sin^{2} \chi (d
\theta^{2} + \sin^{2}\theta d\phi^{2}))
\end{equation}
where $r, \theta, \phi$ are the variables of spherical coordinates
in three dimensions and where $r = a(t)\sin\chi$, $\chi$ varying
from $0$ to $\pi$. Further, we can replace the time variable $t$
by the dimensionless variable $\eta$, defined by $cdt = a d\eta$.
We then obtains :
\begin{equation}
ds^{2} = a^{2}(\eta) (d\eta^{2} - d \chi ^{2} - \sin^{2} \chi (d
\theta^{2} + \sin^{2}\theta d\phi^{2})).
\end{equation}
So we write our equations with variables $x^{0}, x^{1}, x^{2},
x^{3}$ being $\eta, \chi, \phi, \theta$. We have from the previous
equation :
\begin{equation}
g_{00} = a^{2};  g_{11}=- a^{2};  g_{22}= - a^{2} \sin^{2}\chi;
g_{33}= -a^{2}\sin^{2}\chi \sin^{2}\theta
\end{equation}
and all non diagonal terms of $g_{ik}$ vanish, such that the
inverse matrix $g^{ik}$ is straightforward.
\subsection{Closed and open models}
To go from the closed to the open model, we have to apply the
following replacements : $a\rightarrow ia$ ; $\eta\rightarrow
i\eta$ ; $\chi\rightarrow i\chi$. Finally $t\rightarrow-t$ can be
deduced from the former rules, and from the relation $cdt=ad\eta$.
As an example, the metric for the closed model transforms, in the
case of the open model, into :
\begin{equation}
ds^{2} = a^{2}(\eta) (d\eta^{2} - d \chi ^{2} - \sinh^{2} \chi (d
\theta^{2} + \sin^{2}\theta d\phi^{2}))
\end{equation}
In detail, we see that $t\rightarrow-t$ and $dt^{2}$ is invariant.
$a^{2}\rightarrow-a^{2}$, $d\chi^{2}\rightarrow-d\chi^{2}$, and
finally $\sin^{2}\chi\rightarrow-\sinh^{2}\chi$, which establishes
the form of the metric in the open case, from the metric in the
closed case. As for $\lambda$, the relation $\lambda=1/a^{2}$
transforms to
\begin{equation}\lambda=\frac{K}{a^{2}}\end{equation} where $K$
takes the values : $K=+1$ in the closed case, and $K=-1$ in the
open case.
\subsection{The Christoffel symbols}
We first stick to the closed model, at the end of the calculation
we shall deduce, from these results, the formulas for the open
model. We use primes to note the $\eta$-derivation and dots to
note the $t$-derivation. We use the general formula :
\begin{equation}
\Gamma^{i}_{jk} = \frac{1}{2} g^{il} (\partial_{j} g_{lk} +
\partial_{k} g_{jl} - \partial_{l} g_{jk})
\end{equation}
We recall here that latin indices are four dimensional indices and
greek indices are three dimensional ones. In the case of
Einstein's equations, the computation is easier because one only
need the Ricci tensor. To compute $\tilde{\Sigma}_{ik}$, we need
all values of the Riemann tensor, and we do have to compute the
values of $\Gamma^{i}_{jk}$ with care. What makes the computation
easier, is that both $g_{ik}$ and $g^{ik}$ are diagonal. We
compute :
\begin{equation}
\Gamma^{0}_{00} = \frac{1}{2} g^{00} (\partial_{0} g_{00}) =
\frac{1}{2} \frac{1}{a^{2}} 2 aa' = \frac{a'}{a}
\end{equation}
and
\begin{equation}
\Gamma^{0}_{\alpha
\beta}=\frac{1}{2}\frac{1}{a^{2}}(\partial_{\alpha} g_{0
\beta}+\partial_{\beta} g_{0 \alpha}-\partial_{0}g_{\alpha \beta})
=-\frac{a'}{a^{3}} g_{\alpha \beta}
\end{equation}
We further have :
\begin{equation}
\Gamma^{\alpha}_{0 \beta} = \frac{1}{2} g^{\alpha \alpha}
(\partial_{0} g_{\alpha \beta} + \partial_{\beta} g_{0 \alpha} -
\partial_{\alpha} g_{0 \beta}) = \frac{1}{2} g^{\alpha \alpha}
(\partial_{0} g_{\alpha \beta}) = \frac{1}{2} g^{\alpha \alpha}
\frac{2a'}{a} g_{\alpha \beta} = \frac{a'}{a}
\delta_{\beta}^{\alpha}
\end{equation}
As well can we see that :
\begin{equation}
\Gamma^{0}_{0 \alpha} = \Gamma^{\alpha}_{00}= 0
\end{equation}
We will see that we do not need the components
$\Gamma^{\alpha}_{\beta \gamma}$.
\section{The Riemann tensor}
Next, we compute all Riemann tensor components, using the
classical formula :
\begin{equation}
R_{iklm}=\frac{1}{2}\left[\partial_{kl}^{2}
g_{im}+\partial_{im}^{2} g_{kl} - \partial_{il}^{2} g_{km} -
\partial_{km}^{2} g_{il}\right] + g_{np}\left(\Gamma^{n}_{kl}
\Gamma^{p}_{im} - \Gamma^{n}_{km} \Gamma^{p}_{il}\right)
\end{equation}
We separate cases, according to the number of space indices a
component of a tensor possesses. For example, we say that
$R_{\alpha \beta \gamma \delta}$ is a four space indices
component, since it does not possess any time index. As another
example, $R_{\alpha 0 \beta \gamma}$ is named a three space
indices component. We need to precise that the indices are only
counted when they all are down.
\subsection{Riemann : four space indices}
Now, if we compute $R_{\alpha \beta \gamma \delta}$ using the
previous formula, we first find that the Riemann tensor contains
only odd products of the metric tensor, and changing the metric
into its opposite transforms the Riemann tensor into its opposite.
Second, the previous formula contains a sum over indices $p$ and
$n$, which means, for both indices, a sum over the three space
indices and also over the $0$ time index. If we put together all
terms from the summation over $n$ and $p$, only when $n$ an $p$
vary over space indices, as well as the first four terms of
equation (28.1), we get minus the Riemann tensor in three
dimensions. Concerning the sum over $p$ or $n$ when one of them
equals $0$, taken into account that $g_{0 \alpha}= 0$, we only
obtain the term corresponding to the case $n=p=0$ :
$$R_{\alpha \beta \gamma \delta} = -  P_{\alpha \beta \gamma
\delta} + g_{00}(\Gamma^{0}_{\beta \gamma} \Gamma^{0}_{\alpha
\delta} - \Gamma^{0}_{\beta \delta} \Gamma^{0}_{\alpha \gamma})$$
$$ = - \frac{1}{a^{2}} \left(g_{\alpha \gamma} g_{\beta \delta} -
g_{\alpha \delta} g_{\beta \gamma}\right) + a^{2} \left[\left((-
\frac{a'}{a^{3}})g_{\beta \gamma}(- \frac{a'}{a^{3}})g_{\alpha
\delta}\right) - \left((- \frac{a'}{a^{3}}) g_{\beta \delta}(-
\frac{a'}{a^{3}})g_{\alpha \gamma}\right)\right]$$ So finally :
\begin{equation}
R_{\alpha \beta \gamma \delta} =
\frac{a^{2}+a'^{2}}{a^{4}}(g_{\beta \gamma} g_{\alpha \delta} -
g_{\beta \delta} g_{\alpha \gamma})
\end{equation}
\subsection{Riemann : three space indices}
Now we compute $ R_{\alpha \beta \gamma 0}$ : using (28.1) and the
fact the metric tensor is diagonal we find : $$R_{\alpha \beta
\gamma 0} = \frac{1}{2} \left(\partial_{\alpha 0}^{2} g_{\beta
\gamma} - \partial_{\beta 0}^{2} g_{\gamma \delta}\right) +
g_{np}\left(\Gamma^{n}_{\beta \gamma} \Gamma^{p}_{\alpha 0} -
\Gamma^{n}_{\beta 0} \Gamma^{p}_{\alpha \gamma}\right)$$ If we try
to sum the terms over $n$ and $p$, we see that the term
corresponding to $n=p=0$ contains only products containing one
$\Gamma^{0}_{\alpha 0} = 0$, and thus vanishes. We are left with a
sum over $\lambda$ and $\mu$ :
$$g_{\lambda \mu}(\Gamma^{\lambda}_{\beta \gamma}
\Gamma^{\mu}_{\alpha 0} - \Gamma^{\lambda}_{\beta 0}
\Gamma^{\mu}_{\alpha \gamma}) = \frac {a'}{a} \left(g_{\lambda
\alpha} \Gamma^{\lambda}_{\beta \gamma} - g_{\beta \mu}
\Gamma^{\mu}_{\alpha \gamma}\right) = \frac {a'}{a} \left(
\Gamma_{\alpha \beta \gamma} - \Gamma_{\beta \alpha
\gamma}\right)$$ the first equality coming from
$$\Gamma^{\mu}_{\alpha 0} = \frac{a'}{a} \delta^{\mu}_{\alpha}$$ So
we can use, because $g_{\alpha 0}=0$ :
$$\Gamma_{\alpha\beta\gamma}=g_{\alpha n}\Gamma^{n}_{\beta\gamma}
=g_{\alpha\lambda}\Gamma^{\lambda}_{\beta\gamma}=\frac{1}{2}
(\partial_{\gamma}g_{\alpha \beta} +
\partial_{\beta}g_{\alpha \gamma} - \partial_{\alpha}g_{\beta
\gamma})$$ and collecting all terms, without forgetting the two
derivatives of the metric tensor : $$R_{\alpha \beta \gamma 0} =
\frac{1}{2} \left(\partial_{\alpha 0}^{2} g_{\beta \gamma} -
\partial_{\beta 0}^{2} g_{\gamma \delta}\right) + \frac{a'}{a}
\left(\partial_{\beta }g_{\gamma \delta} -
\partial_{\alpha} g_{\beta \gamma}\right)$$
We notice that if we write $g_{\alpha \beta} =
a^{2}\tilde{g}_{\alpha \beta}$, then $\tilde{g}_{\alpha \beta}$
does not depend on $\eta = x^{0}$, and we can write
$$\partial_{0}g_{\alpha \gamma} = 2 \frac{a'}{a}g_{\alpha
\gamma}$$ We thus obtain
$$\frac{1}{2}\partial^{2}_{\alpha 0}g_{\beta \gamma} = \frac{a'}{a}\partial_{\alpha}g_{\beta \gamma}$$
because $a=a(\eta)$ does not depend on any space variable. We
finally obtain :
\begin{equation}
R_{\alpha \beta \gamma 0} = 0
\end{equation}
\subsection{Riemann : two space indices}
We have now to compute $R_{\alpha 0 \beta 0}$, which actually are
the last components of the Riemann tensor which may not vanish.
Indeed, any component of the Riemann tensor containing three or
more indices equal to zero vanishes, because of the antisymmetry
of the two first indices, and also because of the antisymmetry of
the two last indices. Using again the general formula (28.1) for
the Riemann tensor, we write :
$$R_{\alpha 0 \beta 0}
=\frac{1}{2}\left(-\partial_{00}^{2} g_{\alpha \beta} -
\partial_{\alpha \beta}^{2}g_{00}\right) +
g_{np}\left(\Gamma^{n}_{0 \beta} \Gamma^{p}_{\alpha 0} -
\Gamma^{n}_{00} \Gamma^{p}_{\alpha \beta}\right)$$ and since
$$\partial_{0}g_{\alpha \beta} = 2 \frac{a'}{a}g_{\alpha \beta}$$ we
compute : $$\partial_{00}^{2}g_{\alpha \beta}
=\left[\left(2\frac{\dot{a}}{a}\right)'+4\frac{a'^{2}}{a^{2}}\right]g_{\alpha
\beta}=
2\left(\frac{\ddot{a}a-a'^{2}}{a^{2}}+2\frac{a'^{2}}{a^{2}}\right)g_{\alpha
\beta}$$ $$= 2\left( \frac{aa"+a'^{2}}{a^{2}}\right)g_{\alpha
\beta}$$ We also have :
$$\partial_{\alpha \beta}^{2}g_{00} = 0$$
because $\partial_{\alpha}a=0$, since $a=a(\eta)$ does not depend
on any space variable. The first term of
$$g_{00}(\Gamma^{0}_{0 \beta} \Gamma^{0}_{\alpha 0} -
\Gamma^{0}_{00} \Gamma^{0}_{\alpha \beta})$$ vanishes, because
$\Gamma^{0}_{\alpha 0}=0$. Using the values of $\Gamma^{i}_{jk}$,
we find that its second term, taking into account the minus sign,
equals :
$${\left(\frac{a'}{a}\right)}^{2} g_{\alpha \beta}$$
Finally, the second term of : $$g_{\lambda
\mu}\left(\Gamma^{\lambda}_{0 \beta} \Gamma^{\mu}_{\alpha 0} -
\Gamma^{\lambda}_{00} \Gamma^{\mu}_{\alpha \beta}\right)$$
vanishes, because $\Gamma^{\lambda}_{00}=0$, whereas its first
term equals :
$$g_{\lambda \mu} \left(\frac{a'}{a}\right)
\delta_{\beta}^{\lambda}\left(\frac{a'}{a}\right)
\delta_{\alpha}^{\mu} = {\left(\frac{a'}{a}\right)}^{2} g_{\alpha
\beta}$$ Collecting all terms, we obtain :
\begin{equation}
R_{\alpha 0 \beta 0} = \frac{a'^{2} - aa"}{a^{2}} g_{\alpha \beta}
\end{equation}
The Riemann tensor completely computed, we can now contract
indices to find the Ricci tensor, the scalar curvature, and
finally $\Sigma_{ik}$.
\section{The Ricci tensor}
\subsection{Two time indices}
We recall that we count indices in tensors when they are all down,
according to the exact number of time indices $0$ which appear, or
equivalently according to the exact number of space indices which
appear. For the Ricci tensor, there is only one component with two
time indices, which is $R_{00}$. From
$${R^{\alpha}}_{0 \beta 0} = \left(\frac{a'^{2}-aa"}{a^{2}}\right) \delta^{\alpha}_{\beta}$$
we compute
\begin{equation}
R_{00} = {R^{\alpha}}_{0 \alpha 0} = \left(\frac{a'^{2} -
aa"}{a^{2}}\right) \delta^{\alpha}_{\alpha} = 3\left( \frac{a'^{2}
- aa"}{a^{2}}\right)
\end{equation}
and
\begin{equation}
R_{0}^{0} = g^{00}R_{00} = \frac{3 (a'^{2} - aa")}{a^{4}} = b
\end{equation}
which defines $b$.
\subsection{One space index}
We notice that there is no component of the Riemann tensor
containing an odd number of indices equal to $0$, that is to say
time indices. We thus compute :
\begin{equation}
R_{\alpha 0}={R^{n}}_{\alpha n 0}= {R^{\gamma}}_{\alpha \gamma 0}
+ {R^{0}}_{\alpha 0 0} = 0
\end{equation}
\subsection{Two space indices and scalar curvature}
Finally, from
$$ {R^{\alpha}}_{\beta \gamma \delta} =
\left(\frac{a^{2}+a'^{2}}{a^{4}}\right)\left(\delta^{\alpha}_{
\delta} g_{\beta \gamma} - \delta^{\alpha}_{\gamma} g_{\beta
\delta}\right)$$ we find $$ R_{\beta \delta} ={R^{n}}_{\beta n
\delta}= {R^{\gamma}}_{\beta \gamma \delta} + {R^{0}}_{\beta 0
\delta} = \left(\frac{a^{2}+a'^{2}}{a^{4}}\right)( g_{\beta
\delta} - 3 g_{\beta \delta}) + g^{00}R_{\beta 0 \delta 0}$$
$$=\left(-\frac{2}{a^{4}}(a^{2}+a'^{2})+\frac{1}{a^{4}}(a'^{2}-aa")\right)g_{\beta\gamma}$$
We find :
\begin{equation}
R_{\beta \delta} = - \frac{1}{a^{4}} \left(2 a^{2} + a'^{2}
+aa"\right) g_{\beta \delta} = c g_{\beta \delta}
\end{equation}
which defines $c$. We then have the scalar curvature :
\begin{equation}
R = b + 3 c = - \frac{6}{a^{3}}(a+a")=d
\end{equation}
which defines $d$. We are ready now to compute the tensor
$\Sigma_{ik}$ using all our previous calculations.
\section{The Gauss-Bonnet tensor}
\subsection{Ricci-times-scalar curvature and Ricci-times-Ricci}
We start with $R_{ik}R$ and $R_{ia}{R_{k}}^{a}$. From equations
$g_{00}=a^{2}$ of (27.11), from (29.1), (29.2), (29.4) and (29.5),
we obtain :
\begin{equation}
R_{\alpha \beta } R = cd g_{\alpha \beta } = c(b+3c) g_{\alpha
\beta}
\end{equation}
and
\begin{equation}
R_{00} R = bd g_{00} = b(b+3c) g_{00}
\end{equation}
and also from (29.3) :
\begin{equation}
R_{\alpha 0} R =0
\end{equation}
 We have from (29.3), $R_{\alpha 0}=0$ :
\begin{equation}
R_{\alpha a}{R_{\beta}}^{a}=R_{\alpha
\gamma}{R_{\beta}}^{\gamma}+R_{\alpha
0}{R_{\beta}}^{0}=c^{2}g_{\alpha \gamma }
\delta_{\beta}^{\gamma}=c^{2}g_{\alpha \beta}
\end{equation}
and
\begin{equation}
R_{\alpha a}R_{0}^{a}=R_{\alpha 0}R_{0}^{0}+ R_{\alpha
\gamma}R_{0}^{\gamma}=0
\end{equation}
We also have :
\begin{equation}
R_{0a}R_{0}^{a}=R_{00}R_{0}^{0}+R_{0\gamma}R_{0}^{\gamma}=b^{2}g_{00}
\end{equation}
\subsection{Riemann-times-Ricci}
We compute then $R_{iakb}R^{ab}$. In $R_{\alpha a0b}R^{ab}$, we
know that $R^{ab}=0$ if an odd number of the indices among $a$ and
$b$ are $0$, and $R_{\alpha a0b}$ is $0$ if an odd number of
indices among $\alpha, a, 0$ and $b$ are zero, which means a even
number of indices among $a$ and $b$ are $0$. So all terms cancel,
and :
\begin{equation}
R_{\alpha a0b}R^{ab}=0
\end{equation}
Furthermore, we have :
\begin{equation}
R_{\alpha a \beta b}R^{ab}=R_{\alpha 0 \beta 0}R^{00}+R_{\alpha
\gamma \beta \delta}R^{\gamma \delta} =b{R_{\alpha 0
\beta}}^{0}+c{R_{\alpha \gamma \beta}}^{\gamma}
\end{equation}
The first equality results from (29.3), and the second from
(29.2), (29.3) and (29.4), and also because :
$$R_{\alpha\gamma\beta\delta}R^{\gamma\delta}=cR_{\alpha\gamma\beta\delta}g^{\gamma\delta}=R_{\alpha\gamma\beta n}g^{\gamma
n}$$ since $g^{\gamma 0}=0$. Now, we have : $${R_{\alpha \gamma
\beta}}^{\gamma}={R_{\alpha a \beta}}^{a}-{R_{\alpha 0
\beta}}^{0}=R_{\alpha \beta}-{R_{\alpha 0 \beta}}^{0}$$ so :
$$R_{\alpha a \beta b}R^{ab}=(b-c)g^{00}R_{\alpha 0 \beta
0}+cR_{\alpha \beta}$$ We find from (28.4) and (29.2) that :
\begin{equation}
R_{\alpha 0 \beta 0}=\frac{ba^{2}}{3}g_{\alpha \beta}
\end{equation}
We also have $g^{00}=\frac{1}{a^{2}}$, and using again (29.4) :
\begin{equation}
R_{\alpha a \beta b}R^{ab}=\frac{b(b-c)}{3}g_{\alpha
\beta}+c^{2}g_{\alpha \beta}=\frac{b^{2}-bc+3c^{2}}{3}g_{\alpha
\beta}
\end{equation}
Now, using (29.3) : $$R_{0a0b}R^{ab}=R_{0000}R^{00}+R_{0 \alpha 0
\beta}R^{\alpha \beta}$$ However, $R_{0000}=0$, since it possesses
its first two indices in antisymmetric positions and equal, so :
$$R_{0a0b}R^{ab}=R_{0 \alpha 0 \beta}R^{\alpha \beta}=cg^{\alpha
\beta}R_{0 \alpha 0 \beta}=cg^{ab}R_{0a0b}=cg^{ab}R_{a0b0}$$ since
$g^{\alpha 0}=g^{0 \beta}=0$, and $R_{0000}=0$. Finally, we obtain
:
\begin{equation}
R_{0a0b}R^{ab}=cR_{00}=bc g_{00}
\end{equation}
from (29.2).
\subsection{Riemann-times-Riemann}
We need also to compute ${R_{i}}^{abc}R_{kabc}$.\\We first have :
\begin{equation}
{R_{\alpha}}^{abc}R_{0abc}=0
\end{equation}
since the first term in the product vanishes for an odd number of
$0$ among the indices $a, b$ and $c$, whereas the second term
vanishes, for an even number of them. In ${R_{0}}^{abc}R_{0abc}$,
we have $a\neq0$ otherwise $R_{0abc}=0$, and we can put
$a=\alpha$. We recall that $R_{ijkl}=0$ for an odd number of $0$
among the indices $i, j, k$ and $l$. Thus, exactly one index
between $b$ and $c$ is $0$, if we impose to $R_{0\alpha bc}$ being
nonzero.
$${R_{0}}^{abc}R_{0abc}={R_{0}}^{\alpha bc}R_{0\alpha bc}={R_{0}}^{\alpha 0 \gamma}R_{0 \alpha 0
\gamma}+{R_{0}}^{\alpha \beta 0}R_{0 \alpha \beta 0}$$ and since
the last two indices in $R_{ijkl}$ are antisymmetric :
$${R_{0}}^{abc}R_{0abc}=2{R_{0}}^{\alpha 0 \gamma}R_{0 \alpha 0
\gamma}=\frac{2b}{3}a^{2}g_{\alpha \gamma}{R_{0}}^{\alpha 0
\gamma}=\frac{2b}{3}a^{2}R_{0}^{0}$$ because
$$R_{0\alpha0\gamma}=R_{\alpha0\gamma0}=\frac{a'^{2}-aa"}{a^{2}}g_{\alpha\gamma}=\frac{ba^{2}}{3}g_{\alpha\gamma}$$
Finally :
\begin{equation}
{R_{0}}^{abc}R_{0abc}=\frac{2b^{2}}{3}a^{2}=\frac{2b^{2}}{3}g_{00}
\end{equation}
Here, we recall that $g_{00}=a^{2}$. We also have to evaluate
${R_{\alpha}}^{abc}R_{\beta abc}$. In this expression, when it
does not vanish, in each term, the number of $0$ among $a,b$ and
$c$ is even. So, this number can only be $0$ or $2$. Furthermore
$b$ and $c$ cannot be equal, so cannot be $0$ at the same time. In
these conditions, when the number of $0$ between $a, b$ and $c$ is
$2$, we must have $a=0$. Because of this, we obtain :
$${R_{\alpha}}^{abc}R_{\beta abc}={R_{\alpha}}^{\gamma \delta
\epsilon}R_{\beta \gamma \delta \epsilon}+
{R_{\alpha}}^{00\gamma}R_{\beta 00\gamma}+{R_{\alpha}}^{0 \gamma
0}R_{\beta 0 \gamma 0}$$
\begin{equation}
={R_{\alpha}}^{\gamma \delta \epsilon}R_{\beta \gamma \delta
\epsilon}+2{R_{\alpha}}^{0 \gamma 0}R_{\beta 0 \gamma 0}
\end{equation}
From (30.9) : $$2{R_{\alpha}}^{0 \gamma 0}R_{\beta 0 \gamma 0}=
\frac{2b}{3}a^{2}g_{\beta \gamma}{R_{\alpha}}^{0 \gamma 0}
=\frac{2b}{3}g_{00}{{{R_{\alpha}}^{0}}_{\beta}}^{0}$$
\begin{equation}
=\frac{2b}{3}{R_{\alpha 0 \beta}}^{0}=\frac{2b}{3}g^{00}R_{\alpha
0 \beta 0}=\frac{2b^{2}}{9}g_{\alpha \beta}
\end{equation}
from (30.9) again. Now we evaluate ${R_{\alpha}}^{\gamma \delta
\epsilon}R_{\beta \gamma \delta \epsilon}$. We have already
computed the term :
$$R_{\alpha \gamma \delta \epsilon}= \tilde{\lambda} \left(
g_{\gamma \delta}g_{\alpha \epsilon}-g_{\gamma \epsilon}g_{\alpha
\delta}\right)$$ where $$\tilde{\lambda} =
\frac{a^{2}+a'^{2}}{a^{4}}$$ We find :
\begin{equation}
{R_{\alpha}}^{\gamma \delta \epsilon}R_{\beta \gamma \delta
\epsilon}= {\widetilde{\lambda}}^{2}\left(g_{\gamma
\delta}g_{\beta \epsilon}-g_{\gamma \epsilon}g_{\beta
\delta}\right) \left(g^{\gamma
\delta}\delta_{\alpha}^{\epsilon}-g^{\gamma
\epsilon}\delta_{\alpha}^{\delta}\right)=4
{\widetilde{\lambda}}^{2}g_{\alpha \beta}
\end{equation}
Indeed, we have :
$$g_{\gamma\delta}g_{\beta\epsilon}g^{\gamma\delta}\delta^{\epsilon}_{\alpha}=g_{\gamma\delta}g^{\gamma\delta}g_{\alpha\beta}
=\delta^{\gamma}_{\gamma}g_{\alpha\beta}=3g_{\alpha\beta}$$ and
$$-g_{\gamma\delta}g_{\beta\epsilon}g^{\gamma\epsilon}\delta^{\delta}_{\alpha}=-g_{\gamma\alpha}g_{\beta\epsilon}g^{\gamma\epsilon}
=-g_{\gamma\alpha}\delta_{\beta}^{\delta}=-g_{\beta\alpha}=-g_{\alpha\beta}$$
We know that
$$\frac{b}{3}=\frac{a'^{2}-aa"}{a^{4}}$$ and also
$$c=-\frac{2a^{2}+a'^{2}+aa"}{a^{4}}$$
We thus can compute :
\begin{equation}\frac{b}{3}-c=\frac{2a^{2}+2a'^{2}}{a^{4}}\end{equation}
 We finally obtain the value of
$\widetilde{\lambda}$ :
\begin{equation}
\widetilde{\lambda}=\frac{b-3c}{6}
\end{equation}
Putting together all these results, we arrive at :
$${R_{\alpha}}^{abc}R_{\beta
abc}=\left(\frac{2b^{2}}{9}+4{\widetilde{\lambda}}^{2}\right)g_{\alpha\beta}=
\left(\frac{2b^{2}}{9}+\frac{1}{9}(b-3c)^{2}\right)g_{\alpha
\beta}$$ so we obtain :
\begin{equation}
{R_{\alpha}}^{abc}R_{\beta
abc}=\left(\frac{b^{2}-2bc+3c^{2}}{3}\right)g_{\alpha \beta}
\end{equation}
\subsection{The Gauss-Bonnet tensor}
We write (26.3) :
$${\tilde{\Sigma}}_{ik}={R_{i}}^{abc}R_{kabc}- 2 R_{iakb}R^{ab}
- 2 R_{ia}{R_{k}}^{a} + R_{ik}R$$ as well as (26.1):
$$\tilde{\Sigma}=R^{(4)} -4 R^{(2)}+ R^{2}$$
where $\tilde{\Sigma}$ is the trace of ${\tilde{\Sigma}}_{ik}$ and
we have (26.2):
$$\Sigma_{ik}={\tilde{\Sigma}}_{ik}-\frac{1}{4}\tilde{\Sigma}
g_{ik}$$
\\So picking up the terms in (30.3),(30.5),(30.7) and (30.12) we obtain :
\begin{equation}
{\tilde{\Sigma}}_{\alpha 0}=0
\end{equation}
Picking up the terms in (30.2), (30.6), (30.11) and (30.13), we
obtain :
\begin{equation}
{\tilde{\Sigma}}_{00}=\left(\frac{2b^{2}}{3}-2bc-2b^{2}+b(b+3c)\right)g_{00}=\frac{b}{3}(3c-b)g_{00}
\end{equation}
Picking up the terms in (30.1), (30.4), (30.10) and (30.19), we
obtain :
$${\tilde{\Sigma}}_{\alpha
\beta}=\left[\left(\frac{b^{2}-2bc+3c^{2}}{3}\right)-2\left(\frac{b^{2}-bc+3c^{2}}{3}\right)-2c^{2}+c(b+3c)\right]g_{\alpha
\beta}$$ So
\begin{equation}
{\tilde{\Sigma}}_{\alpha \beta}=\frac{b}{3}(3c-b)g_{\alpha \beta}
\end{equation}
And finally, we see that ${\tilde{\Sigma}}_{ik}$ is diagonal. Thus
$\Sigma_{ik}$ also is diagonal, being at the same time of
vanishing trace. So $\Sigma_{ik}=0$, as a check of all the
computations of this part.

\vspace{10mm} Email address : cristobal.real@hotmail.fr

 \vspace{10mm}

I first of all thank Odilia BOUTRY, whose so clever contribution
has been essential. I would like to thank my parents for the
values they have transmitted to us. I especially thank my father
for the spirit of philosophy and freedom he transmitted to us, for
having always believed in me, for its encouragements and help. I
thank Ana BEDOUELLE for its years longing encouragements, and for
having always believed in me. I thank Jean-Pierre COHEN and M.
Jean-Fran\c{c}ois LEMOTEUX for their presence and encouragements
along all these years. I thank Gayle BURSTEIN for her help in
regards to the english version.

\end{document}